\documentclass[aps,pra,superscriptaddress,reprint,a4paper,longbibliography,floatfix, footinbib]{revtex4-2}

\usepackage[utf8]{inputenc}
\usepackage[T1]{fontenc}
\usepackage{amsmath,amssymb,amsfonts,amsthm}
\usepackage{mathtools}
\usepackage{pifont}

\usepackage{graphicx}
\usepackage{dcolumn}
\usepackage{bm}
\usepackage{tikz} 
\usepackage{lipsum}
\usepackage{float}
\usepackage{subfigure}  
\usepackage{makecell}
\usepackage{tabularx} 
\usepackage[percent]{overpic} 
\usepackage{multirow}
\usepackage{booktabs}
\usepackage{braket}

\usepackage[normalem]{ulem}

\graphicspath{ {./Figs/} }

\definecolor{myurlcolor}{rgb}{0,0,0.7}
\definecolor{myrefcolor}{rgb}{0.8,0,0}
\usepackage{hyperref}
\hypersetup{
    unicode=true, bookmarksnumbered=false, bookmarksopen=false, breaklinks=false, pdfborder={0 0 0},
    colorlinks=true,
    linkcolor=myrefcolor,citecolor=myurlcolor,urlcolor=myurlcolor
}

\theoremstyle{plain}

\makeatletter

\newcommand{\mcl}[1]{\mathcal{#1}}

\newcommand{\mrm}[1]{\mathrm{#1}}

\newcommand{\trm}[1]{\textrm{#1}}

\newcommand{\dd}{\mrm{d}}

\newcommand{\ii}{\mrm{i}}
\newcommand{\id}{\mathbb{I}} 
\newcommand{\TT}{\mathsf{T}}
\newcommand{\xmark}{\ding{55}}

\usepackage{xparse}
\newcommand{\Tr}{\mrm{Tr}} 
\NewDocumentCommand\tr{g}{%
  \IfNoValueTF{#1}
    {\Tr}
    {\Tr\!\left\{#1\right\}}
}
\newcommand{\E}{\mathbb{E}} 
\NewDocumentCommand\EE{ m g }{%
  \IfNoValueTF{#2}
    {\E\!\left[#1\right]}     
    {\E_{#1}\!\left[#2\right]} 
}

\newcommand{\eref}[1]{(\ref{#1})}
\newcommand{\eqnref}[1]{Eq.~(\ref{#1})}
\newcommand{\eqnsref}[2]{Eqs.~(\ref{#1}-\ref{#2})}
\newcommand{\figref}[1]{Fig.~\ref{#1}}

\newcommand{\tabref}[1]{Table~\ref{#1}}
\newcommand{\secref}[1]{Sec.~\ref{#1}}
\newcommand{\appref}[1]{App.~\ref{#1}}

\newcommand{\citeref}[1]{Ref.~\cite{#1}}
\newcommand{\refcite}[1]{Ref.~\cite{#1}}

\newcommand{\nsamples}{\nu}
\newcommand{\est}[1]{\tilde{#1}}

\renewcommand{\vec}[1]{\bm{#1}}

\newcommand{\estpar}{\varphi}
\newcommand{\contpar}{\mu}
\newcommand{\drivepar}{\Omega}
\newcommand{\estparV}{\vec{\estpar}}
\newcommand{\contparV}{\vec{\contpar}}

\newcommand{\kd}{\kappa^{\mrm{d}}}
\newcommand{\kl}{\kappa^{\mrm{l}}}

\newcommand{\aMSE}{\trm{aMSE}}
\newcommand{\MSE}{\trm{MSE}}
\newcommand{\NLL}{\trm{NLL}}
\newcommand{\CE}{\trm{CE}}

\newcommand{\cS}{\mathcal{S}}
\newcommand{\cT}{\mathcal{T}}

\newcommand{\FI}{\mathbb{F}}

\definecolor{lime}{HTML}{A6CE39}
\DeclareRobustCommand{\orcidicon}{
	\begin{tikzpicture}
	\draw[lime, fill=lime] (0,0) 
	circle [radius=0.16] 
	node[white] {{\fontfamily{qag}\selectfont \tiny ID}};
	\draw[white, fill=white] (-0.0625,0.095) 
	circle [radius=0.007];
	\end{tikzpicture}
	\hspace{-2mm}
}
\foreach \x in {A, ..., Z}{\expandafter\xdef\csname orcid\x\endcsname{\noexpand\href{https://orcid.org/\csname orcidauthor\x\endcsname}
			{\noexpand\orcidicon}}
}
\makeatother

\begin{document}

\title{Unlocking photodetection for quantum sensing with Bayesian likelihood-free methods and deep learning}

\author{Mateusz Molenda}
\email{mmolenda@ifpan.edu.pl}
\affiliation{Institute of Physics, Polish Academy of Sciences, Aleja Lotnik\'{o}w 32/46, 02-668 Warsaw, Poland.}

\author{Lewis A. Clark\orcidB{}}
\affiliation{Quantum Innovation Centre (Q.InC), Agency for Science Technology and Research (A*STAR), 4 Fusionopolis Way, Kinesis \#05-01, Singapore 138635, Republic of Singapore}

\author{Marcin P\l{}odzie\'{n}\orcidC{}}
\affiliation{Qilimanjaro Quantum Tech, Carrer de Veneçuela 74, 08019
Barcelona, Spain
}

\author{Jan Ko\l{}ody\'{n}ski\orcidA{}}
\email{jankolo@ifpan.edu.pl}
\affiliation{Institute of Physics, Polish Academy of Sciences, Aleja Lotnik\'{o}w 32/46, 02-668 Warsaw, Poland.}


\date{\today}

\begin{abstract}
To operate quantum sensors at their quantum limit in real time, it is crucial to identify efficient data inference tools for rapid parameter estimation. In photodetection, the key challenge is the fast interpretation of click-patterns that exhibit non-classical statistics-the very features responsible for the quantum enhancement of precision. We achieve this goal by comparing Bayesian likelihood-free methods with ones based on deep learning (DL). While the former are more conceptually intuitive, the latter, once trained, provide significantly faster estimates with comparable precision and yield similar predictions of the associated errors, challenging a common misconception that DL lacks such capabilities. We first verify both approaches for an analytically tractable, yet multiparameter, scenario of a two-level system emitting uncorrelated photons. Our main result, however, is the application to a driven non-linear optomechanical device emitting non-classical light with complex multiclick correlations.
Although both methods facilitate inference, only DL provides the execution speed required for real-time performance while naturally learning complex multi-photon click-patterns. Our results pave the way for real-time sensing and dynamical control of quantum devices that leverage non-classical effects in photodetection.
\end{abstract}

\maketitle

\section{Introduction}
\label{sec:introduction}
The observation that non-classical resources, specifically, entanglement and squeezing~\cite{Giovannetti2001,Giovannetti2006,Dowling2008,Giovannetti2011,Pezze2018}, allow for measurement precision significantly surpassing the classical limit has ignited a wave of breakthroughs in quantum sensing. Over the past two decades, quantum-enhanced metrology has enabled significant advances across a variety of platforms, including optical clocks~\cite{Leroux2010,Hosten2016, PedrozoPenafielN2020,Colombo2022}, magnetometers~\cite{Sewell2012, MartinCiurana2017,Wu2025,Gao2025}, interferometers~\cite{Gross2010,DemkowiczDobrzanski2015,Bongs2021}, and gravitational-wave detectors~\cite{Tse2019,LIGO2023}.

The quantum sensors demonstrated in these works predominantly employ interferometric or homodyne detection architectures~\cite{Wieczorek2015,Rossi2019,Magrini2021,Tebbenjohanns2021}, where the physical signal is linearly encoded as a phase shift and extracted by interference with a strong local oscillator or reference field~\cite{Braginsky1995,Wiseman2009}. This often allows the system degrees of freedom, as well as the measurement outcomes, to be described by employing the Gaussian formalism~\cite{Ferraro2005,Weedbrook2012}. Crucially, this enables the integration of fast signal-processing solutions in real time for both estimation, e.g., Kalman or Wiener filters~\cite{Wieczorek2015,Rossi2019}, and feedback, e.g., linear-quadratic-Gaussian (LQG) control~\cite{Magrini2021, Tebbenjohanns2021}.

In contrast, \emph{time-resolved photodetection}~\cite{Hadfield2009}-ubiquitous in time-correlated single-photon counting (TCSPC)~\cite{Becker2005} and widely employed across quantum platforms~\cite{Cohen2015, Riedinger2018,Galinskiy2020,Fiaschi2021,Ortolano2021,Galinskiy2024}-remains uncommon in sensing applications. This is because real-time parameter estimation from photoclick patterns poses a major challenge~\cite{Kiilerich2014}. The complexity of the quantum dynamical description-including environmental noise, detection imperfections, and the stochastic nature of the measurement process~\cite{Plenio1998}-renders direct likelihood evaluation computationally prohibitive~\cite{Clark2025}. Except for systems of very small dimension~\cite{Rinaldi2024,Radaelli2024,Anteneh2025,Khan2025}, standard Bayesian inference becomes then intractable in real time~\cite{Gebhart2023}. In particular, established techniques such as particle filters (sequential) or Markov-Chain Monte Carlo (batch)~\cite{Gilks1995,Murphy2012,Granade2017}, which rely on frequent likelihood evaluations, fail to bypass the underlying computational bottleneck.

In this work, we show that this bottleneck can be overcome via two distinct approaches:~(i) avoiding likelihood computation by resorting to likelihood-free Bayesian inference, specifically \emph{Approximate Bayesian Computation} (ABC)~\cite{Sisson2018, Turner2012, Beaumont2002};~and (ii) employing pre-trained \emph{deep learning} (DL) models~\cite{Goodfellow-et-al-2016}. While ABC was only recently proposed for quantum inference~\cite{Clark2025}-following earlier applications in Hamiltonian learning~\cite{Granade2012, Sergeevich2011, Wiebe2016}-DL has emerged as a powerful framework in quantum metrology for developing estimation~\cite{Khanahmadi2021, Nolan2021, HuertaAlderete2022} and control~\cite{Khalid2023, Nikoloska2025, Zhou2024} strategies. These include techniques for maximising the Quantum Fisher Information~\cite{Xu2021, Xiao2022} and implementing adaptive Bayesian parameter estimation~\cite{Fiderer2021} in photonic interferometry~\cite{Lumino2018,Cimini2019,Cimini2023}. 


Here, we demonstrate that both ABC and DL methods can extract informative representations from fixed-length photoclick trajectories by exploiting the quantum correlations encoded in the detection patterns. Although ABC is too computationally demanding for real-time implementation, and would become even more costly when extended to higher-order correlations, such as $n=3$, it provides a valuable benchmark for interpreting and validating the DL results, thereby improving their interpretability. DL therefore represents the more promising route towards real-time quantum-enhanced parameter estimation in photodetection-based quantum sensors.

The manuscript is organised as follows. In \secref{sec:system}, we describe a general quantum sensing scheme that involves a driven quantum system being measured via photodetection. In \secref{sec:inference}, we detail the inference methods employed (ABC in \secref{sec:ABC} and DL in \secref{sec:DL}); we verify these techniques using a two-level atomic model before focusing on a non-linear optomechanical system. Section~\ref{sec:results} presents the main results of our work, and we offer concluding remarks and an outlook in \secref{sec:conclusions}.

\begin{figure}[t]
    \centering
    \includegraphics[width=.99\columnwidth]{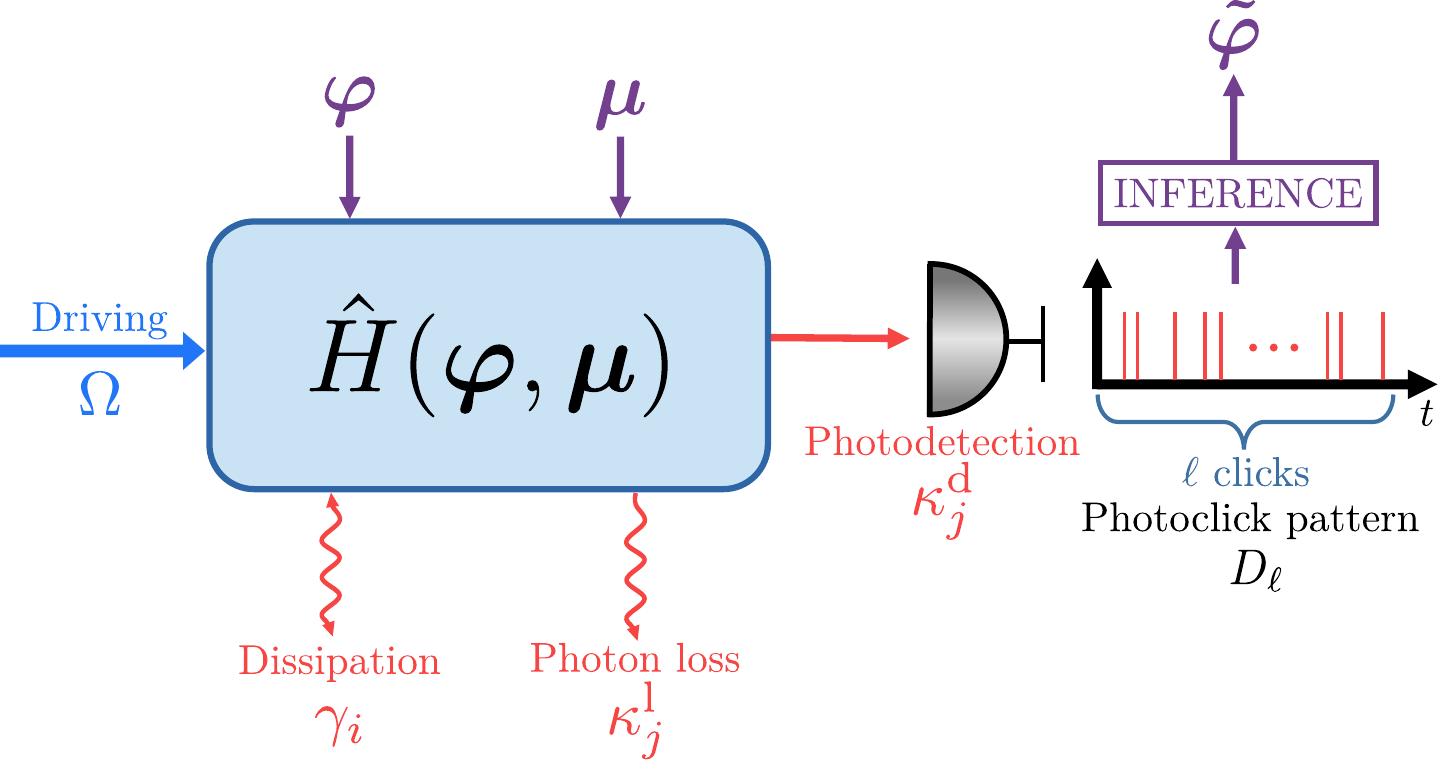}
    \caption{\textbf{Quantum sensing scheme with photodetection.} A quantum system is governed by a Hamiltonian $\hat{H}(\estparV, \contparV)$, where $\estparV$ and $\contparV$ are vectors of estimated and control parameters, respectively. The system is driven by an external (classical, \emph{blue}) field (e.g., strong light-beam) of strength $\Omega$ and dissipates through multiple (quantum, \emph{red}) channels with rates $\gamma_i$. Photon emission constitutes a separate dissipation process, being further divided into detected and lost channels with rates $\kd_j$ and $\kl_j$, respectively. Upon registering $\ell$ photons, the most accurate estimator of parameters $\est{\estparV}$ should be constructed based on the recorded photoclick pattern $D_\ell$-a vector of waiting times between consecutive clicks.}
    \label{fig:system_setup}
\end{figure}

\section{Quantum sensing with photodetection}
\label{sec:system}
The relevant sensing scenario is depicted in~\figref{fig:system_setup}, in which a quantum sensor is monitored by probing the light it emits; in particular, the arrival times of photons being continuously detected. By $\estparV=\{\estpar_1,\estpar_2,\dots,\estpar_d\}$ we denote the vector containing the \emph{estimated parameters}, whose values shall be inferred from a particular \emph{photoclick pattern}, i.e., a vector $D_\ell\coloneqq\{\Delta t_1,\Delta t_2,\dots,\Delta t_\ell\}$ which contains time intervals (waiting times) between consecutive $\ell$ photoclicks. However, within the framework we allow other \emph{control parameters} $\contparV$, whose value is known but should be adjusted for best estimation performance. Moreover, the sensor is considered to be classically driven by some external field of strength $\Omega$. As a result, although the sensing process is modelled dynamically, the instantaneous and steady-state behaviour of the sensor, given a particular form of its dissipation ($\gamma$-channels) and detection ($\kappa$-channels), can be modified by varying $\Omega$.

The goal is to estimate most accurately parameters $\estparV$ based on the detected photoclick data $D_\ell$. In the idealistic scenario, this should be performed in real time as the successive clicks are recorded. On one hand, this opens door for sensing parameters $\estparV$ that change in time, with their time-variation being then tracked ``on the fly''. On the other, this allows for adjusting the control parameters $\contparV$ in between clicks, so that the sensing performance is optimised. All this, however, takes for granted that the inference part can be done sufficiently fast. Here, we make the first important step by proposing necessary estimation methods and test their accuracy when sensing \emph{constant parameters} $\estparV$ for a photoclick trajectories of \emph{fixed length} $D_\ell$. Our work lays the groundwork for developing real-time estimation protocols that can be implemented in future photodetection experiments, eventually allowing for the integration of feedback and control.

\subsection{Quantum system with driving and dissipation}
Considering a general Hamiltonian describing the system used as a sensor, see \figref{fig:system_setup}, which depends on the estimated $\estparV$ and control $\contparV$ parameters, we may decompose it into its free and driving parts, i.e.: 
\begin{equation}
    \hat{H} \equiv \hat{H}(\estparV, \contparV) = \hat{H}_\mrm{free}+\hat{H}_\mrm{drive}(\drivepar)
    \label{eq:sys_Hamiltonian}
\end{equation}
with the latter depending explicitly on the driving strength $\Omega$. Although the form of each is unspecified 
 (and may arbitrarily depend on $\estparV$ and $\contparV$), we are motivated by scenarios in which $\hat{H}_\mrm{free}$ carries non-linear interactions that allow to engineer the statistics of emitted photons~\cite{Chang2014,Aspelmeyer2014}, i.e., control their quantum bunching or anti-bunching properties~\cite{Paul1982,Lee1993}.

Furthermore, to incorporate dissipation of the quantum sensor, we consider its evolution to be described by
the Gorini-Kossakowski-Sudarshan-Lindblad (GKSL) master equation~\cite{Chruscinski2017}:
\begin{align}
    \label{eq:gen_master_eq}
    \frac{\dd  \hat{\rho}}{\dd  t} &= -\frac{\ii}{\hbar} [\hat{H}, \hat{\rho}] + \sum_i \gamma_i \left( \hat{L}_i \hat{\rho} \hat{L}_i^\dagger - \frac{1}{2} \{ \hat{L}_i^\dagger \hat{L}_i, \hat{\rho} \} \right)
\end{align}
where  $\gamma_i$ are the dissipation rates, and $\hat{L}_i$ are the corresponding Lindblad operators. As a result, we obtain a broad description of sensor dynamics, incorporating both the unitary evolution, which contains the estimated and control parameters, and arbitrary irreversible processes that take a semigroup (Markovian) form \eref{eq:gen_master_eq}~\cite{Chruscinski2017}.

\subsection{Photodetection measurement and unravellings}
\label{subsec:photodect}
In order to incorporate the photodetection measurement into the dynamical description, we resort to the continuous measurement theory~\cite{Plenio1998,Wiseman2009} and, as depicted also in~\figref{fig:system_setup}, account for distinct detection channels (light modes) labelled by $j$, via which the information is extracted from the quantum sensor. In particular, each of their outputs is measured at a rate $\kd_j$ with a photodetector, whose finite efficiency in turn increases $\kl_j$-the effective rate of photon loss in each channel.

Assuming that the continuous measurement does not alter the form of \eqnref{eq:gen_master_eq}~%
\footnote{
    Alternatively, one may consider \eqnref{eq:true_evol} as a starting point to describe system dynamics.
},
we arrive at a stochastic master equation that describes the \emph{true evolution} of the system in~\figref{fig:system_setup} when conditioned on a particular photoclick record (trajectory)~\cite{Carmichael2008}:
\begin{subequations}
\label{eq:true_evol}
\begin{align}
    & \dd  \hat{\rho} =-\frac{\ii}{\hbar}[\hat{H}, \hat{\rho}] \dd t + \sum_i \gamma_i \!\left(\!\hat{L}_i \hat{\rho} \hat{L}_i^\dagger - \frac{1}{2} \{ \hat{L}_i^\dagger \hat{L}_i, \hat{\rho} \} \!\right)\! \dd  t \label{eq:diss_part}\\
    & \!\!\!\!\!
    + \sum_{j} \kl_j \!\left( \hat{c}_j \hat{\rho} \hat{c}_j^\dagger - \frac{1}{2} \{ \hat{c}_j^\dagger \hat{c}_j, \hat{\rho} \} \right)\!\dd  t 
    - \frac{\kd_j}{2} \{ \hat{c}_j^\dagger \hat{c}_j, \hat{\rho} \} \dd t
    + \mathcal{J}_{\kd_j}[\hat{\rho}],
    \label{eq:stoch_obs}
\end{align}
\end{subequations}
where $\hat{c}_j$ is the photonic annihilation operator in the $j$th detection channel, and each photodetection superoperator $\mathcal{J}_{\kd_j}$ is defined as:
\begin{equation}
    \mathcal{J}_{\kd_j}[\hat{\rho}] \coloneq \kd_j \tr{\hat{c}_j^\dagger \hat{c}_j \hat{\rho}} \hat{\rho} \, \dd t + \left( \frac{\hat{c}_j \hat{\rho} \hat{c}_j^\dagger}{\tr{\hat{c}_j^\dagger \hat{c}_j \hat{\rho}}} - \hat{\rho} \right) \dd N_t^{\kd_j},
\end{equation}
where $\dd N_t^{\kd_j}$ denotes a Poisson increment taking values 0 or 1~\cite{Gardiner1985}, depending whether a photon was detected or not within a time increment $\dd t$. The mean of a Poisson increment, or equivalently $P(\dd N_t^{\kd_j}=1)$, is given by $\left\langle \dd N_t^{\kd_j} \right\rangle = \kd_j\,\tr{\hat{c}_j^\dagger \hat{c}_j \hat{\rho}} \dd t$ for a given time $t$.

In particular, the photoclick pattern $D_\ell$ shown in~\figref{fig:system_setup} describes a particular detection trajectory for the channel $j$, along which the system evolves according to \eqnref{eq:true_evol} with $\dd N_t^{\kd_j}$ taking value zero at all times apart from the moments at which each of the $\ell$ clicks is registered. Although we allow multiple detection channels for generality-each photoclick within a pattern $D_\ell$ also carries information about its type (label $j$)-note that in the extreme case when all the photons are lost within some channel $j$, i.e., $\kd_j=0$, $\kl_j>0$, then, as it introduces solely dissipation, it can be equivalently included as another $\gamma_{i}=\kl_j$ with $\hat{L}_i=\hat{c}_j$ in \eqnref{eq:diss_part}.

We also consider the \emph{full unravelling} of the master equation \eref{eq:gen_master_eq}, which governs the conditional evolution of the system state under the unphysical assumption that all the processes yielding quantum jumps are monitored~\cite{Dalibard1992,Moelmer1993,Hegerfeldt1993}:
\begin{align}
    \label{full_unravel}
    \dd  \hat{\rho} =& -\frac{\ii}{\hbar} \left( \hat{H}_{\mrm{cond}} \hat{\rho} - \hat{\rho} \hat{H}_{\mrm{cond}}^\dagger \right) \dd t + \sum_i  \, \mathcal{J}_{\gamma_i}[\hat{\rho}] \nonumber \\
    &+ \sum_j \, \mathcal{J}_{\kl_j}[\hat{\rho}]+ \, \mathcal{J}_{\kd_j}[\hat{\rho}],
\end{align}
where the conditional (non-Hermitian) Hamiltonian is defined as
\begin{equation}
    \hat{H}_{\mrm{cond}} \coloneq \hat{H} - \frac{\ii \hbar}{2} \sum_i \gamma_i \hat{L}_i^\dagger \hat{L}_i - \frac{\ii\hbar}{2} \sum_j \kappa_j \hat{c}_j^\dagger \hat{c}_j
\end{equation}
with $\kappa_j\coloneqq\kl_j+\kd_j$ for each $j$. In particular, when evolving $\hat{\rho}$ only according to the first term in \eqnref{full_unravel}, the special conditional trajectory is recovered that corresponds to \emph{no jumps} occurring and, hence, none of them being registered. Although such system dynamics is trace non-preserving, the state norm at time $t$ directly provides the probability of no-jump occurring until the given $t$. 

Importantly, in \eqnref{full_unravel} not only jumps associated with unobserved photons are accounted for, but also ones arising due to each dissipative processes labelled by $i$, i.e.:
\begin{equation}
\label{jump-super_ops}
    \mathcal{J}_{\gamma_i}[\hat{\rho}] \coloneqq \gamma_i \tr{\hat{L}_i^\dagger \hat{L}_i \hat{\rho}} \hat{\rho} \, \dd t + \left( \frac{\hat{L}_i \hat{\rho} \hat{L}_i^\dagger}{\tr{\hat{L}_i^\dagger \hat{L}_i \hat{\rho}}} - \hat{\rho} \right) \dd N_t^{\gamma_i}
\end{equation}
with the stochastic Poisson increment now satisfying $\left\langle \dd N_t^{\gamma_i} \right\rangle = \gamma_i\,\tr{\hat{L}_i^\dagger\hat{L}_i\hat{\rho}}\dd t$, which may not possess physical interpretation despite reproducing the correct dissipative dynamics when averaged over.

Although the conditional evolution of the system that is based only on physical observations has to be computed using \eqnref{eq:true_evol}, \eqnref{full_unravel} proves to be useful, as it describes a pure-state evolution along each trajectory. Hence, it is much less demanding computationally, as it involves only the wavefunction description instead of the density matrix~\cite{Dalibard1992,Moelmer1993,Hegerfeldt1993}. In particular, if one is interested in generating examples of photoclick patterns, $D_\ell$, and not the actual evolution of the system, one may generate these with help of \eqnref{full_unravel} by \emph{disregarding} the inaccessible (unphysical) clicks (events) in a record-these arise when any of $\dd N_t^{\gamma_i}$ or $\dd N_t^{\kl_j}$ takes the value 1. 

This does not conflict the fact that \emph{averaging} over inaccessible events-specifically, all the state trajectories generated with \eqnref{full_unravel} that yield the same record of accessible detection events-consistently recovers the conditional state dynamics \eref{eq:true_evol}. Furthermore, by averaging over all potential trajectories without fixing the pattern of detection events ($\dd N_t^{\kd_j}$), the reduced dynamics described by the original master equation \eref{eq:gen_master_eq} is reproduced.

\begin{figure}[t]
    \centering
    \includegraphics[width=\columnwidth]{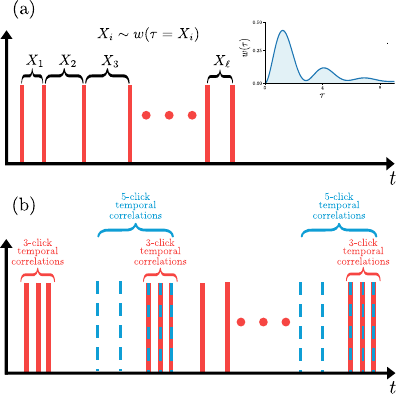}
    \caption{\textbf{Interpreting photodetection as a temporal point process (TPP).} (a) \emph{Renewal TPP}:~Waiting times $X=\{X_1,X_2,\dots,X_\ell\}$ are independently and identically distributed (i.i.d.), sampled from a stationary waiting-time distribution $w(\tau)$;~(b) \emph{History-dependent TPP}:~Waiting times are not i.i.d.~and cannot be described by a single distribution. As illustrated, specific records may exhibit hierarchical temporal correlations-such as three-click patterns nested within larger five-click sequences. Any such multi-time correlations require a non-renewal framework to account for quantum effects like photon (anti)bunching~\cite{Paul1982}, which are essential for quantum-enhanced sensing tasks.}
    \label{fig:TPPs}
\end{figure}

\subsection{Photoclick pattern as a temporal point process}
\label{sec:TPPs}
The record of a photodetection process constitutes a realisation of a stochastic \emph{temporal point process} (TPP), where the recorded data $D_\ell$ consists of a sequence of photon waiting times $\{X_i\}_{i=1}^\ell$-see \figref{fig:TPPs}(a). Broadly speaking, TPPs are categorised as either \emph{renewal} or \emph{history-dependent}~\cite{Daley2005}. In a renewal process, the probability of an event depends solely on the time elapsed since the most recent event. Such processes are characterised by independent and identically distributed (i.i.d.) waiting times, with their statistics thus being fully captured by a single \emph{waiting-time distribution} $w(\tau)$, e.g., the one sketched in~\figref{fig:TPPs}(a). In contrast, a history-dependent TPP cannot be described by a single distribution, as the statistics of subsequent events are conditioned on the history of prior detections in a non-trivial manner.

In the quantum setting, a \emph{two-level system} (TLS) serves as the paradigmatic renewal source; because the system resets to its ground state upon emission, the memory of prior events is effectively erased~\cite{Carmichael1989,Kiilerich2014}. Conversely, most quantum systems possess complex energy-level structures that generate short- or long-lived memory dependencies. These are manifested as temporal correlations in the photoclick pattern that cannot be described by a single distribution~\cite{Landi2024}, as illustrated in~\figref{fig:TPPs}(b). Crucially, it is these multi-time correlations that facilitate quantum-enhanced sensing, as they possess significant additional information about estimated parameters beyond what is available in lower-order statistics~\cite{Burgarth_2015,Kiilerich2016,Clark2019,Clark2022}.

As shown in~\figref{fig:TPPs}(b), non-renewal photoclick patterns can exhibit hierarchical correlations-where fast three-click clusters are nested within broader five-click sequences emergent at larger timescales. Formally, this corresponds to the regime where higher-order correlation functions, specifically third-order $g^{(3)}(\tau_1,\tau_2)$ or higher~\cite{Mandel1995}, exhibit non-classical signatures. This is in stark contrast to the renewal framework, where the i.i.d.~nature of the waiting times ensures that the second-order $g^{(2)}(\tau)$-function, being the Laplace transform of $w(\tau)$, completely determines all higher-order correlations;~in such cases, $g^{(n)}$ provides no information beyond what is already captured by the second-order statistics~\cite{Teich1984,Cao2006}. Consequently, while renewal processes may exhibit non-classical features like antibunching~\cite{Paul1982}, they are fundamentally limited for quantum-enhanced estimation as they cannot exploit the higher-order dependencies that typically drive significant sensitivity gains.

However, extracting this higher-order information poses a formidable challenge;~the likelihood functions for history-dependent processes are computationally intensive to evaluate, making parameter estimation significantly more difficult than for renewal processes~\cite{Radaelli2024,Clark2025}. This overhead currently precludes the use of non-renewal TPPs in real-time applications, such as active feedback or closed-loop quantum control, where estimation must occur on timescales faster than the system dynamics.

\section{Efficient inference from photoclick patterns}
\label{sec:inference}

Having established the general framework for photodetection-based quantum sensing in \secref{sec:system}, we now address the central inference problem:~how to accurately and efficiently estimate the parameters $\estparV$ from a photoclick pattern $D_\ell$. We first introduce the Bayesian formulation of parameter estimation and the corresponding performance measures, including local (frequentist) precision bounds based on the \emph{Fisher information} (FI)~\cite{Kay1993}. We then discuss representations of photoclick data before introducing the two likelihood-free approaches considered in this work:~ABC in \secref{sec:ABC} and DL in \secref{sec:DL}.

\subsection{Global estimator construction}
Under the Bayesian paradigm, the $d$ parameters to be estimated, $\estparV=\{\estpar_i\}_{i=1}^d\in\mathbb{R}^d$,  are treated as random variables rather than fixed constants. Their prior distribution, $p(\estparV)$, represents then the \emph{a priori} knowledge about their potential values before any measurements are taken. Given a set of measurement data $D$, the optimal parameter estimator, $\est{\estparV}(D)$, is typically constructed by minimising the \emph{average Mean Squared Error} (aMSE)~\cite{Kay1993}:
\begin{align}
\label{eq:aMSE}
    \aMSE[\est{\estparV}] 
    &\coloneqq \EE{\estparV}{\EE{D|\estparV}{||\est{\estparV}(D)- \estparV ||^2}}
    \\
    &= 
    \int\!\dd \estparV\, p(\estparV) \int\!\dd D\; p(D|\estparV) \, ||\est{\estparV}(D)- \estparV ||^2,
    \nonumber
\end{align}
where $p(D|\estparV)$ is the likelihood distribution-describing probability of obtaining data $D$ given specific parameter values $\estparV$-and $||\cdot ||$ denotes the Euclidean norm between the true parameter vector $\estparV$ and its estimate $\est{\estparV}(D)$.

The optimal estimator that minimises the aMSE \eref{eq:aMSE} is generally the mean of the posterior distribution~\cite{Kay1993}:
\begin{equation}
    \est{\estparV}_\trm{optim} \coloneqq \EE{\estparV|D}{\estparV} = \int\!\dd\estparV\,p(\estparV|D)\,\estparV,
    \label{eq:opt_estimator}
\end{equation}
where the posterior distribution, representing the updated belief about the parameters $\estparV$ after  observing the data $D$ (\emph{a posteriori}), is given by the Bayes' rule~\cite{Kay1993}:
\begin{equation}
\label{eq:posterior}
    p(\estparV|D) = \frac{p(D|\estparV)p(\estparV)}{p(D)} = \frac{p(D|\estparV)p(\estparV)}{\int\!\dd \estparV \, p(D|\estparV)p(\estparV)}.
\end{equation}

However, computing \eqnref{eq:posterior} is typically intractable, even numerically. While exhaustive integration or sampling over the $\estparV$-parameter space can often be avoided using methods like particle filters (sequential) or Markov-Chain Monte Carlo (batch)~\cite{Gilks1995,Murphy2012,Granade2017}, these techniques remain insufficient if computing the likelihood $p(D|\estparV)$ itself is infeasible. This is particularly relevant for quantum systems with states spanning high-dimensional Hilbert spaces~\cite{Clark2025}. In real-time scenarios, such as the photodetection scheme of \figref{fig:system_setup}, calculating the likelihood requires the system state to be continuously updated in response to real-time measurement data-accounting for every individual click in the photoclick pattern in~\figref{fig:system_setup}.

Given the complexity of non-renewal quantum systems, recall \figref{fig:TPPs}(b), the challenge is not merely to optimise likelihood evaluation, but to bypass it entirely during the inference stage. In what follows, we explore likelihood-free inference methodologies that shift the computational burden away from real-time execution. First, we describe a Bayesian approach to estimation based on ABC, which circumvents the likelihood bottleneck by comparing observed data with simulated samples~\cite{Clark2025}. We then present an alternative approach using DL, where the same simulated data is used to train \emph{Deep Neural Networks} (DNNs) offline. While ABC naturally provides estimation errors and a reconstructed posterior, we demonstrate that DNNs can be similarly configured to provide these metrics. Thus, while DNNs offer a far faster solution with matching capabilities, ABC remains invaluable for interpreting the statistics and temporal patterns the network has learnt.

\subsection{Local estimator performance and the associated biased Cram\'{e}r-Rao bound}
\label{subsec:local_estim}
Although the Bayesian approach ensures that the estimator is constructed independently of particular parameter values, we wish to evaluate its performance locally, i.e.,~at particular estimation points $\estparV$. For this reason, it is also useful to define the \emph{MSE matrix} that is not prior-averaged as in \eqnref{eq:aMSE}, whose entries read~\cite{Kay1993}: 
\begin{equation}
    \label{eq:MSE}
    \MSE[\est{\estparV}|\estparV]_{ij} \coloneqq \EE{D|\estparV}{(\est{\varphi}_i(D)- \estpar_i)(\est{\varphi}_j(D)- \estpar_j)},    
\end{equation}
which, for any \emph{unbiased} estimator $\est{\estparV}$, is lower-bounded by the \emph{Cram\'{e}r-Rao bound} (CRB)~\cite{Kay1993}: 
\begin{equation}
    \MSE[\est{\estparV}|\estparV]
    \;\underset{\trm{unb.}\,\est{\estparV}}{\succeq}\; 
    \FI[D|\estparV]^{-1}.
    \label{eq:CRB}
\end{equation}
The CRB \eref{eq:CRB} is determined by the FI matrix $\FI$, whose entries are given by
\begin{equation}
    \FI[D|\estparV]_{ij} \coloneqq \EE{D|\estparV}{\partial_{\estpar_i}\!\ln p(D|\estparV)\; \partial_{\estpar_j}\!\ln p(D|\estparV)}.
    \label{eq:FImatrix}
\end{equation}

However, because we wish to evaluate the performance of estimators for finite-sized data (i.e.,~finite photoclick numbers) outside the regime of sufficiently large statistics where the Bernstein--von Mises theorem holds~\cite{Vaart1998}, we compute a version of the CRB \eref{eq:CRB} which applies to biased estimators~\cite{Rinaldi2024, Eldar2008}, i.e.~the \emph{biased Cram\'{e}r-Rao bound} (bCRB):
\begin{align}
    \MSE[\est{\estparV}|\estparV] 
    &\succeq 
    (\id+\nabla_{\estparV}\vec{b}(\est{\estparV}))\FI[D|\estparV]^{-1}(\id+\nabla_{\estparV}\vec{b}(\est{\estparV}))^\TT \nonumber\\
    &\qquad +\vec{b}(\est{\estparV})\vec{b}(\est{\estparV})^\TT,
    \label{eq:biased_CRB}
\end{align}
where $\vec{b}(\est{\estparV})=\{\EE{D|\estparV}{\est{\varphi}_i}-\varphi_i\}_{i=1}^d$ is the bias vector of the estimator under consideration. 

Consequently, unlike the standard unbiased CRB \eref{eq:CRB}, the bCRB \eref{eq:biased_CRB} depends directly on the chosen estimator through its bias vector and must be evaluated for each one separately. While not a universal lower bound across all estimators, the bCRB accounts for the bias penalty of a given estimator, establishing its absolute variance limit and verifying whether it operates optimally.

In our work, to quantify the simultaneous estimation of multiple parameters in local estimation, we use the (local) \emph{root-mean-square error} (RMSE):
\begin{equation}
    \mrm{RMSE} \coloneqq \sqrt{\tr{\MSE}}=\sqrt{\EE{D|\estparV}{||\est{\estparV}(D)- \estparV ||^2}},
\end{equation}
which does not prioritise any of the estimated parameters-consistently with the aMSE in \eqnref{eq:aMSE}.

\subsubsection{Evaluation of the FI matrix along a photoclick trajectory}
\label{sec::trajectory_FI}
For our purposes, to evaluate the FI matrix \eref{eq:FImatrix} along a photoclick trajectory-accounting for the fact that its likelihood function may not be computable in practice-we resort to the sampling-based formalism developed in Refs.~\cite{Gammelmark2013,Albarelli2018} and focus on the case of pure states. 

Denoting by $0<t<T^{(k)}$ the time along a particular trajectory $D_\ell^{(k)}$ (indexed by $k$ and containing $\ell$ clicks), the entries of the FI matrix \eref{eq:FImatrix} are obtained by averaging over trajectories as follows:
\begin{equation}
    \FI[D_\ell|\estparV]_{ij} = \EE{(k)}{\tr{\tau_i(T^{(k)})} \tr{\tau_j(T^{(k)})}},
    \label{eq:event_based_FI}
\end{equation}
where $\tau_i(T^{(k)})$ and $\tau_j(T^{(k)})$ correspond to the monitoring operators associated with the $i$th and $j$th parameters, respectively, evaluated at the end of the trajectory $D_\ell^{(k)}$, i.e., under Type-II censoring. Note that under Type-I censoring-where the total trajectory duration rather than the number of events is fixed-one obtains an FI matrix of the exact same functional form~\cite{Lawless2003}.

For a given time $t$ along the trajectory, the monitoring operator associated with the parameter $\estpar_i$ is defined as~\cite{Albarelli2018,Radaelli2024}
\begin{equation}
    \tau_i(t)
    \coloneqq
    \ket{\phi_i^t}\bra{\psi^t}
    +
    \ket{\psi^t}\bra{\phi_i^t},
\end{equation}
where $\ket{\psi^t}$ denotes the normalised conditional pure state of the system at time $t$, and
\begin{equation}
    \ket{\phi_i^t}
    \coloneqq
    \frac{
        \ket{\partial_{\estpar_i}\tilde{\psi}^t}
    }{
        \sqrt{\braket{\tilde{\psi}^t|\tilde{\psi}^t}}
    }
\end{equation}
is the rescaled parameter derivative of the unnormalised pure state $\ket{\tilde{\psi}^t}$. In practice, evaluation along a trajectory relies on propagating two state vectors-the unnormalised pure state $\ket{\tilde{\psi}^t}$ and its derivative $\ket{\partial_{\estpar_i}\tilde{\psi}^t}$-piecewise between measurement jumps under effective non-Hermitian dynamics. At each jump event (photoclick), the state vector must be normalised to apply the jump operator correctly and update both vectors before resuming unnormalised propagation. The state norm at any target time $t$ is then used to construct the normalised vectors $\ket{\psi^t}$ and $\ket{\phi_i^t}$ needed to evaluate $\tau_i(t)$.

To describe this propagation, let $t_k$ denote the time immediately following the $k$-th detection event, at which point the pure state $\ket{\psi^{t_k}}$ is normalised and its rescaled derivative $\ket{\phi_i^{t_k}}$ is known. Between consecutive detection events, the system undergoes continuous \emph{no-jump} evolution generated by the non-Hermitian effective propagator
\begin{equation}
    \hat{U}^{\estparV}_{\mathrm{eff}}(s)
    =
    \exp\left[
        -\frac{\ii}{\hbar} s
        \left(
            \hat{H}(\estparV, \contparV)
            -
            \frac{\ii \hbar}{2}
            \sum_j \kappa_j
            \hat{c}_j^\dagger \hat{c}_j
        \right)
    \right],
\end{equation}
where $\hat{H}$ is the system Hamiltonian \eref{eq:sys_Hamiltonian} encoding the parameters $\estparV$, whereas $\hat{c}_j$ are jump operators associated with photodetection in \eqnref{eq:true_evol}.

Over an elapsed time $s$ since $t_k$, this non-unitary operator generates the evolution of the unnormalised pure state $\ket{\tilde{\psi}^{t_k+s}}$ and its unnormalised parameter derivative $\ket{\tilde{\phi}_i^{t_k+s}} \coloneqq \ket{\partial_{\estpar_i}\tilde{\psi}^{t_k+s}}$:
\begin{align}
    \text{no jump:}\quad
    \ket{\tilde{\psi}^{t_k+s}}
    &=
    \hat{U}^{\estparV}_{\mathrm{eff}}(s)
    \ket{\psi^{t_k}},
    \\
    \text{no jump:}\quad
    \ket{\tilde{\phi}_i^{t_k+s}}
    &=
    \left(
        \partial_{\estpar_i}
        \hat{U}^{\estparV}_{\mathrm{eff}}(s)
    \right)
    \ket{\psi^{t_k}}
    +
    \hat{U}^{\estparV}_{\mathrm{eff}}(s)
    \ket{\phi_i^{t_k}}.
\end{align}
This linear propagation continues until the next detection event at time $t_{k+1} = t_k + s_k$. At this instant, a detection associated with the jump operator $\hat{c}_j$ triggers an instantaneous update, re-normalising both state vectors to reset the initial conditions for the subsequent interval:
\begin{align}
    \text{jump:}\quad
    \ket{\psi^{t_{k+1}}}
    &=
    \frac{
        \hat{c}_j\ket{\tilde{\psi}^{t_k+s_k}}
    }{
        \sqrt{
            \bra{\tilde{\psi}^{t_k+s_k}}
            \hat{c}_j^\dagger \hat{c}_j
            \ket{\tilde{\psi}^{t_k+s_k}}
        }
    },
    \\
    \text{jump:}\quad
    \ket{\phi_i^{t_{k+1}}}
    &=
    \frac{
        \hat{c}_j\ket{\tilde{\phi}_i^{t_k+s_k}}
    }{
        \sqrt{
            \bra{\tilde{\psi}^{t_k+s_k}}
            \hat{c}_j^\dagger \hat{c}_j
            \ket{\tilde{\psi}^{t_k+s_k}}
        }
    }.
\end{align}
Thus, trajectory evolution reduces to simple linear propagation of two unnormalised state vectors between detections, punctuated by norm resets at each jump time.

Finally, the trajectory-sampling procedure outlined above is not required in the special case where the photoclick pattern constitutes a \emph{renewal} TPP (recall \secref{sec:TPPs})~\cite{Kiilerich2014}. Because successive waiting times are independent and identically distributed, the joint probability density factorises into a product of single-interval waiting-time distributions $w(\tau|\estparV)$, i.e.~$p(D_\ell|\estparV)=w(\tau|\estparV)^\ell$. Consequently, the total FI matrix simplifies to $\ell$ times the FI matrix of $w(\tau|\estparV)$:
\begin{equation}
    \FI\!\left[D_\ell|\estparV\right] = \ell \;\FI\!\left[w(\tau|\estparV)\right].
\end{equation}
Beyond eliminating the need for trajectory sampling, this relation implies that any deviation from linear scaling in the click count $\ell$ serves as a direct signature that the underlying TPP is \emph{history dependent}.

\subsubsection{Upper-bounding the FI matrix by considering the full unravelling}
\label{subsec:UB_FI_matrix}
When the system dynamics are fully unravelled, the complete measurement record comprises both observed jumps (photoclicks) and unobserved jumps (e.g., lost photons or virtual phonons). Given a parameter vector $\estparV$, $p(D_\ell,D_{\ell'}'|\estparV)$ denotes the joint probability density function of arrival times for $\ell$ observed and $\ell'$ unobserved jumps. Crucially, because the stopping time $T_{\ell}$ is set strictly by the observed trajectory, the unobserved record $D'$ can contain any number of jumps $\ell'$ occurring within $[0,T_{\ell}]$.

Marginalising over these inaccessible outcomes, the likelihood of the observed trajectory reads
\begin{align}
    p(D_\ell|\estparV)
    &= 
    \int_{\Omega(T_\ell)}\!\!\mu_{D_\ell}(\dd D^\prime)\;
    p(D_\ell,D'|\estparV) \nonumber\\
    &=
    \sum_{\ell^\prime = 0}^\infty \; \int_{\Omega_{\ell^\prime}(T_\ell)}\!\!\mu_{D_\ell}(\dd D^\prime_{\ell^\prime})\; 
    p(D_\ell,D_{\ell^\prime}'|\estparV),
    \label{eq:margin}
\end{align}
where $\Omega(T_\ell) = \bigsqcup_{\ell^\prime=0}^\infty \Omega_{\ell^\prime}(T_\ell)$ is the grand configuration space encompassing potential arrival times of unobserved jumps across all counts $\ell'$ within $T_\ell$.

As shown in \appref{app:FI_ineq}, this marginalisation implies that the FI matrix of the fully unravelled dynamics upper-bounds that of the observed record in the Loewner order:
\begin{equation}
    \FI[D_\ell|\estparV] \preceq \FI[D_\ell,D'|\estparV].
    \label{eq:full_unrav_ineq}
\end{equation}

\subsection{Data preprocessing and sufficient statistics}

Effective \emph{data preprocessing} can significantly improve the accuracy and computational efficiency of estimation tasks. Here, we treat preprocessing as the construction of \emph{summary statistics}-a mapping $\cS$ that transforms the raw data into a lower-dimensional representation capturing its key features:
\begin{equation}
    \bar{D}\coloneq \cS(D),
    \label{eq:sum_stat}
\end{equation}
where in the case of photodetection $D$ represents a general TPP trajectory of a finite length ($D_\ell$ in~\figref{fig:TPPs}) and $\bar{D}$ denotes the preprocessed data. Within this framework, we define \emph{sufficient summary statistics} as a special preprocessing $\cS$ that retains all information about the parameters $\estparV$ contained in the original data; formally, this occurs if the posterior can be expressed as:
\begin{equation}
    p(\estparV|\cS(D))=p(\estparV|D).
    \label{eq:suff_sum_stat}    
\end{equation}

Selecting an adequate sufficient summary statistic is non-trivial and relies heavily on the physical intuition of the problem. For instance, in any renewal TPP of \figref{fig:TPPs}(a), the \emph{histogram of waiting times} serves as a sufficient summary statistic. Because the waiting-time distribution, $w(\tau)$ in~\figref{fig:TPPs}(a), uniquely and completely defines a renewal process, its empirical representation-the histogram-captures all the information available for estimation~\cite{Clark2025}.

\begin{figure*}[t]
    \centering
    \includegraphics[width=\textwidth]{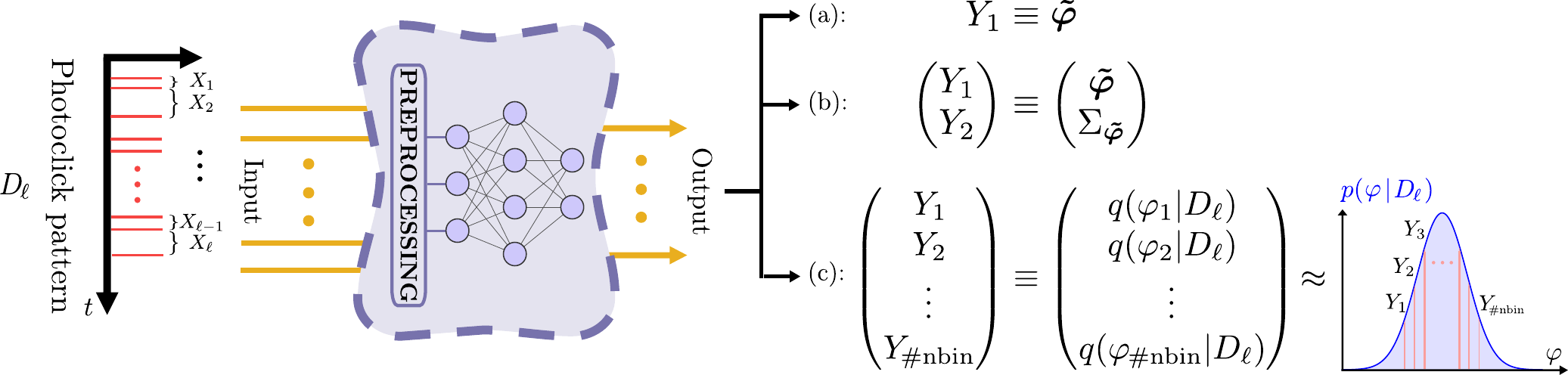}
    \caption{\textbf{Deep learning (DL) inference frameworks considered.} The input of a neural network, $X=\{X_1,X_2,\dots,X_\ell\}$, is set to the sequence of inter-event waiting times for a given photoclick pattern $D_\ell$. The output $Y$ is adapted to a specific statistical task: (a) Within the \emph{regression} framework, the network provides point estimates of the parameter vector $\est{\estparV}$ by minimising the Mean Squared Error (MSE). (b) In the \emph{probabilistic regression} framework, the network is assumed to yield a Gaussian posterior; thus, the Negative Log-Likelihood (NLL) is employed as the loss function, with the output providing both the parameter estimates and the corresponding errors covariance matrix, $\Sigma_{\est{\estparV}}$. (c) In  the \emph{classification} framework, which uses Cross-Entropy (CE) as the loss function, the output directly provides a discrete grid-representation of the posterior distribution.}
    \label{fig:NN_models}
\end{figure*}

\subsection{Approximate Bayesian Computation}
\label{sec:ABC}
To overcome the bottleneck of likelihood computation, we turn our attention first to likelihood-free inference, specifically \emph{Approximate Bayesian Computation} (ABC). This framework bypasses direct evaluation of the likelihood function by leveraging data simulations~\cite{Turner2012,Sisson2018}. The core logic of ABC is straightforward:~rather than constructing $p(D|\estparV)$ from the model, we use the model as a generative engine. We draw parameter values $\estparV$ from the prior distribution $p(\estparV)$ and, for each $\estparV$, simulate a synthetic trajectory $D'$ (or draw it from a precomputed library). We then compare the simulated dataset $D'$ with the observed data $D$ using a predefined distance measure. If $D'$ is close enough to $D$ to meet a set tolerance $\epsilon$, we retain the specific parameters $\estparV$ used to generate it. By aggregating these accepted parameter values, we construct an empirical distribution that faithfully approximates the posterior-provided that the number of simulated trajectories is large enough and the distance criterion is sufficiently stringent~\cite{Turner2012,Sisson2018}.

Mathematically, ABC yields an approximation of the posterior distribution in the form:
\begin{equation} \label{Eq:ABC_post1}
    q_\epsilon(\estparV|D) \propto p(\estparV) \int\!\dd D'\; \id[\rho(D, D') \le \epsilon] \; p(D'|\estparV),
\end{equation}
where $\rho(D, D')\ge 0$ is a proper distance function quantifying dissimilarity between datasets, and $\id[\cdot]$ is an indicator function that equals 1 when the condition is met and 0 otherwise. As $\epsilon\to 0$, the approximation converges to the true posterior, albeit at the cost of diminishing acceptance rate-requiring more and more samples $D'$ to be drawn with less and less of them being accepted.

In practice, it is usually infeasible to compare entire datasets directly, especially when they are high-dimensional or continuous. Instead, we identify summary statistics, $\vec{\cS}=\{\cS_i\}_i$ each corresponding to a particular preprocessing in \eqnref{eq:sum_stat}, designed together to be nearly sufficient such that $p(\estparV|\vec{\cS}(D))\approx p(\estparV|D)$. As a result, \eqnref{Eq:ABC_post1} can be rewritten as
\begin{equation}
    q_\epsilon(\estparV|D) \propto p(\estparV) \int\!\dd D'\; \id[\delta(\vec{\cS}(D), \vec{\cS}(D')) \le \epsilon] \; p(D'|\estparV),
    \label{eq:ABC_post_suff_stat}
\end{equation}
where $\delta$ is now the measure quantifying overall distance between multiple summary statistics evaluated for the two datasets, $D$ and $D'$. 

As before, for a sufficiently small $\epsilon$, the results remain accurate within a marginal error. By employing summary statistics-which can be precomputed and cached for each dataset $D'$ stored-the computation time of $\delta(\vec{\cS}(D), \vec{\cS}(D'))$ becomes negligible. Furthermore, as relevant summary statistics may be shared across multiple datasets, the acceptance rate is improved, thereby accelerating the sampling process. Balancing the acceptance rate against approximation accuracy is a central challenge in any ABC implementation~\cite{Clark2025}. 

Note that acceptance criteria can be applied either to each summary statistic individually or jointly through a common, weighted threshold, as in \eqnref{eq:ABC_post_suff_stat}. In this work, however, we set the thresholds individually for convenience whenever multiple summary statistics are used. In particular, we replace the joint indicator in \eqnref{eq:ABC_post_suff_stat} with a product of individual ones, $\id[\delta(\vec{\cS}(D), \vec{\cS}(D')) \le \epsilon]\to\prod_i\id[\delta(\cS_i(D), \cS_i(D')) \le \epsilon_i]$, allowing us to calibrate each $\epsilon_i$ separately~\cite{Clark2025}.

\subsection{Deep learning inference frameworks}
\label{sec:DL}
In recent years, the rapid evolution of \emph{Deep Learning} (DL) algorithms has enabled their successful deployment across a wide spectrum of problems in quantum physics~\cite{Dawid2025}, ranging from quantum state reconstruction~\cite{Palmieri2024} to phase classification and anomaly detection~\cite{Huembeli2018,Kottmann2020}. 

As for the use of DL in quantum sensing, previous applications have included magnetic-field tracking~\cite{Khanahmadi2021,Duan2025}, DNN assisted Bayesian inference based on individual measurement outcomes~\cite{Cimini2023,Nolan2021}, and feedback or control of quantum sensors~\cite{Cimini2019,Fallani2022,Vaidhyanathan2024}. Here, we address a different setting: history-dependent TPPs, in which correlations between successive detection events themselves carry parameter information and can enhance sensitivity. In addition, we focus on achieving fast inference, which is a key requirement for potential real-time sensing applications.

Here, we investigate the performance of three specific DL inference frameworks based on both \emph{regression} and \emph{classification} architectures~\cite{Goodfellow-et-al-2016}. While regression is utilised for direct point-parameter estimation and uncertainty quantification, the classification framework allows for a non-parametric reconstruction of the full posterior distribution. A comparative analysis of these schemes allows us to assess the trade-offs between computational efficiency and the fidelity of the inferred statistics in the quantum scenarios considered.

As presented in~\figref{fig:NN_models}, here the input to the neural network is always a TPP trajectory $D_\ell$. To ensure compatibility with the ABC and enhance learning efficiency, we allow for preprocessing the raw data via a summary statistic, i.e., $\bar{D}_{\ell} = \cS(D_\ell)$. Such a transformation can significantly improve DL performance if $\cS$ is chosen to distil the key dynamical features of photoclick patterns $D_\ell$. For instance, in scenarios where $D_\ell$ follow a renewal process, constructing a histogram of waiting times can be highly beneficial~\cite{Kiilerich2014,Radaelli2024,Rinaldi2024}, as it provides a sufficient statistic for the data representation.

\subsubsection{MSE-based estimation as a regression problem}
\emph{Regression}, a fundamental task in supervised machine learning, involves learning a mapping from input features to continuous-valued target variables~\cite{Goodfellow-et-al-2016}.
The general model description is depicted in Fig.~\ref{fig:NN_models}(a).
The model is trained by minimising the MSE between the predicted and true values across a training set $\cT=\{D^{(k)}\}_{k=1}^\nsamples$ of $\nsamples$ independent realisations:
\begin{equation}
    \mcl{L}_\MSE 
    \coloneqq \frac{1}{\nsamples} \sum_{k=1}^\nsamples \left|\left|\est{\estparV}(D^{(k)}) - \estparV^{(k)}\right|\right|^2,
    \label{eq:loss_MSE}
\end{equation}
where $\estparV^{(k)}$ refers to the vector of true parameter values for the $k$-th sample and $\est\estparV(D^{(k)})$ are the corresponding model’s predictions based on the input photoclick dataset $D^{(k)}$. Here, $||\cdot||$ denotes the Euclidean distance between them. The MSE loss penalises larger errors more heavily and is standard in regression tasks due to its smoothness and convexity properties.

Moreover, as demonstrated in~\refcite{Mingard2020}, for Stochastic Gradient Descent (SGD)-based Deep Neural Networks (DNNs), the probability of converging to a generalising function $f$  given a training set $\cT$, denoted $P_{\trm{SGD}}(f|\cT)$, closely approximates the Bayesian posterior over functions $P(f | \cT)$, particularly in the overparametrised regime. This similarity also holds for variants of SGD, such as Adam, Adagrad, Adadelta, and RMSprop~\cite{Ruder2016}. In cases where the Bayesian posterior is highly biased toward simple functions, SGD-like optimisers tend to sample these simple, better-generalising functions with high probability, behaving-to first order-like Bayesian samplers. This behaviour is consistent across various architectures, including Multi-Layer Perceptrons (MLP), Convolutional Neural Networks (CNN), and Long Short-Term Memory (LSTM) networks~\cite{Goodfellow-et-al-2016}. 

Consequently, the output of a regression DNN, $\est{\estparV}(D)=f(D)$, can be interpreted as an estimate derived from a learned mapping that structurally reflects the Bayesian posterior distribution over the function space. Hence, given that the aMSE \eref{eq:aMSE} is used as the training objective \eref{eq:loss_MSE}, the network converges to a point estimator that, as discussed around \eqnsref{eq:aMSE}{eq:posterior}, effectively approximates the posterior mean~\cite{Mingard2020}.

\subsubsection{Probabilistic regression under Gaussian approximation}
While standard regression models provide point estimates, typically corresponding to the posterior mean, they do not capture the structure of the posterior distribution. As a result, these models cannot quantify uncertainty or provide error estimates. A natural approach to address this limitation is through \emph{probabilistic regression} methods~\cite{Kneissl2025}, in which the network is forced to output also other parameters of the posterior distribution, whose form is assumed to belong to a particular class. A common and often effective way is to model the posterior distribution as approximately Gaussian~\cite{Nix1994,Lakshminarayanan2017}. 

Under this assumption, the network can be trained to predict both the parameters $\estparV$ and their covariance $\Sigma_{\estparV}$, as shown in~\figref{fig:NN_models}(b), by using the \emph{Negative Log-Likelihood} (NLL) as loss function, which encourages the model to output not only the best-fit prediction but also an estimate of the associated uncertainty:
\begin{align}
\mcl{L}_\NLL
  &\coloneqq \frac{1}{\nsamples} \sum_{k=1}^{\nsamples}
     \biggl[
       \bigl( \est{\estparV}(D^{(k)}) - \estparV^{(k)} \bigr)^{T}
       (\Sigma^{(k)})^{-1} \times \nonumber\\
  &\qquad
       \times \bigl( \est{\estparV}(D^{(k)}) - \estparV^{(k)} \bigr) + \frac{1}{2}\,\ln\bigl|\Sigma^{(k)}\bigr|
     \biggr],
     \label{eq:loss_NLL}
\end{align}
where $\est{\estparV}(D^{(k)})$ are the predicted parameters for the $k$-th data point $D^{(k)}$, and $\Sigma^{(k)}$ is their predicted covariance matrix. 
The true value of the parameter is denoted $\estparV^{(k)}$, and $\nsamples$ is the total number of samples. In order to efficiently compute the loss function, the Cholesky decomposition for the covariance is used, $\Sigma = AA^T$, where $A$ is a lower triangular matrix that can be parametrised by $d(d+1)/2$ values. Thus, strictly speaking, the network outputs the vector of predicted parameter values, $\est{\estparV}$, and the predicted entries of $A$, from which the covariance matrix, $\Sigma_{\est{\estparV}}$, is straightforwardly reconstructed.

\begin{figure*}[t]
    \centering
    \includegraphics[width=1.0\textwidth]{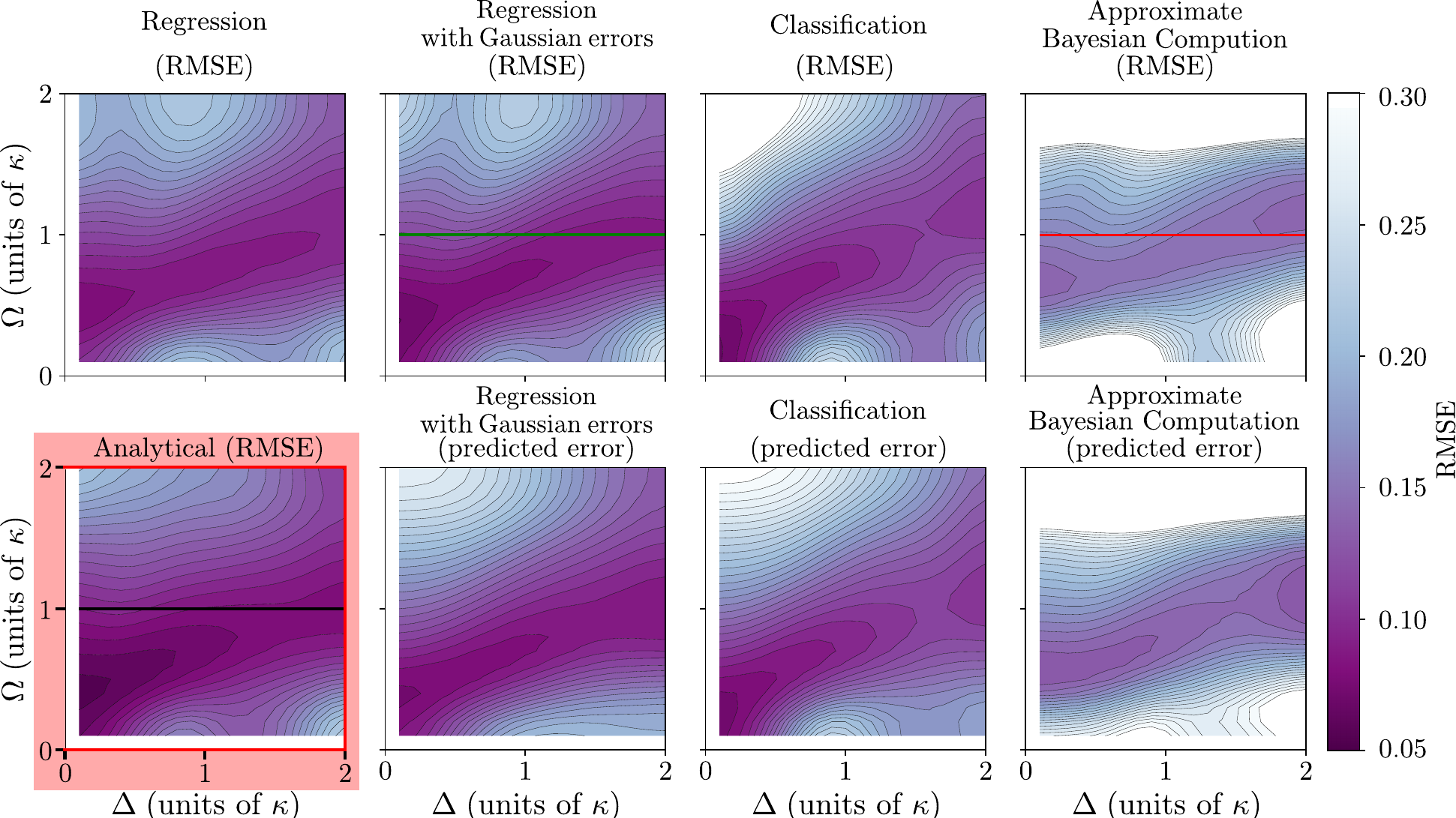}
    \caption{\textbf{Simultaneous estimation of the driving frequency $\Omega$ and the laser detuning $\Delta$ for a two-level atom.} We compare the joint RMSEs,
    $\sqrt{\MSE[\est{\Omega}]+\MSE[\est{\Delta}]}$, 
    achieved by the DL and ABC methods with the true RMSE (\emph{red frame}) computed analytically, as well as the errors predicted by the methods:~the variances of reconstructed posteriors for ABC and classification, and the covariance predicted in probabilistic regression. In each case a 2D plot of error (RMSE) is shown as a function of estimated parameter values $\Delta,\Omega\in[0,2]$ (in units of $\kappa$). \emph{Top row (from left to right)}:~RMSEs attained by the histogram-based regression network, Negative Log-Likelihood (NLL) histogram-based network, CNN classification network and by the ABC method. \emph{Bottom row}:~The true RMSE (in red) evaluated based on the waiting-time distribution followed by (\emph{from left to right}) the error predicted by the NLL histogram-based network, CNN classification network and by the ABC method. The horizontal lines, shown in black for the analytical result, green for Gaussian-error regression, and red for ABC, mark the $\Omega=1$ slice that is plotted in the lower panel of~\figref{fig:fisher_tls}.}
    \label{fig:two_level_atom_res}
\end{figure*}

\subsubsection{Posterior reconstruction by means of classification}
Finally, to enable the DL model to reconstruct the posterior distribution directly without prior assumptions regarding its functional form (e.g., Gaussianity), we employ the supervised \emph{classification} framework~\cite{Goodfellow-et-al-2016}. In this setting, the continuous parameter space is discretised into a $d$-dimensional grid of potential values. During training, each input photoclick pattern $D^{(k)}$ is associated with a discrete class label corresponding to its ground-truth parameters $\estparV^{(k)}$; however, the trained network predicts a probability distribution across the entire grid, providing a discrete approximation of the full posterior, as illustrated in~\figref{fig:NN_models}(c) (shown in one dimension for simplicity).

The model is trained using the \emph{Cross-Entropy} (CE) loss:
\begin{equation}
    \mcl{L}_\CE 
    \coloneqq -\sum_{k=1}^\nsamples \sum_{n=1}^{\#\trm{nbin}} \delta_{\estparV^{(k)},\estparV_n}\log\left[q(\estparV_n|D^{(k)})\right]
    \label{eq:loss_CE}
\end{equation}
where $q(\estparV_n|D^{(k)})$ is the output of the DNN associated with the bin labelled by $n$ evaluated for the input $D^{(k)}$, whereas $\estparV_n$ are the parameter values corresponding to that bin. In \appref{app:class_proof}, we show explicitly that minimisation of the CE \eref{eq:loss_CE} is equivalent to minimising the Kullback-Leibler (KL) divergence between the true posterior $p(\estparV|D)$ and the DL-based estimate $q(\estparV|D)$. 

Consequently, the resulting network output is a discrete probability mass function over the parameter grid, which directly approximates the posterior distribution. Importantly, this scheme bypasses the intractable evidence integral in \eqnref{eq:posterior}, as the posterior is learned directly from a synthetic dataset sampled from the joint distribution $p(D,\estparV)$ (recall \secref{subsec:photodect}).

\subsubsection{Neural network architectures used}
We evaluated several neural network architectures to identify the optimal configuration for each estimation task; further technical details are provided in \appref{app:DNN_details}. Additionally, all computational costs associated with data generation, training and evaluation are presented in \appref{app:comp_cost}. For \emph{regression} applied to renewal TPPs, we implement an architecture similar to that in \citeref{Rinaldi2024} with preprocessing, where the initial layer computes an empirical histogram of the input waiting times. For history-dependent TPPs, we employ a network composed of LSTM layers~\cite{Hochreiter1997}, followed by a fully connected network. The LSTM cells are a form of recurrent neural network (RNN) architecture specifically designed to process sequential data~\cite{Goodfellow-et-al-2016}. By maintaining an internal memory that enables the model to learn complex temporal dependencies, they are well-suited for analysing history-dependent TPP trajectories. 

For point estimation using the MSE loss, the output layer provides a single prediction for each parameter. In the case of \emph{probabilistic regression} (NLL loss), the output layer includes additional neurons to parametrise the covariance matrix. Finally, for the \emph{classification} framework, we use one-dimensional convolutional layers (CNN) with max-pooling, followed by a fully connected network. The output layer represents a probability mass function over the discretised parameter grid, from which we derive point estimates by calculating either the mode or the mean of the resulting posterior distribution.
Each of these approaches relies on supervised learning, in which all system parameters other than those being estimated are held fixed.

\subsection{Verification:~Multiparameter sensing with a two-level atom}
\label{sec:two_level_atom}
To verify the ABC and DL approaches to estimation, we first test them in the simple setting of a laser-driven two-level atom, described by the Hamiltonian~\cite{Breuer2007}:
\begin{equation}
    H^{\text{atom}} = -\hbar \Delta\hat{\sigma}^+ \hat{\sigma}^- + \frac{\hbar \Omega}{2} (\hat{\sigma}^- + \hat{\sigma}^+),
\end{equation}
where $\hat{\sigma}^+ (\hat{\sigma}^-)$ is the atomic raising (lowering) operator. Our goal is to estimate both the laser detuning $\Delta$ and the driving frequency $\Omega$. 

For the purpose of verification, we assume a single light detection channel of unit efficiency, as described in \secref{subsec:photodect}, with the $j$-index being redundant and $\kl=0$. Crucially, as the atom resets to the ground state after every emission event,  photoclick trajectories form a renewal TPP;~hence, their statistics are fully characterised by the waiting-time distribution \cite{Carmichael1989}. Consequently, in this special case, both the likelihood and the posterior distribution can be determined analytically~\cite{Kiilerich2014, Clark2025}, providing an ideal benchmark to verify the accuracy of our likelihood-free methods.

\subsubsection{Local estimation performance}

The estimators are constructed globally, by minimising the averaged error over the full parameter domain, however we quantify their performance from a local-estimation perspective.
In~\figref{fig:two_level_atom_res}, we present the two-parameter estimation performance for the proposed likelihood-free methods. Since the photoclicks in this model constitute a renewal process, the inference problem can be simplified by pre-processing the trajectories into histograms of waiting times without loss of information. We show the joint RMSEs attained by each method in the top row of~\figref{fig:two_level_atom_res}, which should be compared directly with the analytical solution provided in the bottom-left panel. Furthermore, we display the self-estimated error predictions in the bottom row; for both ABC and the CNN classification, the predicted error corresponds to the variance of the reconstructed posterior, whereas for the probabilistic regression, it is given by the covariance output by the network.

When implementing ABC (fourth column in~\figref{fig:two_level_atom_res}), we include the total time as an additional summary statistic alongside the histogram to improve the efficiency of the acceptance criteria~\cite{Clark2025}. In the case of classification (third column in~\figref{fig:two_level_atom_res}) and its CNN implementation, we do not preprocess the data, as the network relies on convolution filters applied directly to the photoclick pattern.

All tested methods show good agreement with the analytical results, for both the estimation of parameters and the prediction of errors, which underlines the multiparameter estimation capabilities of the likelihood-free ABC and DL approaches.

\begin{figure*}[t]
    \centering
    \includegraphics[width=0.8\textwidth]{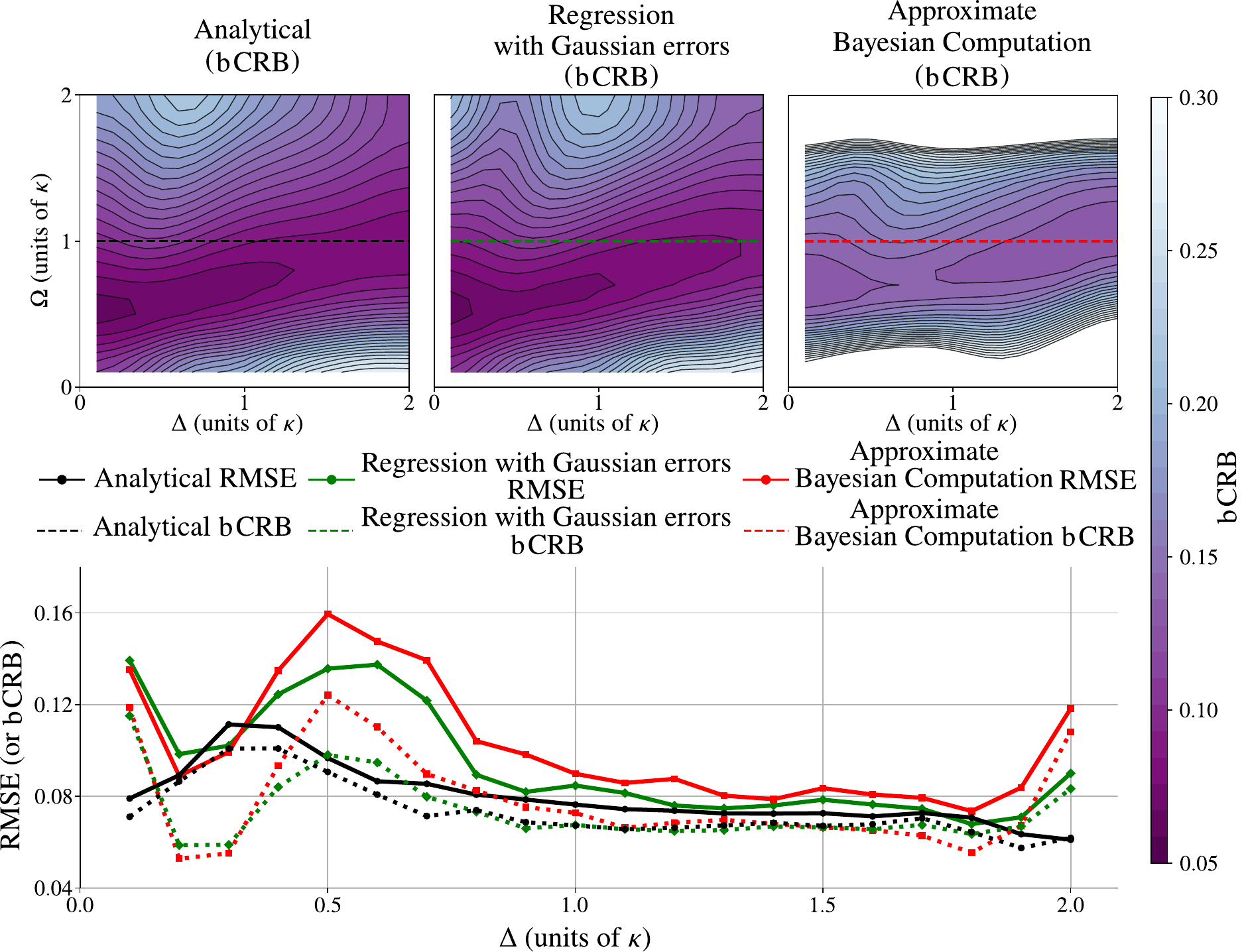}
    \caption{
    \textbf{Biased Cramér-Rao bound (bCRB) for simultaneous estimation of the driving frequency $\Omega$ and laser detuning $\Delta$ for a two-level atom.}
    We compare the analytical bCRB with the bounds obtained from the uncertainty estimates produced by the regression network with Gaussian errors and by the ABC method. 
    The top row shows 2D maps of the joint bound,
    $\sqrt{\mathrm{bCRB}_{\Delta}+\mathrm{bCRB}_{\Omega}}$,
    as a function of the parameter values $\Delta,\Omega\in[0,2]$ in units of $\kappa$.
    The dashed horizontal lines indicate the slice at $\Omega=1$ used in the lower panel.
    From left to right: analytical bCRB, bCRB inferred from the Gaussian-error regression model, and bCRB inferred from the ABC posterior.
    The bottom panel compares the corresponding one-dimensional slices at $\Omega=1$ with the true RMSEs obtained from the analytical calculation, the Gaussian-error regression model, and ABC.
    }
    \label{fig:fisher_tls}
\end{figure*}

\subsubsection{Comparison with the bCRB}

As a further verification of the proposed inference methods, we compare each estimator's error with the bCRB introduced in~\eqnref{eq:biased_CRB}. The estimators considered here are not guaranteed to be locally unbiased. Indeed, both the ABC and the NN are constructed or trained under a global objective, and therefore may exhibit a non-zero local bias. In the top row of~\figref{fig:fisher_tls} we show the resulting comparison for the simultaneous estimation of bCRB for the detuning $\Delta$ and the driving strength $\Omega$. We compare it with the RMSE estimations produced by ABC and NNs.

The agreement between the analytical bCRB and the bounds obtained from the inferred posterior errors confirms that the developed models do not merely provide accurate point estimates, but also capture the local parameter dependence of the estimation uncertainty. In particular, in the bottom row of~\figref{fig:fisher_tls}, we show one-dimensional slices at $\Omega=1$ that both ABC and the probabilistic-regression network reproduce. They match the main dependence of the bound on $\Delta$, including the regions where the estimation problem becomes locally harder. Small discrepancies are expected because the learned or reconstructed posteriors are obtained from finite datasets and, in the case of ABC, from a finite acceptance tolerance.

This analysis completes the verification stage for the two-level atom. Since this system generates a renewal photoclick process, both the likelihood and the FI are verified with numerical integration~\cite{Kiilerich2014}. The observed agreement therefore validates the use of ABC and DL not only as estimators, but also as tools for uncertainty quantification.

\begin{figure}[t]
    \centering
    \includegraphics[width=1.0\columnwidth]{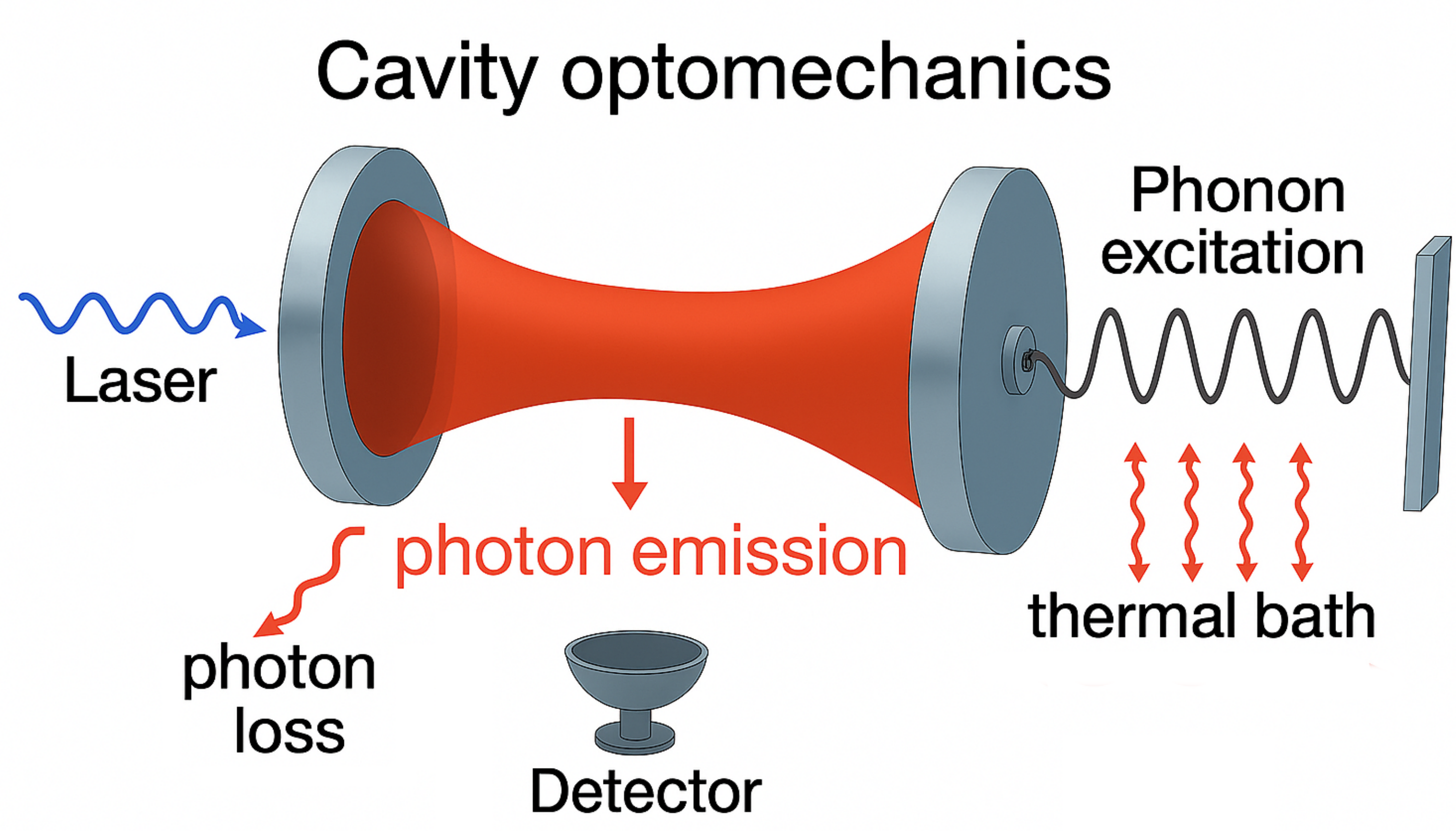}
    \caption{\textbf{Schematic of a generic optomechanical sensor.} An external laser drive populates the optical cavity mode, which couples to a mechanical resonator (e.g., a vibrating mirror, a levitated nanoparticle, or a photonic crystal mode) via radiation pressure. This interaction imprints the mechanical dynamics onto the cavity field, which is then probed by monitoring the photons leaking out via photodetection. The resulting photoclick patterns provide the data for our likelihood-free inference methods.}
    \label{fig:optomech_system}
\end{figure}

\section{Results:~non-linear optomechanical sensor}
\label{sec:results}
%
We now focus on the main application motivating our work, depicted schematically in~\figref{fig:optomech_system}:~a laser-driven optomechanical device~\cite{Aspelmeyer2014} whose output light is measured by photodetection. Crucially, when operated in the non-linear regime-i.e.~when the coupling between its optical (cavity photons) and mechanical (phonons) degrees of freedom is significant relative to other parameters-the system yields non-classical multiclick correlations that can be leveraged for sensing. Moreover, these correlations can be directly controlled by varying the detuning $\Delta$ of the laser driving the cavity~\cite{Kronwald2013,Clark2022}. 

This allows us to study the capabilities of both ABC and DL inference frameworks in capturing correlations in photoclick patterns that must be described by history-dependent TPPs, as introduced in \secref{sec:TPPs}. However, in contrast to the case of the two-level atom which yields a renewal TPP, an analytic form for the likelihood or the posterior distribution-and, by extension, the minimal aMSE \eref{eq:aMSE} or the optimal estimator \eref{eq:opt_estimator}-is here intractable. Furthermore, the numerical complexity involved in computing these precludes a direct analysis of optimal estimation performance, as it would require evaluating averages over a vast number of photoclick trajectories. In this regard, the two likelihood-free approaches, ABC and DL, offer the first practical solutions.

Applications of optomechanical devices range from external force and acceleration sensing~\cite{Guha2020,Haelg2021} to broadband displacement and force measurements at and below the standard quantum limit~\cite{Kampel2017,Mason2019}. More recently, schemes involving multiple sensors for distributed force sensing were demonstrated~\cite{Xia2023} and proposed for dark-matter field searches~\cite{Brady2023}. These applications are implemented across various platforms, including optically levitated nanoparticles~\cite{Hebestreit2018,Tseng2025} and integrated photonic crystals~\cite{Krause2012,Guo2017}, the latter of which frequently rely on discrete photodetection and photon-correlation statistics~\cite{Cohen2015,Hong2017}.

Typically, these scenarios involve estimating parameters ($\estparV$ in~\figref{fig:system_setup}) that enter the mechanical Hamiltonian in a linear fashion, while control parameters ($\contparV$ in~\figref{fig:system_setup}) are properties of the driving laser. However, to evaluate the capacity of ABC and DL for extracting information from multiphoton correlations, we test them in a scenario where the drive detuning, $\Delta$, is the parameter to be estimated. While $\Delta$ could be estimated in the absence of mechanics, its variations here directly translate into higher-order correlations emerging via the non-linear interaction with the mechanics-features we intend to capture using ABC and DL inference frameworks.

\subsubsection{Optomechanical dynamics}
The generic optomechanical Hamiltonian relevant for the setup of \figref{fig:optomech_system} is given by~\cite{Aspelmeyer2014}:
\begin{align}
    \hat{H} & = - \hbar \Delta\, \hat{a}^\dagger \hat{a} + \hbar \omega_M\, \hat{b}^\dagger \hat{b}       \label{OM_hamiltonian}\\
      & \quad + \frac{\hbar \Omega}{2} (\hat{a} + \hat{a}^\dagger) + \hbar g\, \hat{a}^\dagger \hat{a} (\hat{b} + \hat{b}^\dagger), \nonumber
\end{align}
where $\hat{a}$ and $\hat{b}$ are the bosonic annihilation operators corresponding to the photonic and phononic modes, respectively. The parameter $\Delta$ is the laser detuning, $\omega_M$ is the mechanical oscillator frequency, $\Omega$ is the laser driving strength, and $g$ characterises the optomechanical coupling.

To capture decoherence effects, we model dissipation through several decay channels. Photons leak out of the optical cavity at a rate $\kappa$, while the mechanical oscillator is thermally coupled to a bath, with a damping rate $\gamma$, allowing phonons to be spontaneously created or annihilated. The resulting evolution is described by the GKSL master equation~\cite{Breuer2007}: 
\begin{equation}
\begin{aligned}
\label{ensamble_master_eq}
    \frac{\dd  \hat{\rho}}{\dd  t} = & - \frac{\ii}{\hbar} [\hat{H}, \hat{\rho}] + \kappa \left( \hat{a} \hat{\rho} \hat{a}^\dagger - \frac{1}{2} \{ \hat{a}^\dagger \hat{a}, \hat{\rho} \} \right) \\
    & + \gamma(\bar{m} + 1) \left( \hat{b} \hat{\rho} \hat{b}^\dagger - \frac{1}{2} \{ \hat{b}^\dagger \hat{b}, \hat{\rho} \} \right) \\
    & + \gamma \bar{m} \left( \hat{b}^\dagger \hat{\rho} \hat{b} - \frac{1}{2} \{ \hat{b} \hat{b}^\dagger, \hat{\rho} \} \right),
\end{aligned}
\end{equation}
where $\bar{m} = \left[\exp\left( \hbar \omega_M / k_B T \right) - 1 \right]^{-1}$ is the mean phonon occupation number, dictated by the mechanical frequency $\omega_M$ and the temperature $T$ of the bath.

\subsubsection{Simulating photoclick patterns}
Following the unravelling procedure described in \secref{subsec:photodect}, we define the conditional Hamiltonian with multiple decay channels:
\begin{equation}
\label{cond_H}
    \hat{H}_{\mrm{cond}} = \hat{H} - \frac{\ii\hbar}{2}(\kappa \hat{a}^\dagger \hat{a} + \gamma(\bar{m}+1)\hat{b}^\dagger \hat{b} + \gamma \bar{m} \hat{b} \hat{b}^\dagger).
\end{equation}
This leads to the fully unravelled stochastic master equation~\eqref{full_unravel} of the form:
\begin{align}
\label{unravelling_SME}
    \dd \hat{\rho} 
    &= -\frac{\ii}{\hbar} [\hat{H}_{\mrm{cond}} \hat{\rho} - \hat{\rho} \hat{H}^\dagger_{\mrm{cond}}]  \ \dd t \\
    &\qquad + \mcl{J}_{\kd}(\hat{\rho}) + \mcl{J}_{\kl}(\hat{\rho}) + \mcl{J}_{\gamma_-}(\hat{\rho}) + \mcl{J}_{\gamma_+}(\hat{\rho}), \nonumber
\end{align}
where we partition loss into detected ($\kd$) and undetected ($\kl$) channels ($\kd + \kl = \kappa$), and obtain also the jump processes, as defined in \eqnref{jump-super_ops}, responsible for phonon creation ($\gamma_+$) and annihilation ($\gamma_-$). We generate  waiting-time trajectories $D_\ell$ by simulating click patterns until $\ell$ photon detections are registered, while ignoring jumps of other types.

In the analysis that follows, we generate photoclick patterns by setting the optomechanical parameters to values consistent with \citeref{Clark2025}, as summarised in \tabref{tab:optomech_pars}. These values correspond to operation in the sideband-resolved regime~\cite{Aspelmeyer2014}. However, we stress that while sideband-resolved dynamics are not strictly required for the inference tasks considered~\cite{Clark2022}, they provide a realistic framework for generating the non-renewal photoclick patterns analysed herein. Furthermore, this regime and the choice of red (negative) detuning, see below, align with the standard physical intuition from linear optomechanics:~such a configuration is typically required in sensing applications to avoid heating the mechanical resonator during the measurement process~\cite{Aspelmeyer2014}.

\begin{table}[t!]
\centering
    \begin{tabular}{lccc}
        \hline
        \hline
        Parameter & Symbol & Value & Units \\
        \hline
        Cavity decay rate               & $\kappa$          & $1$       & (scaled) \\
        Photon detection rate           & $\kd$        & $0.9$    & $\kappa$ \\
        Mechanical frequency            & $\omega_M$        & $4\sqrt{2}$    & $\kappa$ \\
        Mechanical damping              & $\gamma$          & $10^{-3} \omega_M$ & $\kappa$ \\
        Optomechanical coupling         & $g$             & $4$     & $\kappa$ \\
        Mean phonon number              & $\bar{m}$ & $1$       & -      \\
        Laser driving strength          & $\Omega$               & $0.3\omega_M$    & $\kappa$ \\
        Detuning range                  & $\Delta$          & $[-10, 0]$  & $\kappa$ \\
        \hline
        Dark count rate (DCR)          & $\lambda_{\text{DCR}}$ & $10^{-2}$ & $\kappa$ \\
        \hline
        \hline
    \end{tabular}
    \caption{\textbf{Optomechanical parameters} assumed in photoclick pattern simulations following \citeref{Clark2025}, supplemented by the dark count rate (DCR) consistent with \citeref{Galinskiy2020}.}
    \label{tab:optomech_pars}
\end{table}

\subsubsection{Controlling statistics of emitted photons}
In the non-linear regime, radiation pressure induces an energy shift proportional to the square of the cavity photon number, $n_{\trm{cav}}$. This shift is reflected in the system's energy spectrum~\cite{Kronwald2013, Aspelmeyer2014}:
\begin{equation}
    E(n_{\trm{cav}}, n_{\trm{mech}}) = -\hbar \Delta\,n_{\trm{cav}} + \hbar \omega_M n_{\trm{mech}} - \hbar \frac{g^2}{\omega_M}n_{\trm{cav}}^2,
\end{equation}
where $n_{\trm{cav}}$ and $n_{\trm{mech}}$ are the cavity and mechanical occupation numbers, respectively. 

By selecting specific detuning regimes, $\Delta_n = -n \frac{g^2}{\omega_M}$ (where $n \in \mathbb{Z}$), one can cancel the non-linear term for a target photon number $n$. For $n=1$, single-photon transitions are energetically favoured. For higher integers $n \geq 2$, the cavity resonance is restored specifically for the $n$-photon manifold, forcing photons to be emitted in correlated groups of size $n$, resulting in bunched light in the form of pairs ($n=2$), triplets ($n=3$), or larger $n$-photon clusters~\cite{Aspelmeyer2014}. 

These cases are a direct manifestation of photoclick patterns constituting a history-dependent TPP. The delayed mechanical response ensures that once an initial photon is emitted, the cavity remains in a favourable state for $n-1$ subsequent emissions. This mechanism generates the temporal multiclick correlations that contain the information about system parameters we aim to capture using ABC and DL inference frameworks.

\begin{figure*}[t]
    \centering
    \includegraphics[width=\textwidth]{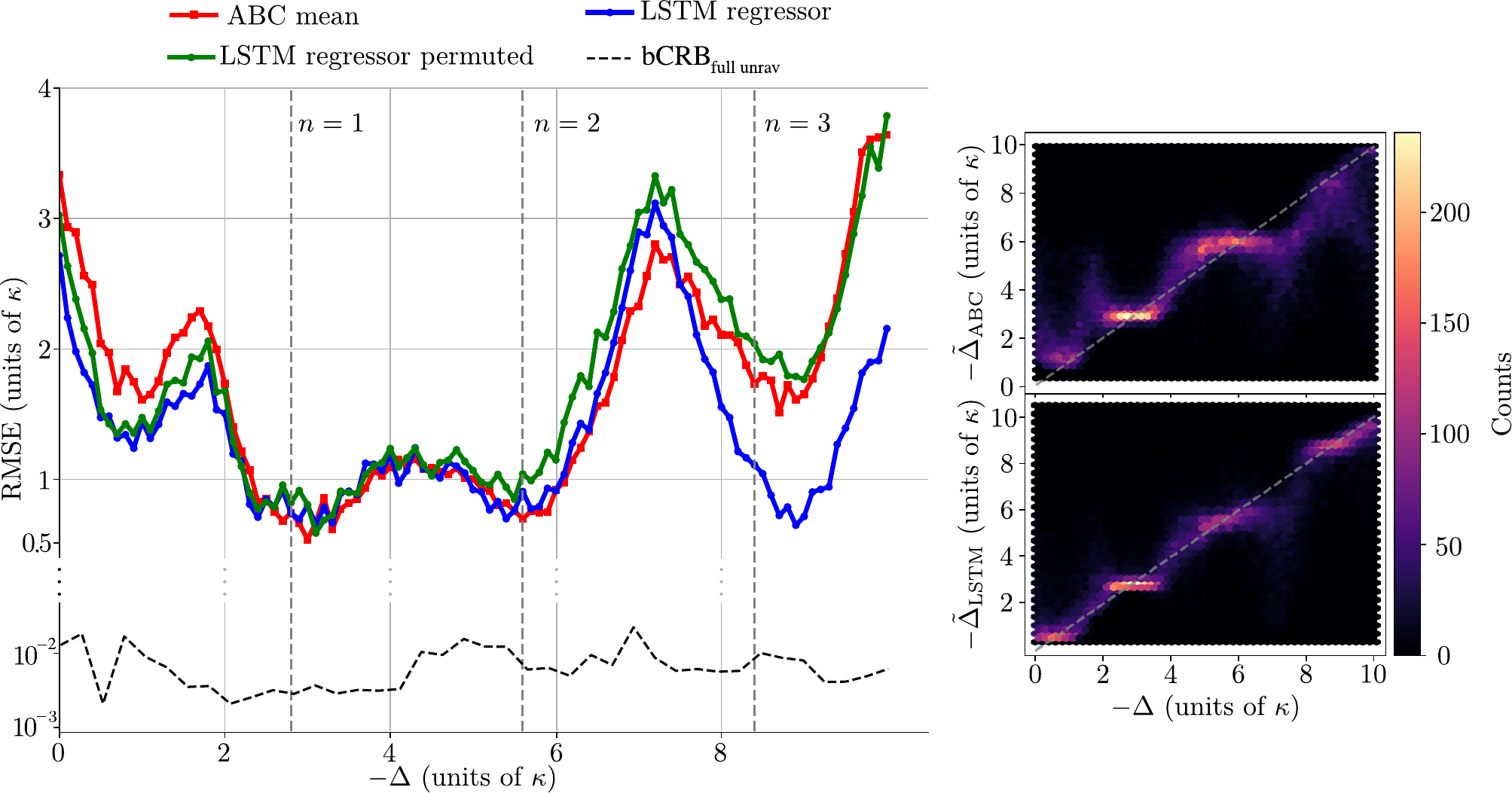}
    \caption{\textbf{Root Mean Squared Error (RMSE) attained with ABC and DL-regression frameworks} as a function of the true detuning $\Delta$. In the \emph{main plot}, the ABC estimator is constructed as the mean of the reconstructed posterior (\emph{red}), while the regression framework relies on an LSTM architecture trained on raw waiting-time trajectories (\emph{blue}). The performance of the LSTM regressor when applied to randomly permuted input trajectories is also shown (\emph{green}). Vertical dashed grey lines indicate the special detuning regimes ($n = 1, 2, 3$) associated with resonance photon statistics. The \emph{right panels} display one-to-one plots of the estimated values $\est{\Delta}_{\text{ABC}}$ (top) and $\est{\Delta}_{\text{LSTM}}$ (bottom) against the true parameter $\Delta$, where colour intensity represents prediction density. The regression framework demonstrates superior accuracy, particularly in the $n = 3$ regime; here, the performance degradation under permutation confirms the model's ability to exploit history-dependent temporal correlations.
    For comparison with \figref{fig:fisher_tls}, we also display the bCRB for the fully unravelled dynamics (\emph{dashed black}, shown on a logarithmic scale). As follows directly from \eqnref{eq:full_unrav_ineq}, this curve provides a fundamental lower bound to the bCRB of the marginalised photodetection record. As discussed in the main text, this is because a significant amount of information about the detuning is obtained from the otherwise unobserved phonon jumps (excitations and relaxations) and, since these jumps render the conditional dynamics pure, the Fisher information may also scale differently with the measurement time (the numerical variance of the bCRB from trajectory averaging is of order $10^{-6}$ and hence negligible).
    }
    \label{fig:RMSE_LSTM}
\end{figure*}

\subsection{Estimation performance}
The performance of our inference frameworks-Approximate Bayesian Computation (ABC) and Deep Learning (DL) regression-is compared in~\figref{fig:RMSE_LSTM}. Our primary objective is to evaluate their ability to approximate the posterior mean, which theoretically minimises the aMSE \eref{eq:aMSE}, averaged over the prior. To ensure a consistent and fair comparison, we assume a uniform (flat) prior over the parameter range, matching the distribution used to generate the training data. Crucially, the same dataset of waiting-time trajectories serves both as the training set for the neural networks and as the reference library for the ABC algorithm. Details of computational costs and average estimation times for our methods can be found in~\appref{app:comp_cost}.

While these estimators are constructed globally over the prior, we evaluate their local performance (recall \secref{subsec:local_estim}) pointwise across the parameter range. Specifically, we quantify the precision of our predictions as a function of the true detuning $\Delta$ using the Root Mean Squared Error (RMSE), alongside the one-to-one estimation plots in~\figref{fig:RMSE_LSTM}. We observe the emergence of plateaus in one-to-one estimation and subsequently address them through bias compensation.

\subsubsection{Approximate Bayesian Computation}
We perform estimation over the range $\Delta \in [-10,0]$ by constructing the ABC posterior for each dataset $D_\ell$ and extracting its mean as the parameter estimator. The ABC algorithm utilises two summary statistics applied to each dataset $D'_\ell$  (each consisting of $\ell=80$ photoclicks) drawn from the precomputed library:~the average waiting time $\bar{t}_\ell = \frac{1}{\ell}\sum_{i=1}^{\ell} t_i$, and the $L^2$-norm between waiting-time histograms:
\begin{equation}
    \delta(\cS(D_\ell), \cS(D'_\ell)) = \sqrt{\sum_i(x_i - x^\prime_i)^2},
\end{equation}
where $\cS(D_\ell)$ and $\cS(D'_\ell)$ are the time-binned histograms of the waiting times for the observed and library-drawn datasets, respectively, with $x_i$ and $x^\prime_i$ denoting the corresponding bin counts. 

As seen in~\figref{fig:RMSE_LSTM} (red), the RMSE attained by ABC exhibits distinct dips near all resonance values. Although the average waiting time statistic is sufficient for the $n=1$ regime, in the $n=2$ and $n=3$ regimes, rapid successive emissions manifest as a shift in the histogram density toward shorter intervals~\cite{Clark2025}. Notably, the RMSE dip for $n=3$ is significantly higher than for $n=1$ or $n=2$. This limitation suggests that while the chosen summary statistics extract relevant physical information, their reliance on aggregated data fails to capture the information contained within the temporal ordering of photon emissions that becomes critical at $n \ge 3$.

This gap could, in principle, be closed by incorporating more sophisticated summary statistics capable of capturing higher-order structural dependencies, such as the Wasserstein metric~\cite{Arjovsky2017}. However, computing such optimal transport distances across large datasets would make ABC completely intractable. Alternatively, one could resort to hybrid strategies like Neural ABC~\cite{Fearnhead2010, Wong2018, Aakesson2020}, which automatically learns optimal low-dimensional summary statistics from the data. However, as presented here, DL models perform far more efficiently when deployed on their own directly.

\begin{table*}[t]
    \centering
    \footnotesize
    \begin{tabular}{c|c c c||c c}
    \toprule
    \multirow{2}{*}{\textbf{\makecell{Detuning \\ Regime}}} 
    & \multicolumn{3}{c||}{\textbf{ABC}} 
    & \multicolumn{2}{c}{\textbf{DL}} \\
    \cmidrule(lr){2-4} \cmidrule(lr){5-6}
     & \textbf{Summary Statistics} & \textbf{Sufficiency} & \textbf{Explanation} 
     & \textbf{\makecell{Summary Statistics \\ Learnt}} & \textbf{Verification} \\
    \midrule
    $n = 1$ & total time & $\checkmark$ & \makecell{Effectively \\ a renewal TPP.} & total time & PCA analysis, see \appref{pcs_analysis}. \\
    $n = 2$ & total time + histogram & $\checkmark$ & \makecell{Photon-pair emission \\ captured by the histogram.} & total time + histogram & \makecell{Invariance to random \\ permutations,  see \figref{fig:RMSE_LSTM}.} \\
    $n = 3$ & total time + histogram & \xmark & \makecell{A history-dependent TPP \\ with third order correlations.} & beyond histogram & \makecell{Strong sensitivity to \\ trajectory ordering, see \figref{fig:RMSE_LSTM}.} \\
    \bottomrule
    \end{tabular}
    \caption{\textbf{Summary statistics crucial to each of the $n=1,2,3$ detuning regimes}. The sufficiency of the summary statistics employed by the ABC algorithm is verified. These are compared with ones learnt by the DL model for each $n$.}
    \label{tab:summary_n_regimes}
\end{table*}

\subsubsection{Neural Network Regression}
To evaluate the DL-based regression approach, we trained and evaluated models based on LSTM, CNN, and Transformer architectures. Here, we present results only for the architecture that achieved the highest performance:~the LSTM regressor. Further details regarding the network hyperparameters and training procedures are provided in \appref{app:DNN_details}.

The LSTM regressor (\figref{fig:RMSE_LSTM}, blue) operates directly on raw trajectories without prior physical knowledge. Its performance matches that of ABC at lower $\Delta$, particularly in the $n=1$ and $n=2$ regimes, but it significantly outperforms ABC at $n=3$, where it yields an RMSE dip comparable to the other resonances. The equivalence at $n=1$ is clarified by the Principal Component Analysis (PCA) in \appref{pcs_analysis}, which demonstrates that the total arrival time is the most informative feature for the network to learn in this regime. 

To explicitly study the role of the temporal structure learned by the LSTM, we evaluate its performance on \emph{randomly permuted} trajectories (\figref{fig:RMSE_LSTM}, green). Comparing the blue (raw) and green (permuted) curves reveals the extent to which the network extracts information from higher-order temporal correlations. While the improvement near $n=2$ is negligible-confirming that the histogram statistic is largely sufficient for pair-bunching-the LSTM achieves significantly lower RMSE in the $n=3$ regime. Here, random permutations destroy the specific ordering critical to the $n=3$ manifold, where the LSTM (blue) exploits history-dependent correlations that the ABC summary statistics discard. Conversely, since the LSTM under random permutations (green) is only mildly outperformed by the ABC (red) at $n=3$, this suggests that the neural network successfully captures the detuning information contained within the histogram statistics even when temporal ordering is lost.

These results are summarised in \tabref{tab:summary_n_regimes}, which lists the summary statistics used by the ABC algorithm alongside those deduced for the LSTM regressor. It evaluates their respective sufficiency for the $n=1, 2, \text{ and } 3$ detuning regimes, providing an explanation for whether the chosen features are adequate to capture the underlying photoclick dynamics.

\subsubsection{Comparison with the fully unravelled biased Cram\'{e}r-Rao bound}

In addition to the RMSEs of the different inference frameworks, in \figref{fig:RMSE_LSTM} we show the lower bound on them obtained by computing the bCRB \eref{eq:biased_CRB} of the fully unravelled dynamics for the non-linear optomechanical system, as described in \secref{subsec:UB_FI_matrix}. The corresponding FI cannot be computed as simply as for the two-level atom in \secref{sec:two_level_atom}, as the jump processes are non-renewal, so the FI of a fixed-click trajectory cannot be reduced to $\ell$ independent single-waiting-time contributions, i.e., $\FI[D_\ell|\estparV] \neq \ell\,\FI[w(\tau|\estparV)]$, and the prescription of \secref{sec::trajectory_FI} must be followed. 

As is clear from \figref{fig:RMSE_LSTM}, the fully unravelled FI is much larger than that available from only the observed photodetection record. For the parameters in \tabref{tab:optomech_pars}, the inaccessible processes of phonon jumps (relaxations and excitations) carry a significant amount of additional information owing to the larger number of events: for $80$ detected photoclicks we get on average between $\approx\!200$ (small detunings) and $\approx\!25$ (large detunings) virtual phonon (creation and annihilation) jumps. Furthermore, since all jumps are then perfectly monitored, including those associated with undetected photons, the conditional dynamics is pure, which implies that the fully unravelled Fisher information may grow even quadratically with the measurement time~\cite{Albarelli2018}, instead of the linear scaling emerging in dissipative dynamics~\cite{Macieszczak2016}. Together, see \figref{fig:RMSE_LSTM}, these two effects cause the so-obtained bCRB to lie well below the RMSEs achieved by ABC- and DL-based estimators. Tightening the bCRB by tracing out the inaccessible degrees of freedom is a topic of ongoing research~\cite{Khan2025,Yang2025}, but beyond the scope of our work.

\begin{figure*}[t]
    \centering
    \includegraphics[width=0.85\linewidth]{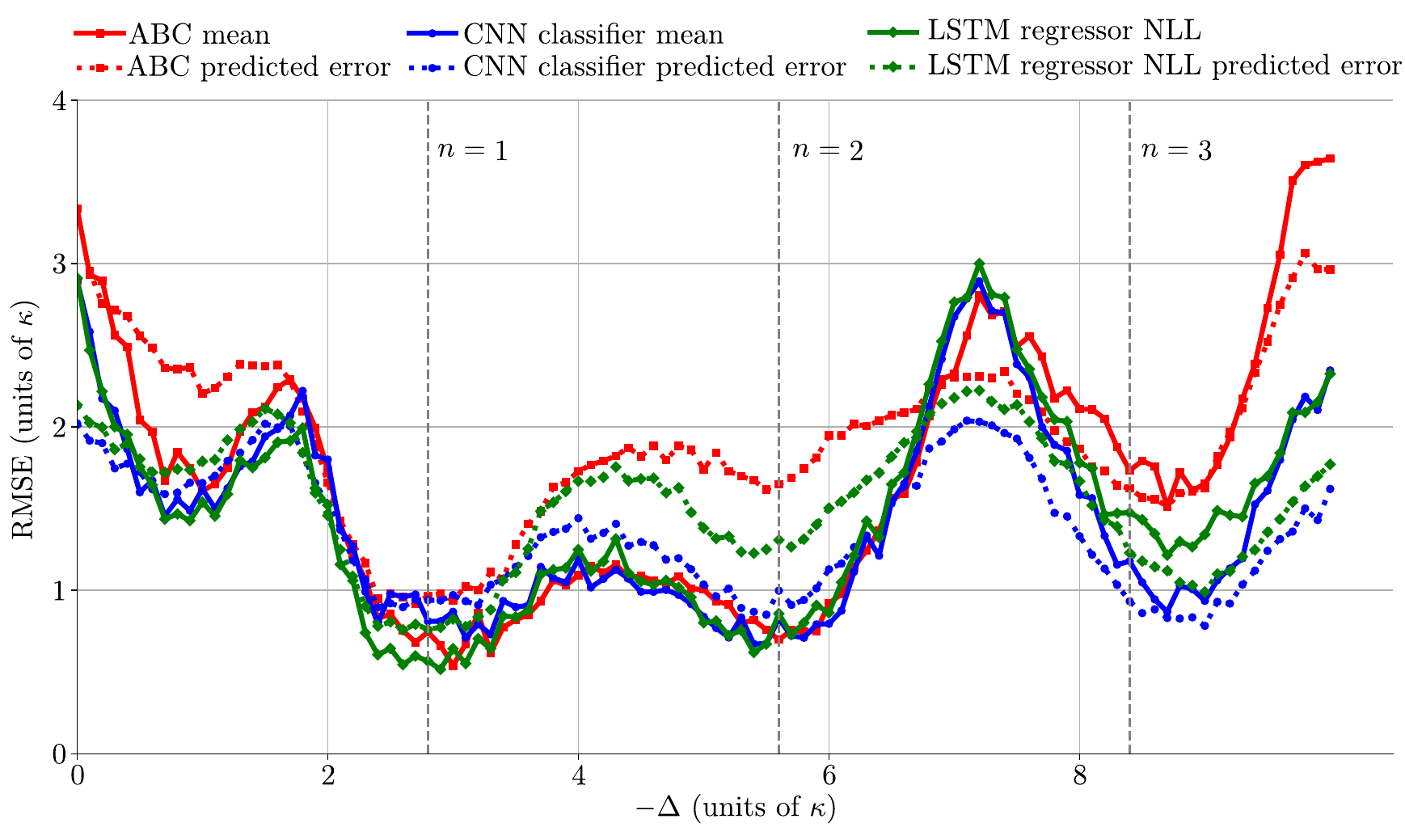}
    \caption{\textbf{Root Mean Squared Errors (RMSEs) attained with ABC, DL-based probabilistic regression, and classification frameworks, presented alongside their predicted uncertainties} as a function of the true detuning $\Delta$. Vertical dashed grey lines indicate the special detuning regimes ($n = 1, 2, 3$). \emph{Solid lines} represent the RMSE for:~ABC (\emph{red}, same curve as in~\figref{fig:RMSE_LSTM}), the LSTM regressor trained with NLL loss (\emph{green}), and the CNN classifier (\emph{blue}). \emph{Dashed lines} indicate the predicted uncertainty for each method. The CNN classifier consistently provides the lowest RMSE while maintaining high calibration across the entire parameter range; in particular, at $n=2$-where all three methods achieve comparable RMSE-the ABC framework (\emph{red}) heavily overestimates its RMSE. Similarly, the LSTM regressor based on NLL also fails to accurately predict its error in this regime, albeit less severely, whereas the CNN continues to provide a reasonable self-estimate of its error. Overall, the LSTM regressor (\emph{solid green}) outperforms the ABC framework in terms of both RMSE and uncertainty calibration across the entire parameter range. However, it is surpassed by the CNN classifier at $n=3$, where the Gaussian approximation of the posterior fails to capture the non-Gaussian statistics associated with photons being emitted in triplets.}
    \label{fig:RMSE_all_with_errors}
\end{figure*}
\subsubsection{Mitigating Estimation Bias}
The right panels of \figref{fig:RMSE_LSTM} show the one-to-one comparison between estimated and true parameters. Both frameworks exhibit plateau-like behaviour, though the spread is markedly smaller for the LSTM. The significant degradation of ABC near $n=3$ highlights the limitation of histogram-based statistics in capturing high-order temporal structures.

These plateaus arise because RMSE minimisation inherently prioritises low estimator variance over low bias. While this bias can be reduced-either by weighting the loss function or by using the posterior mode as the estimate-such improvements generally come at the cost of increased overall RMSE, as discussed in \appref{app:red_bias}.

\subsection{Estimation with predicted uncertainty}
To further assess the performance and reliability of our inference frameworks, we employ estimation procedures that not only provide point estimates but also meaningful measures of predictive uncertainty. These quantifiers enable us to evaluate the \emph{calibration} of each framework-defined by how closely the predicted uncertainty aligns with the actual spread of estimation errors (RMSE). By analysing the interplay between estimates and their associated uncertainties, we identify where specific frameworks fail to capture the underlying physics of the photoclick trajectories.

\begin{figure*}[t]
    \centering
    \includegraphics[width=.85\textwidth]{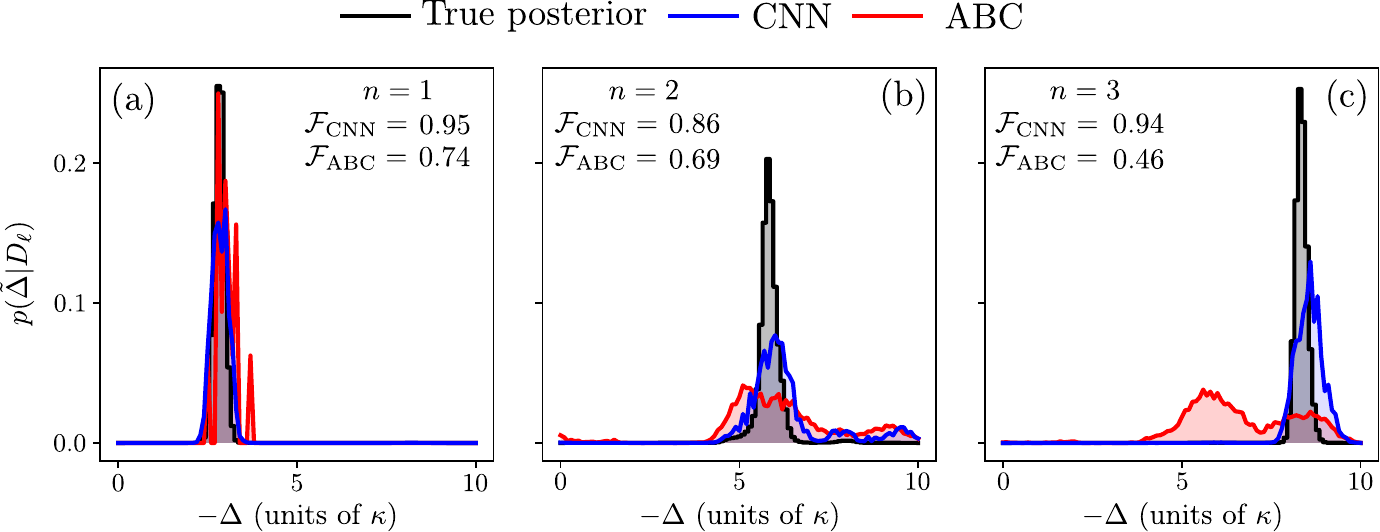}
    \caption{\textbf{Posterior distribution reconstruction} for representative photoclick trajectories across the $n = 1, 2, \text{ and } 3$ resonance regimes. Each subplot compares the ground-truth posterior distribution (\emph{black}) with the distributions reconstructed using ABC (\emph{red}) and the CNN classifier (\emph{blue}). The subplots also display the Bhattacharyya coefficient $\mathcal{F}$ for both methods, quantifying the reconstruction fidelity. The comparison visualises the limitations of ABC at higher resonances-specifically the overestimation of variance at $n=2$ (cf.~\figref{fig:RMSE_all_with_errors}) and the failure to concentrate on the true parameter value at $n=3$ (cf.~\figref{fig:RMSE_LSTM})-which arise because ABC relies on aggregated statistics. In contrast, the CNN closely approximates the true posterior in all regimes, including $n=3$ at which it is able to learn the history-dependent features of the full photoclick trajectories.}
    \label{fig:posterior_recon}
\end{figure*}

\subsubsection{Approximate Bayesian Computation}
In~\figref{fig:RMSE_all_with_errors}, we supplement the ABC RMSE curve (solid red, same as in~\figref{fig:RMSE_LSTM}) with the average variance of its reconstructed posterior (dashed red). Ideally, if the true posterior were exactly reproduced, these two curves would coincide, as the posterior variance should accurately represent the expected estimation error.

We observe that at the $n=2$ resonance, the ABC framework heavily overestimates its uncertainty. This discrepancy indicates that the reconstructed posterior is significantly broader than the actual distribution of estimation errors. Such a mismatch suggests that while the chosen histogram statistics are sufficient to minimise the average error, they have not been sufficiently fine-tuned to fully capture the true shape and width of the posterior. This failure to reproduce the posterior width demonstrates that aggregated summary statistics discard higher-order information (here, two-photon correlations) necessary for a precise probabilistic characterisation of the system, even in regimes where the point-estimate accuracy is relatively high.

\subsubsection{Probabilistic regression assuming Gaussianity}
Estimating predictive uncertainty via probabilistic regression is often based on the assumption that the posterior distribution is approximately Gaussian~\cite{Kneissl2025}. Under this framework, the model is trained using a Negative Log-Likelihood (NLL) loss to provide both a predicted mean and variance, as described in~\figref{fig:NN_models}(b).

In~\figref{fig:RMSE_all_with_errors}, we present the results for the NLL-based LSTM regressor in green. While the RMSEs, as well as the uncertainty calibration, are improved compared to the ABC method across the entire parameter range, the Gaussian assumption introduces inherent limitations. At $n=2$, the predicted error (dashed green) is closer to the true RMSE than the ABC estimate, yet it remains significantly overestimated, indicating that the Gaussian constraint is still too rigid to perfectly match the posterior width. More notably, in the $n=3$ regime, while the results appear relatively well-calibrated, the attained RMSE (solid green) is significantly higher than that of the MSE-trained LSTM regressor shown in~\figref{fig:RMSE_LSTM} in blue. This demonstrates that forcing the model to produce a Gaussian-calibrated uncertainty estimate at $n=3$ comes at the cost of point-estimate accuracy, as the Gaussian form fails to capture non-Gaussian statistics associated with three-photon correlations.

\subsubsection{Posterior reconstruction by classification}
By utilising the classification framework, we perform non-parametric posterior reconstruction by treating the network's binned output as a probability distribution over the estimated parameter, as introduced in~\figref{fig:NN_models}(c). The results, shown in blue in~\figref{fig:RMSE_all_with_errors}, demonstrate that this approach indeed avoids the accuracy-calibration trade-off encountered in probabilistic regression at the cost of high-dimensional network output.

Notably, the CNN classifier provides point estimates that achieve almost the same RMSE as the MSE-based LSTM regressor (blue curve in~\figref{fig:RMSE_LSTM}) across the entire range of the estimated detuning $\Delta$. Unlike the previous methods, the predicted error (dashed blue) for the classifier closely tracks the attained RMSE (solid blue) even at the $n=2$ and $n=3$ resonances. This high degree of calibration highlights the model's success in capturing the non-Gaussianity of the posterior, particularly for $\Delta \in (3.5, 6.5)$. By bypassing the restrictive Gaussian assumption, the classification framework successfully exploits the three-photon statistics embedded in the photoclick trajectories, delivering a more reliable measure of uncertainty than both ABC and Gaussian regression.

\begin{figure*}[t]
    \centering
    \includegraphics[width=0.85\linewidth]{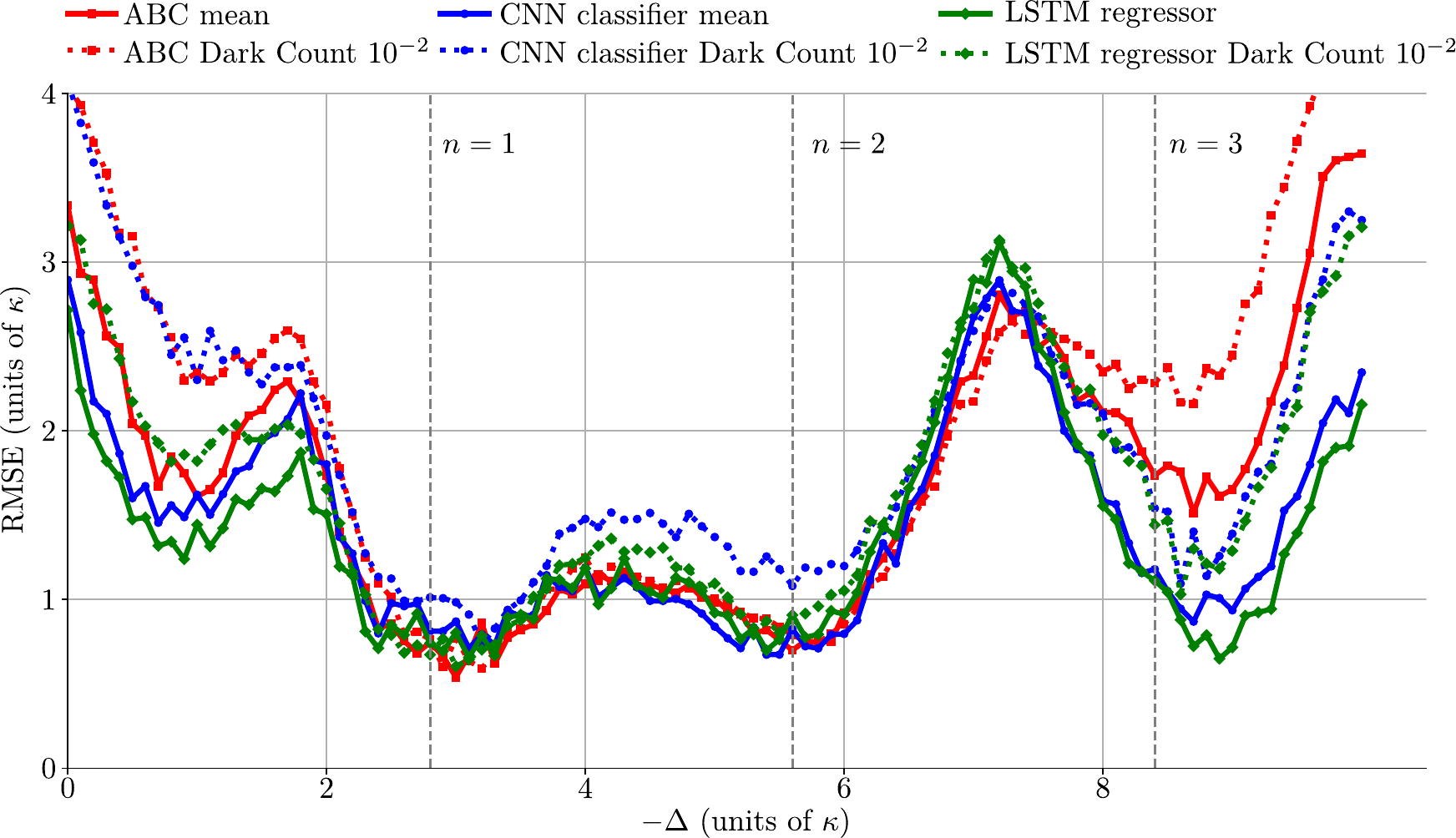}
    \caption{\textbf{Root Mean Squared Error (RMSE) estimation with and without dark counts} at a relatively high rate (see the main text) of $\lambda_{\text{DCR}} = 10^{-2}$ (in units of $\kappa$) as a function of the parameter value $\Delta$ for ABC (\emph{red}), the CNN classifier (\emph{blue}), and the LSTM regressor (\emph{green}). \emph{Solid lines} represent the original noise-free datasets, while \emph{dashed lines} indicate results including dark counts. For ABC and the LSTM, performance degradation is almost exclusively confined to the $n=3$ regime, whereas the CNN exhibits a slight increase in RMSE across all regimes. This suggests that the CNN is currently the least robust of the three;~however, this should be attributed to the fact that its training has not yet saturated for the photoclick library considered, the size of which is kept constant across all methods to ensure a fair comparison.}
    \label{fig:RMSE_DCR}
\end{figure*}
\subsection{Posterior reconstruction analysis}
Having estimated the posterior distribution via both ABC and the CNN classifier, we now evaluate how effectively these methods reconstruct the true posterior density. To establish a ground truth for comparison, we compute the exact posterior distributions using a brute-force simulation approach~\cite{Clark2022} for a small number of photoclick trajectories. The results are visualised in~\figref{fig:posterior_recon}(a--c), where we present a representative trajectory for each of the $n = 1, 2, \text{ and } 3$ regimes, respectively.

Similarly to \citeref{Clark2025}, we quantify the fidelity of the posterior reconstruction using the \emph{Bhattacharyya coefficient}:
\begin{equation}
\mathcal{F}\left[p, q|D_{\ell}\right]
:= \int\!\dd\Delta\,
\sqrt{p\!\left(\Delta | D_{\ell}\right) q\!\left(\Delta | D_{\ell}\right)},
\end{equation}
where $p\!\left(\Delta | D_{\ell}\right)$ represents the ground-truth posterior and $q\!\left(\Delta | D_{\ell}\right)$ is the posterior reconstructed by either ABC or the CNN classifier.

We observe that the CNN consistently outperforms ABC across all three resonance regimes ($n=1,2,3$), capturing the shape of the true posterior with high accuracy and achieving superior values of $\mathcal{F}$. In the $n=1$ regime, \figref{fig:posterior_recon}(a), both methods produce posteriors that are very similar to the ground truth. However, at the $n=2$ resonance in~\figref{fig:posterior_recon}(b), while the ABC framework provides a reasonably accurate point estimate, it is visually apparent from the shape of the distribution that the posterior variance is significantly overestimated. The disparity is most pronounced at $n=3$, shown in~\figref{fig:posterior_recon}(c), where ABC fails to capture both the true parameter value and the width of the distribution due to its lack of access to higher-order temporal correlations. In contrast, the CNN closely approximates then the true posterior by learning the history-dependent features within each specific photoclick pattern. All the above observations are entirely consistent with the RMSE and predicted uncertainty results presented previously in~\figref{fig:RMSE_all_with_errors}.

\subsection{Robustness to dark counts}
To assess the practical applicability of our inference frameworks, we evaluate their performance in the presence of \emph{dark counts}~\cite{Cohen2015, Fiaschi2021, Galinskiy2020}. These random detections follow a Poisson process at a rate $\lambda_{\text{DCR}}$ dictated by thermal or electronic fluctuations. This is an independent process, where the rate is an extra parameter, which is \emph{not} known by the DL model. In order to consider a significant noise scenario, we follow the experimental case of \citeref{Galinskiy2020}, in which only the photons associated with the anti-Stokes sideband are detected. This leads to an increased effective dark-count rate (DCR) of $15.5 \pm 0.5$~Hz, which for the $\kappa/2\pi = 2.1$~kHz applicable to that setup, yields $\lambda_{\text{DCR}} \approx 10^{-2}\kappa$. Depending on the estimated detuning $\Delta$, this results in approximately 10 to 30 dark counts per each 80-click trajectory $D_\ell$ in our library.

In \figref{fig:RMSE_DCR}, we compare the resulting RMSE degradation for ABC (\emph{red}), the CNN classifier (\emph{blue}), and the LSTM regressor (\emph{green}). Both ABC and the LSTM are remarkably resilient in the $n=1$ and $n=2$ regimes, with performance degradation almost exclusively confined to the $n=3$ resonance. In contrast, the CNN exhibits a slight increase in RMSE across all regimes. This relative sensitivity is attributed to the fact that the CNN training has not yet saturated given the size of the photoclick library considered; while the LSTM has already converged, the more demanding CNN architecture would require a larger dataset to achieve similar noise resilience. We maintain a constant library size across all methods to ensure a fair comparison.

Despite the introduction of significant noise, we observe a clear dip in the RMSE at $n=3$ for all methods, while the DL frameworks continue to outperform the ABC approach there by learning higher-order click correlations. Overall, the methods thus exhibit similar robustness to dark counts, while the DL methods retain their advantage where the summary statistics used by ABC are insufficient ($n=3$), as well as in inference speed. Furthermore, we emphasise that for typical experiments employing modern photodetectors where $\lambda_{\text{DCR}} \le 1$~Hz, the increase in estimation error due to dark counts is negligible. Indeed, our numerical tests at lower rates of $10^{-3}\kappa$ and $10^{-4}\kappa$ showed no observable change in the RMSE curves, rendering the impact of noise effectively imperceptible on the scale shown in \figref{fig:RMSE_DCR}.

\section{Conclusions and outlook}
\label{sec:conclusions}
%
In this work, we introduced a novel framework for parameter estimation in non-linear quantum sensors involving photodetection by leveraging likelihood-free inference. Specifically, we developed and compared two distinct approaches: Approximate Bayesian Computation (ABC) and various Deep Learning (DL) architectures. By systematically evaluating MSE-based regression, NLL-based regression with uncertainty estimation, and CE-based classification, we demonstrated that the optimal neural network design depends non-trivially on the specific estimation task and the statistical structure of the photoclick patterns observed.

Our results show that while both likelihood-free methods are effective, the DL models match or exceed the ABC approach in estimation accuracy. Moreover, in regimes exhibiting bunching of more than two photons---where the summary statistics used by ABC prove insufficient---the DL networks achieve superior precision by autonomously extracting and exploiting relevant history-dependent quantum correlations from raw measurement data, without requiring explicit prior knowledge of the system's underlying non-renewal dynamics.

Although one could attempt to improve the accuracy of ABC by employing, e.g., Neural ABC methods~\cite{Fearnhead2010, Wong2018, Aakesson2020} to construct better summary statistics, the trained DL models deliver a 100-fold speed-up in inference time compared to ABC ($\sim 0.17~\text{ms}$ versus $\sim 18~\text{ms}$ per trajectory). Furthermore, our robustness tests confirm that DL maintains high precision even under significant, uncharacterised dark-count noise in photon-bunching regimes, validating its applicability to realistic experimental conditions. At the same time, because the summary statistics in ABC are selected based on the underlying physics, they remain valuable, providing an interpretable benchmark to identify which temporal patterns the DL model learns directly from the data.

The primary motivation for employing quantum sensors, particularly in the context of optomechanics, is the high-precision estimation of external parameters and forces. Optomechanical platforms, ranging from cryogenic membranes to levitated nanospheres, have already proved to be formidable tools for sensing displacements, masses, and accelerations~\cite{Barzanjeh2021, Cripe2019, Guha2020, Haelg2021, Kampel2017, Mason2019, Xia2023}. Our work represents a critical first step in this direction; by identifying the likelihood-free tools best suited to capturing non-classical photon statistics, we provide the necessary toolkit for real-time sensing applications.

Looking ahead, we aim to transition from fixed-length trajectory analysis to \emph{real-time, event-by-event} estimation. This could be achieved by developing Reinforcement Learning (RL) frameworks that update parameter estimates with each detected photon, enabling truly adaptive quantum sensing~\cite{Cimini2023, Belliardo2024pra, Belliardo2024quantum}---a capability particularly valuable for tracking time-varying forces or weak gravitational fields~\cite{Qvarfort2018}. Such an approach would allow for the estimation of coupled optomechanical dynamics where mechanical displacement encodes external force information, even in the presence of back-action effects~\cite{Cripe2019}.

While we have demonstrated the practical efficacy of ABC- and DL-based estimators, our comparison against fundamental precision limits remains partial. For the two-level atom, the simplicity of the system (the absence of decoherence and its renewal-process characteristics) allowed us to verify their performance in multiparameter local estimation against biased Cram\'{e}r-Rao bounds~\cite{Eldar2008}. In contrast, for the optomechanical sensor, we evaluated the bCRB only for the fully unravelled dynamics, which is not tight: it accounts for the information carried by the inaccessible phonon and photon jumps, and assumes fully purified conditional dynamics throughout. 

Tighter benchmarks, including those assessing global (Bayesian) performance~\cite{Trees1968}, remain an active subject of research and could potentially be derived via Gillespie-type algorithms for jump processes~\cite{Radaelli2024, Radaelli2024b}. The bottleneck, however, remains the computation of Fisher information along photoclick trajectories, which requires monitoring both the conditional reduced state and its derivative~\cite{Gammelmark2013, Albarelli2018}.

Furthermore, the question of whether photodetection constitutes the optimal measurement for a given quantum sensing task remains to be addressed. In the ideal case of pure states and fully monitored channels, this could be explored via the global Quantum Fisher Information (QFI)~\cite{Gammelmark2014, Albarelli2018, Ilias2022, Yang2023}. A physically more relevant QFI, however, corresponds to the scenario in which only the outgoing photons are accessible (given a realistic detection efficiency), but where arbitrary measurements on them are permitted---including collective measurements requiring quantum memory. Computing such bounds is nevertheless significantly more complex for high-dimensional systems (such as optomechanical platforms), especially when accounting for unobserved degrees of freedom. Notably, this task has only recently been approached by leveraging Matrix Product Operator (MPO) representations~\cite{Cirac2021} to describe the combined state of the sensor and emitted light with some unobserved degrees of freedom traced out~\cite{Khan2025, Yang2025}. We leave the integration of these theoretical benchmarks into our framework for future investigation.

\section*{Code Availability}
We provide the complete source code and data used for this study at~\cite{github}.

\begin{acknowledgments}
We would like to thank Pierre Rouchon, Patrick P.~Potts, Piotr Grochowski, Marco Genoni and Klaudia Dilcher for insightful comments and suggestions. The "Novel applications of signal processing methods in quantum sensing" project is carried out within the FIRST TEAM programme of the Foundation for Polish Science co-financed by the European Union under the European Funds for Smart Economy 2021-2027 (FENG) [MM, JK].  This research is supported by the National Research Foundation, Singapore, under its Competitive Research Programme  (CRP30-2023-0033) and A*STAR under its Quantum Innovation Centre (Q.InC) Strategic Research and Translational Thrust. Any opinions, findings and conclusions or recommendations expressed in this material are those of the author(s) and do not reflect the views of National Research Foundation, Singapore and A*STAR [LAC].
\end{acknowledgments}

\appendix

\begin{table}[t!]
    \centering
    \setlength{\tabcolsep}{12pt} 
    \begin{tabular}{@{}lll@{}}
        \toprule
        \multicolumn{3}{c}{\textbf{CNN Classifier}} \\
        \midrule
        \textbf{Layer} & \textbf{Output shape} & \textbf{Activation} \\
        \midrule
        Conv1D      & (80,~32) & LeakyReLU \\
        Batch Norm  & -        & - \\
        MaxPool1D   & (26,~32) & - \\
        Conv1D      & (26,~64) & LeakyReLU \\
        Batch Norm  & -        & - \\
        MaxPool1D   & (8,~64)  & - \\
        Conv1D      & (8,~64)  & LeakyReLU \\
        Batch Norm  & -        & - \\
        MaxPool1D   & (2,~64)  & - \\
        Flatten     & 128      & - \\
        Dense       & 200      & LeakyReLU \\
        Dense       & 101      & SoftMax \\
        \midrule
        \multicolumn{2}{l}{\textit{Trainable params.}} & \textbf{83,637} \\
        \bottomrule
            \toprule
        \multicolumn{3}{c}{\textbf{LSTM Regressor}} \\
        \midrule
        \textbf{Layer} & \textbf{Output shape} & \textbf{Activation} \\
        \midrule
        LSTM        & (80,~40)       & - \\
        LSTM        & 40            & LeakyReLU \\
        Batch Norm  & -             & - \\
        Dense       & 64            & LeakyReLU \\
        Dense       & 16            & LeakyReLU \\
        Dense       & 1             & - \\
        \midrule
        \multicolumn{2}{l}{\textit{Trainable params.}} & \textbf{23,441} \\
        \bottomrule
    \end{tabular}
    \caption{Architecture details of the LSTM regressor and CNN classifier models.}
    \label{tab:arch_details}
\end{table}

\begin{figure*}[t!]
 \centering
 \begin{overpic}[width=0.99\textwidth, grid=False]{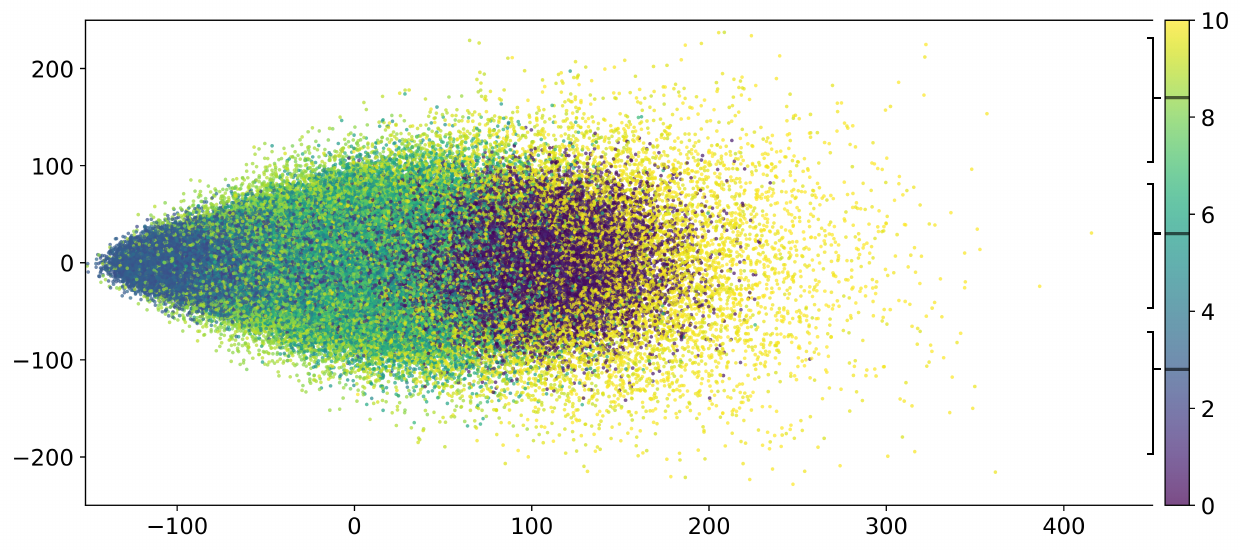}
   \put(72, 5.5){\includegraphics[width=0.20\textwidth]
   {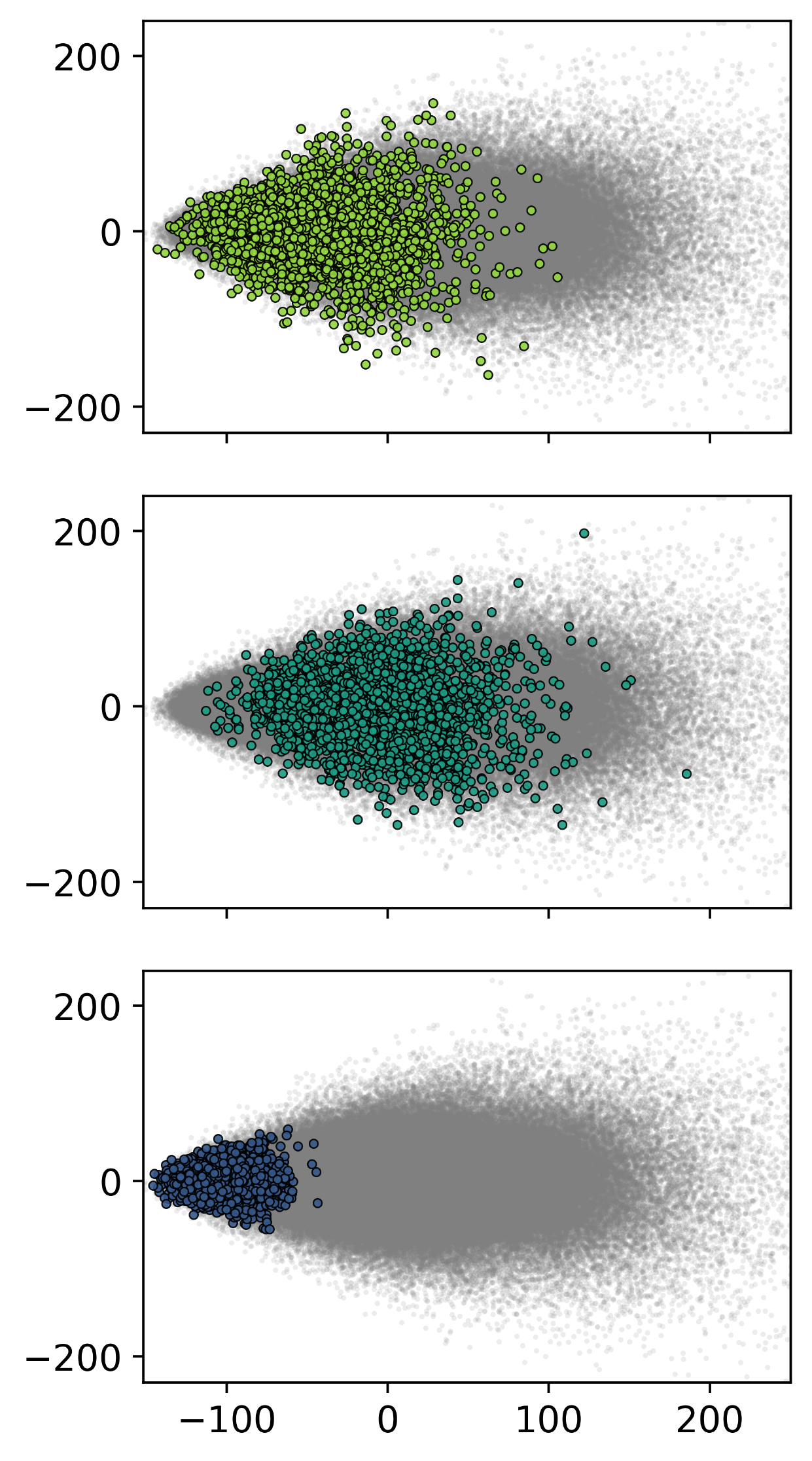}}
   \put(79.5, 4.5){\scriptsize PCA Comp 1}
   \put(41, -1.5){\large PCA Component 1}
   \put(-1, 14){\rotatebox{90}{\large PCA Component 2}}
   \put(71, 8){\rotatebox{90}{\scriptsize PCA Comp 2}}
   \put(71, 20){\rotatebox{90}{\scriptsize PCA Comp 2}}
   \put(71, 31.5){\rotatebox{90}{\scriptsize PCA Comp 2}}
   \put(76, 39.5){\footnotesize $n=3$}
   \put(76, 27.5){\footnotesize $n=2$}
   \put(76, 16){\footnotesize   $n=1$}
   \put(99, 15){\rotatebox{90}{\Large $\Delta$ (units of $\kappa$)}}
   \put(76, 8.5){\footnotesize (a)}
   \put(76, 20){\footnotesize (b)}
   \put(76, 32){\footnotesize (c)}
 \end{overpic}
 \vspace{0.8em}
\caption{\textbf{Principal Component Analysis (PCA) of the optomechanical dataset.}  
The main panel shows the dataset projected onto the first two principal components, with points colour-coded by the detuning parameter~$\Delta$. The first principal component (PC1) predominantly represents the total trajectory time, as the system quickly reaches a steady state where all entries contributing to PC1 become nearly identical. The spread along the PC1 axis thus reflects variations in total time for varying~$\Delta$. Insets~(a),(b) and (c) display selected regions corresponding to special detuning regimes $n = 1,2$~and~$3$, respectively. Distinct clustering patterns, particularly visible for $n=1$, highlight how total time serves as a discriminative feature across regimes. The exact detuning values yielding $n = 1,2$~and~$3$ are indicated in black on the colour bar.}
\label{fig:pca}
\end{figure*}

\section{Contractivity of Fisher information under marginalisation of unobserved jumps}
\label{app:FI_ineq}

Let us define the \emph{score vector} of any $\vec{\theta}$-parametrised likelihood $p(X|\vec{\theta})$ as:
\begin{equation}
    S_{\vec{\theta}}[X] \coloneqq \nabla_{\vec{\theta}} \ln p(X|\vec{\theta}).
    \label{eq:score_function}
\end{equation}
For the joint trajectory distribution introduced in \secref{subsec:UB_FI_matrix}, we specify $S_{\estparV}[D_\ell,D']$ and $S_{\estparV}[D_\ell]$ as the score vectors of the fully unravelled likelihood $p(D_\ell,D^\prime|\estparV)$ and its marginalised counterpart $p(D_\ell|\estparV)$, respectively.

Assuming \emph{regularity} of these likelihoods, such that $\nabla_{\estparV}$ can be freely interchanged with integration over the configuration space $\Omega(T_\ell)$~\cite{Grimmett2001}, the score associated with the marginalised likelihood is the conditional expectation of the full score given the observed jump record $D_\ell$:
\begin{align}
    S_{\estparV}[D_\ell]
    &= \nabla_{\estparV} \ln p(D_\ell|\estparV)
    =
    \frac{
        \nabla_{\estparV} p(D_\ell|\estparV)
    }{
        p(D_\ell|\estparV)
    }
    \\
    &=
    \frac{1}{p(D_\ell|\estparV)}
    \int_{\Omega(T_\ell)}\!\!\mu_{D_\ell}(\dd D^\prime)\;
    \nabla_{\estparV}
    p(D_\ell,D^\prime|\estparV)
    \nonumber\\
    &=
    \int_{\Omega(T_\ell)}\!\!\mu_{D_\ell}(\dd D^\prime)\;
    \frac{
        p(D_\ell,D^\prime|\estparV)
    }{
        p(D_\ell|\estparV)
    }
    \frac{
        \nabla_{\estparV}
        p(D_\ell,D^\prime|\estparV)
    }{
        p(D_\ell,D^\prime|\estparV)
    }    
    \nonumber\\
    &=
    \int_{\Omega(T_\ell)}\!\!\mu_{D_\ell}(\dd D^\prime)\; p(D^\prime|D_\ell,\estparV) \; S_{\estparV}[D_\ell,D'] 
    \nonumber\\
    &=
    \mathbb{E}_{D'|D_\ell}\!\left[S_{\estparV}[D_\ell,D']\right].
    \label{eq:cond_expectation_score}
\end{align}

Since regularity implies that any score function has zero mean, $\mathbb{E}_{X}\!\left[S_{\vec{\theta}}[X]\right]=\int\!\mu(\dd X)\,\nabla_{\vec{\theta}} p(X|\vec{\theta})=\nabla_{\vec{\theta}}(1)=0$, the FI matrix \eref{eq:FImatrix} can be expressed as the covariance matrix of its score vector:
\begin{align}
    \FI[D_\ell,D^\prime|\estparV]
    &=
    \mrm{Cov}_{D_\ell,D'}\!\left[S_{\estparV}[D_\ell,D']\right] \\
    \FI[D_\ell|\estparV]
    &=
    \mrm{Cov}_{D_\ell}\!\left[S_{\estparV}[D_\ell]\right] \nonumber \\
    &=
    \mrm{Cov}_{D_\ell}\!\left[\EE{D'|D_\ell}{S_{\estparV}[D_\ell,D']}\right],
\end{align}
where the last equality follows from \eqnref{eq:cond_expectation_score}.

Applying the law of total covariance~\cite{Grimmett2001} to $S_{\estparV}[D_\ell,D']$ conditioned on $D_\ell$, we obtain
\begin{align}
    \mrm{Cov}_{D_\ell,D'}\!
    \left[
        S_{\estparV}[D_\ell,D']
    \right]
    &=
    \mrm{Cov}_{D_\ell}\!
    \left[
        \EE{D'|D_\ell}{S_{\estparV}[D_\ell,D']}
    \right]
    \nonumber\\
    &\qquad+
    \EE{D_\ell}{\mrm{Cov}_{D^\prime|D_\ell}\!\left[S_{\estparV}[D_\ell,D']\right]}.
\end{align}
Substituting the corresponding FI matrix expressions yields:
\begin{equation}
    \FI[D_\ell,D^\prime|\estparV] = \FI[D_\ell|\estparV]+
    \EE{D_\ell}{\mrm{Cov}_{D^\prime|D_\ell}\!\left[S_{\estparV}[D_\ell,D']\right]}.
    \label{eq:total_covariance_applied}
\end{equation}

Because any covariance matrix is positive semidefinite, the conditional covariance matrix $\mrm{Cov}_{D^\prime|D_\ell}$ is positive semidefinite for every trajectory realisation $D_\ell$, and so is its expectation over $D_\ell$. Consequently,
\begin{equation}
    \FI[D_\ell|\estparV]
    \preceq
    \FI[D_\ell,D^\prime|\estparV].
\end{equation}
which establishes \eqnref{eq:full_unrav_ineq} of the main text.

\begin{figure*}[t!]
    \centering
    \includegraphics[width=0.99\linewidth]{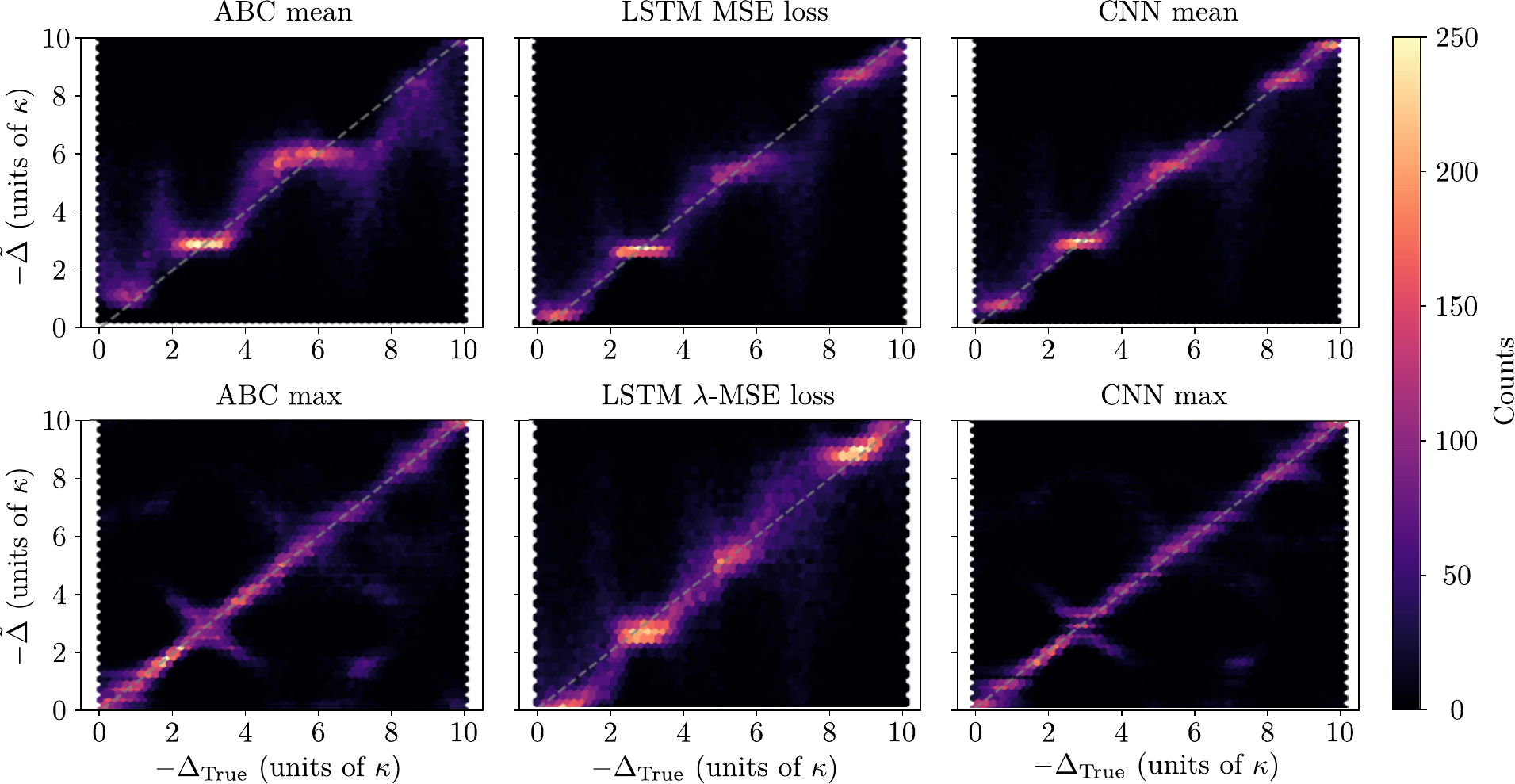}
    \caption{\textbf{Parameter estimation across various inference frameworks.} Comparison of parameter estimation for different detunings, $\Delta_{\text{True}} \in [0, 10]$, when leveraging ABC (left), regression (middle), or classification (right) methods. Each subplot displays the joint distribution of the true parameter $\Delta_{\text{True}}$ (horizontal axis) and its estimator $\est{\Delta}$ (vertical axis) provided by a given method. The colour intensity indicates the density of estimates, with lighter regions denoting higher frequencies; the dashed diagonal line represents ideal estimation. \emph{Left column (ABC)}:~Estimator $\est{\Delta}$ provided by either the mean (top) or mode (bottom) of the reconstructed posterior, using total time and the waiting-time histogram as summary statistics. \emph{Middle column (regression)}:~LSTM-based regression employing a standard MSE loss (top) or a modified loss function designed to mitigate bias (bottom). \emph{Right column (classification)}:~CNN-based classification with point estimates being provided either by the mean (top) or the mode (bottom) of the predicted posterior distribution.}
    \label{fig:2d_all_methods}
\end{figure*}

\section{Equivalence between Cross-Entropy and KL divergence minimisation}
\label{app:class_proof}
To quantify the similarity between the ground-truth ``reference'' distribution $p(\varphi)$ and the network-estimated ``model'' distribution $q(\varphi)$, one may employ the forward ($p \to q$) \emph{Kullback-Leibler} (KL) \emph{divergence} defined as:
\begin{equation}
    D_{\text{KL}}(p \parallel q) \coloneqq \mathbb{E}_{p} \left[ \log \frac{p(\varphi)}{q(\varphi)} \right] = \int\!d\varphi\, p(\varphi) \log \frac{p(\varphi)}{q(\varphi)},
    \label{eq:KL_definition}
\end{equation}
where $\mathbb{E}_p$ denotes the expectation value with respect to the reference distribution $p$. Within the classification-based inference framework, $p$ represents the target distribution over the ground-truth labels.

Specifically, given a discrete grid defining the parameter labels $\varphi_i$ and an input dataset $D$ generated with $\varphi=\varphi_{\text{true}}$, the reference is represented by an empirical distribution $p(\varphi_i|D)=\delta_{\varphi_i,\varphi_{\text{true}}}$. On the other hand, $q$ is the network output, i.e., the model's predicted probability mass function $q(\varphi_i|D)$, which aims to approximate the true posterior distribution. 

Omitting the conditioning on the input $D$ for brevity, the empirical KL divergence then reads:
\begin{align}
    D_{\text{KL}}(p \parallel q) 
    &= \sum_{i} \left[ p(\varphi_i) \log p(\varphi_i) - p(\varphi_i) \log q(\varphi_i) \right] \nonumber\\
    &= -H(p) + \mathcal{L}_{\CE}(p, q),
\end{align}
where $H(p) \coloneqq -\sum_i p(\varphi_i) \log p(\varphi_i)$ is the Shannon entropy of the reference distribution, and $\mathcal{L}_{\CE}(p, q) = -\sum_i p(\varphi_i) \log q(\varphi_i)$ is the Cross-Entropy (CE) between  reference and model distributions, defined in \eqnref{eq:loss_CE}.

In our training scheme, the distribution $p(\varphi)$ is fixed by each input dataset $D$. Consequently, $H(p)$ is a constant with respect to the neural network parameters $\vec{\theta}$, and its gradient vanishes, $\nabla_{\vec{\theta}} H(p) = 0$. Minimising the CE loss during training is therefore mathematically equivalent to minimising the KL divergence:
\begin{equation}
    \arg\min_{\vec{\theta}} \mathcal{L}_{\CE}(p, q_{\vec{\theta}}) = \arg\min_{\vec{\theta}} D_{\text{KL}}(p \parallel q_{\vec{\theta}}).
\end{equation}
This confirms that the classification framework yields a network that approximates the true posterior distribution by minimising the information-theoretic distance between the predicted and actual probability masses.

\section{Summary of neural network parameters used for the optomechanical system}
\label{app:DNN_details}
In \tabref{tab:arch_details}, we provide the summary of NN architectures and their specs using the standard convention (see, e.g.,~\citeref{Goodfellow-et-al-2016}) necessary for the reconstruction of the main results concerning the optomechanical system.

\section{PCA of the optomechanical dataset}
\label{pcs_analysis}
To gain insight into the features leveraged by the deep learning model for parameter estimation, we perform a Principal Component Analysis (PCA) on the input data (see Fig.~\ref{fig:pca}). PCA identifies orthogonal directions—principal components—along which the data exhibits maximal variance, providing a compact representation of the data’s dominant structure.

In the main panel of Fig.~\ref{fig:pca}, we project the dataset onto the first two principal components, with each point colour-coded by the detuning parameter $\Delta$. The first principal component (PC1), associated with the largest eigenvalue, closely corresponds to a rescaled sum over all elements of the input vectors. 

Physically, the projection onto PC1 effectively represents the \textit{total time} of each trajectory. This behaviour arises because, after a brief transient period, the system rapidly reaches its steady state, at which point all entries contributing to PC1 become nearly identical. Consequently, the PCA primarily captures variations in the cumulative waiting times that compose each trajectory, making PC1 a natural representation of the total time.

The spread of points along the PC1 axis reflects changes in the trajectory's total time for different detunings $\Delta$. Meanwhile, the inset plots in Fig.~\ref{fig:pca}(a--c) display data subsets corresponding to special photon regimes $n=1, 2$ and $3$. Distinct clustering patterns emerge, particularly for $n=1$, where trajectories are well separated from those in the $n=2$ and $n=3$ regimes. This observation indicates that when estimating detuning values in the $n=1$ regime, the total time can serve as a sufficient statistics.

\begin{table*}[t]
    \begin{ruledtabular}
    \begin{tabular}{cccc}
        \textbf{Dataset generation time} & \textbf{\makecell{Average ABC estimation time\\per single trajectory}} & \textbf{ML training time} & \textbf{\makecell{Average ML estimation time\\per single trajectory}} \\[2.0ex]
        \makecell{$\sim$ Two weeks \\ (400 000 trajectories)} & $\sim 18$ms (400 000 queries)& \makecell{$\sim 8$h CNN Classifier \\$\sim 10$h LSTM Regressor} & \makecell{$\sim 0.17$ms CNN Classifier\\ $\sim 0.8$ms LSTM Regressor} 
    \end{tabular}
    \end{ruledtabular}
    \caption{Summary of computational costs and average estimation times.}
    \label{tab:computational_costs}
\end{table*}
\section{Reducing bias in estimation}
\label{app:red_bias}
As an additional analysis, we compare different inference methods in terms of their parameter estimation behaviour, visualised via 2D density plots in Fig.~\ref{fig:2d_all_methods}. These plots map the relationship between the true detuning value $\Delta$ and the corresponding estimated value $\est{\Delta}$ across multiple trials, allowing for a visual assessment of the inference structure. In Bayesian parameter estimation, one typically chooses between the mean or the mode of the posterior distribution as the point estimator. While the mean is theoretically optimal under the MSE loss, the mode often yields more representative values in cases of skewed or multimodal distributions.

Using ABC to reconstruct the posterior distribution and taking its mean as the parameter estimator (\figref{fig:2d_all_methods}, top-left), we observe distinct horizontal regions-plateaus-specifically around the $n=1,2,$ and $3$ regimes. These plateaus indicate a lack of sensitivity to changes in the true parameter. Interestingly, these artefacts disappear (at the cost of a slightly higher MSE) when the mode of the reconstructed posterior is used instead (\figref{fig:2d_all_methods}, bottom-left). A similar behaviour is observed when reconstructing the posterior with a CNN (\figref{fig:2d_all_methods}, bottom-right). Nonetheless, because the CNN is capable of learning higher-order correlations between photoclicks, its performance significantly exceeds that of ABC in the $n=3$ regime, when interpreting the posterior mean as the estimator $\est{\Delta}$ (\figref{fig:2d_all_methods}, top-right). 

Regarding the LSTM regressor trained with a standard MSE loss (\figref{fig:2d_all_methods}, top-middle), the appearance of plateaus suggests that the network is effectively learning to approximate the posterior mean. To understand and mitigate this effect, we consider the decomposition of the MSE~\cite{Wackerly2001}:
\begin{equation}
    \MSE(\hat{\Delta}) = \text{Var}(\hat{\Delta}) + \left[ \text{Bias}(\hat{\Delta}, \Delta) \right]^2,
\end{equation}
where $\Delta$ is the true parameter and $\hat{\Delta}$ is its estimate. 

The emergence of plateaus in all top plots \figref{fig:2d_all_methods} indicates that these architectures prioritise low variance at the expense of high bias in certain parameter regions. To address this issue in regression, we introduce a modified loss function that explicitly balances bias and variance via a weighting factor $\lambda$, i.e.:
\begin{equation}
    \MSE_\lambda(\hat{\Delta}) = \lambda\;\text{Var}(\hat{\Delta}) + \left[ \text{Bias}(\hat{\Delta}, \Delta) \right]^2.
\end{equation}
 With $\lambda = 0.8$ (favouring bias reduction), the LSTM regressor yields a smoother estimator (\figref{fig:2d_all_methods}, bottom-middle), significantly reducing plateau artefacts and mirroring the behaviour when it is rather the mode of posterior used as the estimator $\est{\Delta}$.

Despite the improved qualitative behaviour, this modification does not lead to a significant reduction in the Root Mean Squared Error (RMSE). Nonetheless, this analysis provides insight into how does the objective function design dictates the structure of learned estimators and demonstrates how different posterior summary statistics (mean vs.\ mode) manifest in learning-based inference.

\section{Computational costs}
\label{app:comp_cost}
In \tabref{tab:computational_costs}, we summarise the computational costs associated with data generation, model training, and estimation. All training procedures and evaluations were performed on a desktop PC equipped with an Intel Core i7 CPU and 32~GB of RAM. We also note that beyond a certain number of queries, the performance improvement of ABC becomes negligible~\cite{Clark2025}, whereas training for too many epochs may lead to overfitting in deep learning models.

\bibliographystyle{myapsrev4-2}
\bibliography{ABC_vs_ML}

\begin{thebibliography}{137}%
\makeatletter
\providecommand \@ifxundefined [1]{%
 \@ifx{#1\undefined}
}%
\providecommand \@ifnum [1]{%
 \ifnum #1\expandafter \@firstoftwo
 \else \expandafter \@secondoftwo
 \fi
}%
\providecommand \@ifx [1]{%
 \ifx #1\expandafter \@firstoftwo
 \else \expandafter \@secondoftwo
 \fi
}%
\providecommand \natexlab [1]{#1}%
\providecommand \enquote  [1]{``#1''}%
\providecommand \bibnamefont  [1]{#1}%
\providecommand \bibfnamefont [1]{#1}%
\providecommand \citenamefont [1]{#1}%
\providecommand \href@noop [0]{\@secondoftwo}%
\providecommand \href [0]{\begingroup \@sanitize@url \@href}%
\providecommand \@href[1]{\@@startlink{#1}\@@href}%
\providecommand \@@href[1]{\endgroup#1\@@endlink}%
\providecommand \@sanitize@url [0]{\catcode `\\12\catcode `\$12\catcode
  `\&12\catcode `\#12\catcode `\^12\catcode `\_12\catcode `\%12\relax}%
\providecommand \@@startlink[1]{}%
\providecommand \@@endlink[0]{}%
\providecommand \url  [0]{\begingroup\@sanitize@url \@url }%
\providecommand \@url [1]{\endgroup\@href {#1}{\urlprefix }}%
\providecommand \urlprefix  [0]{URL }%
\providecommand \Eprint [0]{\href }%
\providecommand \doibase [0]{https://doi.org/}%
\providecommand \selectlanguage [0]{\@gobble}%
\providecommand \bibinfo  [0]{\@secondoftwo}%
\providecommand \bibfield  [0]{\@secondoftwo}%
\providecommand \translation [1]{[#1]}%
\providecommand \BibitemOpen [0]{}%
\providecommand \bibitemStop [0]{}%
\providecommand \bibitemNoStop [0]{.\EOS\space}%
\providecommand \EOS [0]{\spacefactor3000\relax}%
\providecommand \BibitemShut  [1]{\csname bibitem#1\endcsname}%
\let\auto@bib@innerbib\@empty
\bibitem [{\citenamefont {Giovannetti}\ \emph {et~al.}(2001)\citenamefont
  {Giovannetti}, \citenamefont {Lloyd},\ and\ \citenamefont
  {Maccone}}]{Giovannetti2001}%
  \BibitemOpen
  \bibfield  {author} {\bibinfo {author} {\bibfnamefont {V.}~\bibnamefont
  {Giovannetti}}, \bibinfo {author} {\bibfnamefont {S.}~\bibnamefont {Lloyd}},\
  \bibnamefont {and}\ \bibinfo {author} {\bibfnamefont {L.}~\bibnamefont
  {Maccone}},\ }\bibfield  {title} {\emph {\enquote {\bibinfo {title}
  {Quantum-enhanced positioning and clock synchronization},}\ }}\href
  {https://doi.org/10.1038/35086525} {\bibfield  {journal} {\bibinfo  {journal}
  {Nature}\ }\textbf {\bibinfo {volume} {412}},\ \bibinfo {pages} {417}
  (\bibinfo {year} {2001})}\BibitemShut {NoStop}%
\bibitem [{\citenamefont {Giovannetti}\ \emph {et~al.}(2006)\citenamefont
  {Giovannetti}, \citenamefont {Lloyd},\ and\ \citenamefont
  {Maccone}}]{Giovannetti2006}%
  \BibitemOpen
  \bibfield  {author} {\bibinfo {author} {\bibfnamefont {V.}~\bibnamefont
  {Giovannetti}}, \bibinfo {author} {\bibfnamefont {S.}~\bibnamefont {Lloyd}},\
  \bibnamefont {and}\ \bibinfo {author} {\bibfnamefont {L.}~\bibnamefont
  {Maccone}},\ }\bibfield  {title} {\emph {\enquote {\bibinfo {title} {Quantum
  metrology},}\ }}\href {https://doi.org/10.1103/physrevlett.96.010401}
  {\bibfield  {journal} {\bibinfo  {journal} {Phys. Rev. Lett.}\ }\textbf
  {\bibinfo {volume} {96}},\ \bibinfo {pages} {010401} (\bibinfo {year}
  {2006})}\BibitemShut {NoStop}%
\bibitem [{\citenamefont {Dowling}(2008)}]{Dowling2008}%
  \BibitemOpen
  \bibfield  {author} {\bibinfo {author} {\bibfnamefont {J.~P.}\ \bibnamefont
  {Dowling}},\ }\bibfield  {title} {\emph {\enquote {\bibinfo {title} {Quantum
  optical metrology---The lowdown on high-{N00N} states},}\ }}\href
  {https://doi.org/10.1080/00107510802091298} {\bibfield  {journal} {\bibinfo
  {journal} {Contemp. Phys.}\ }\textbf {\bibinfo {volume} {49}},\ \bibinfo
  {pages} {125} (\bibinfo {year} {2008})}\BibitemShut {NoStop}%
\bibitem [{\citenamefont {Giovannetti}\ \emph {et~al.}(2011)\citenamefont
  {Giovannetti}, \citenamefont {Lloyd},\ and\ \citenamefont
  {Maccone}}]{Giovannetti2011}%
  \BibitemOpen
  \bibfield  {author} {\bibinfo {author} {\bibfnamefont {V.}~\bibnamefont
  {Giovannetti}}, \bibinfo {author} {\bibfnamefont {S.}~\bibnamefont {Lloyd}},\
  \bibnamefont {and}\ \bibinfo {author} {\bibfnamefont {L.}~\bibnamefont
  {Maccone}},\ }\bibfield  {title} {\emph {\enquote {\bibinfo {title} {Advances
  in quantum metrology},}\ }}\href {https://doi.org/10.1038/nphoton.2011.35}
  {\bibfield  {journal} {\bibinfo  {journal} {Nat. Photonics}\ }\textbf
  {\bibinfo {volume} {5}},\ \bibinfo {pages} {222} (\bibinfo {year}
  {2011})}\BibitemShut {NoStop}%
\bibitem [{\citenamefont {Pezz{\`e}}\ \emph {et~al.}(2018)\citenamefont
  {Pezz{\`e}}, \citenamefont {Smerzi}, \citenamefont {Oberthaler},
  \citenamefont {Schmied},\ and\ \citenamefont {Treutlein}}]{Pezze2018}%
  \BibitemOpen
  \bibfield  {author} {\bibinfo {author} {\bibfnamefont {L.}~\bibnamefont
  {Pezz{\`e}}}, \bibinfo {author} {\bibfnamefont {A.}~\bibnamefont {Smerzi}},
  \bibinfo {author} {\bibfnamefont {M.~K.}\ \bibnamefont {Oberthaler}},
  \bibinfo {author} {\bibfnamefont {R.}~\bibnamefont {Schmied}},\ \bibnamefont
  {and}\ \bibinfo {author} {\bibfnamefont {P.}~\bibnamefont {Treutlein}},\
  }\bibfield  {title} {\emph {\enquote {\bibinfo {title} {Quantum metrology
  with nonclassical states of atomic ensembles},}\ }}\href
  {https://doi.org/10.1103/revmodphys.90.035005} {\bibfield  {journal}
  {\bibinfo  {journal} {Rev. Mod. Phys.}\ }\textbf {\bibinfo {volume} {90}},\
  \bibinfo {pages} {035005} (\bibinfo {year} {2018})}\BibitemShut {NoStop}%
\bibitem [{\citenamefont {Leroux}\ \emph {et~al.}(2010)\citenamefont {Leroux},
  \citenamefont {Schleier-Smith},\ and\ \citenamefont
  {Vuleti{\'c}}}]{Leroux2010}%
  \BibitemOpen
  \bibfield  {author} {\bibinfo {author} {\bibfnamefont {I.~D.}\ \bibnamefont
  {Leroux}}, \bibinfo {author} {\bibfnamefont {M.~H.}\ \bibnamefont
  {Schleier-Smith}},\ \bibnamefont {and}\ \bibinfo {author} {\bibfnamefont
  {V.}~\bibnamefont {Vuleti{\'c}}},\ }\bibfield  {title} {\emph {\enquote
  {\bibinfo {title} {Implementation of cavity squeezing of a collective atomic
  spin},}\ }}\href {https://doi.org/10.1103/physrevlett.104.073602} {\bibfield
  {journal} {\bibinfo  {journal} {Phys. Rev. Lett.}\ }\textbf {\bibinfo
  {volume} {104}},\ \bibinfo {pages} {073602} (\bibinfo {year}
  {2010})}\BibitemShut {NoStop}%
\bibitem [{\citenamefont {Hosten}\ \emph {et~al.}(2016)\citenamefont {Hosten},
  \citenamefont {Engelsen}, \citenamefont {Krishnakumar},\ and\ \citenamefont
  {Kasevich}}]{Hosten2016}%
  \BibitemOpen
  \bibfield  {author} {\bibinfo {author} {\bibfnamefont {O.}~\bibnamefont
  {Hosten}}, \bibinfo {author} {\bibfnamefont {N.~J.}\ \bibnamefont
  {Engelsen}}, \bibinfo {author} {\bibfnamefont {R.}~\bibnamefont
  {Krishnakumar}},\ \bibnamefont {and}\ \bibinfo {author} {\bibfnamefont
  {M.~A.}\ \bibnamefont {Kasevich}},\ }\bibfield  {title} {\emph {\enquote
  {\bibinfo {title} {Measurement noise 100 times lower than the
  quantum-projection limit using entangled atoms},}\ }}\href
  {https://doi.org/10.1038/nature16176} {\bibfield  {journal} {\bibinfo
  {journal} {Nature}\ }\textbf {\bibinfo {volume} {529}},\ \bibinfo {pages}
  {505} (\bibinfo {year} {2016})}\BibitemShut {NoStop}%
\bibitem [{\citenamefont {Pedrozo-Pe{\~n}afiel}\ \emph
  {et~al.}(2020)\citenamefont {Pedrozo-Pe{\~n}afiel}, \citenamefont {Colombo},
  \citenamefont {Shu}, \citenamefont {Adiyatullin}, \citenamefont {Li},
  \citenamefont {Mendez}, \citenamefont {Braverman}, \citenamefont {Kawasaki},
  \citenamefont {Akamatsu}, \citenamefont {Xiao},\ and\ \citenamefont
  {Vuleti{\'c}}}]{PedrozoPenafielN2020}%
  \BibitemOpen
  \bibfield  {author} {\bibinfo {author} {\bibfnamefont {E.}~\bibnamefont
  {Pedrozo-Pe{\~n}afiel}}, \bibnamefont {et~al.},\ }\bibfield  {title} {\emph
  {\enquote {\bibinfo {title} {Entanglement on an optical atomic-clock
  transition},}\ }}\href {https://doi.org/10.1038/s41586-020-3006-1} {\bibfield
   {journal} {\bibinfo  {journal} {Nature}\ }\textbf {\bibinfo {volume}
  {588}},\ \bibinfo {pages} {414} (\bibinfo {year} {2020})}\BibitemShut
  {NoStop}%
\bibitem [{\citenamefont {Colombo}\ \emph {et~al.}(2022)\citenamefont
  {Colombo}, \citenamefont {Pedrozo-Pe{\~n}afiel},\ and\ \citenamefont
  {Vuleti{\'c}}}]{Colombo2022}%
  \BibitemOpen
  \bibfield  {author} {\bibinfo {author} {\bibfnamefont {S.}~\bibnamefont
  {Colombo}}, \bibinfo {author} {\bibfnamefont {E.}~\bibnamefont
  {Pedrozo-Pe{\~n}afiel}},\ \bibnamefont {and}\ \bibinfo {author}
  {\bibfnamefont {V.}~\bibnamefont {Vuleti{\'c}}},\ }\bibfield  {title} {\emph
  {\enquote {\bibinfo {title} {Entanglement-enhanced optical atomic clocks},}\
  }}\href {https://doi.org/10.1063/5.0121372} {\bibfield  {journal} {\bibinfo
  {journal} {Appl. Phys. Lett.}\ }\textbf {\bibinfo {volume} {121}},\ \bibinfo
  {pages} {210502} (\bibinfo {year} {2022})}\BibitemShut {NoStop}%
\bibitem [{\citenamefont {Sewell}\ \emph {et~al.}(2012)\citenamefont {Sewell},
  \citenamefont {Koschorreck}, \citenamefont {Napolitano}, \citenamefont
  {Dubost}, \citenamefont {Behbood},\ and\ \citenamefont
  {Mitchell}}]{Sewell2012}%
  \BibitemOpen
  \bibfield  {author} {\bibinfo {author} {\bibfnamefont {R.~J.}\ \bibnamefont
  {Sewell}}, \bibinfo {author} {\bibfnamefont {M.}~\bibnamefont {Koschorreck}},
  \bibinfo {author} {\bibfnamefont {M.}~\bibnamefont {Napolitano}}, \bibinfo
  {author} {\bibfnamefont {B.}~\bibnamefont {Dubost}}, \bibinfo {author}
  {\bibfnamefont {N.}~\bibnamefont {Behbood}},\ \bibnamefont {and}\ \bibinfo
  {author} {\bibfnamefont {M.~W.}\ \bibnamefont {Mitchell}},\ }\bibfield
  {title} {\emph {\enquote {\bibinfo {title} {Magnetic sensitivity beyond the
  projection noise limit by spin squeezing},}\ }}\href
  {https://doi.org/10.1103/physrevlett.109.253605} {\bibfield  {journal}
  {\bibinfo  {journal} {Phys. Rev. Lett.}\ }\textbf {\bibinfo {volume} {109}},\
  \bibinfo {pages} {253605} (\bibinfo {year} {2012})}\BibitemShut {NoStop}%
\bibitem [{\citenamefont {Martin~Ciurana}\ \emph {et~al.}(2017)\citenamefont
  {Martin~Ciurana}, \citenamefont {Colangelo}, \citenamefont {Slodi{\v{c}}ka},
  \citenamefont {Sewell},\ and\ \citenamefont {Mitchell}}]{MartinCiurana2017}%
  \BibitemOpen
  \bibfield  {author} {\bibinfo {author} {\bibfnamefont {F.}~\bibnamefont
  {Martin~Ciurana}}, \bibinfo {author} {\bibfnamefont {G.}~\bibnamefont
  {Colangelo}}, \bibinfo {author} {\bibfnamefont {L.}~\bibnamefont
  {Slodi{\v{c}}ka}}, \bibinfo {author} {\bibfnamefont {R.~J.}\ \bibnamefont
  {Sewell}},\ \bibnamefont {and}\ \bibinfo {author} {\bibfnamefont {M.~W.}\
  \bibnamefont {Mitchell}},\ }\bibfield  {title} {\emph {\enquote {\bibinfo
  {title} {Entanglement-enhanced radio-frequency field detection and waveform
  sensing},}\ }}\href {https://doi.org/10.1103/physrevlett.119.043603}
  {\bibfield  {journal} {\bibinfo  {journal} {Phys. Rev. Lett.}\ }\textbf
  {\bibinfo {volume} {119}},\ \bibinfo {pages} {043603} (\bibinfo {year}
  {2017})}\BibitemShut {NoStop}%
\bibitem [{\citenamefont {Wu}\ \emph {et~al.}(2025)\citenamefont {Wu},
  \citenamefont {Davis}, \citenamefont {Hughes}, \citenamefont {Ye},
  \citenamefont {Wang}, \citenamefont {Kufel}, \citenamefont {Ono},
  \citenamefont {Meynell}, \citenamefont {Block}, \citenamefont {Liu},
  \citenamefont {Yang}, \citenamefont {Bleszynski~Jayich},\ and\ \citenamefont
  {Yao}}]{Wu2025}%
  \BibitemOpen
  \bibfield  {author} {\bibinfo {author} {\bibfnamefont {W.}~\bibnamefont
  {Wu}}, \bibnamefont {et~al.},\ }\bibfield  {title} {\emph {\enquote {\bibinfo
  {title} {Spin squeezing in an ensemble of nitrogen--vacancy centres in
  diamond},}\ }}\href {https://doi.org/10.1038/s41586-025-09524-8} {\bibfield
  {journal} {\bibinfo  {journal} {Nature}\ }\textbf {\bibinfo {volume} {646}},\
  \bibinfo {pages} {74} (\bibinfo {year} {2025})}\BibitemShut {NoStop}%
\bibitem [{\citenamefont {Gao}\ \emph {et~al.}(2025)\citenamefont {Gao},
  \citenamefont {Martin}, \citenamefont {Hughes}, \citenamefont {Leitao},
  \citenamefont {Put}, \citenamefont {Zhou}, \citenamefont {Koyluoglu},
  \citenamefont {Meynell}, \citenamefont {Bleszynski~Jayich}, \citenamefont
  {Park},\ and\ \citenamefont {Lukin}}]{Gao2025}%
  \BibitemOpen
  \bibfield  {author} {\bibinfo {author} {\bibfnamefont {H.}~\bibnamefont
  {Gao}}, \bibnamefont {et~al.},\ }\bibfield  {title} {\emph {\enquote
  {\bibinfo {title} {Signal amplification in a solid-state sensor through
  asymmetric many-body echo},}\ }}\href
  {https://doi.org/10.1038/s41586-025-09452-7} {\bibfield  {journal} {\bibinfo
  {journal} {Nature}\ }\textbf {\bibinfo {volume} {646}},\ \bibinfo {pages}
  {68} (\bibinfo {year} {2025})}\BibitemShut {NoStop}%
\bibitem [{\citenamefont {Gross}\ \emph {et~al.}(2010)\citenamefont {Gross},
  \citenamefont {Zibold}, \citenamefont {Nicklas}, \citenamefont {Est{\`e}ve},\
  and\ \citenamefont {Oberthaler}}]{Gross2010}%
  \BibitemOpen
  \bibfield  {author} {\bibinfo {author} {\bibfnamefont {C.}~\bibnamefont
  {Gross}}, \bibinfo {author} {\bibfnamefont {T.}~\bibnamefont {Zibold}},
  \bibinfo {author} {\bibfnamefont {E.}~\bibnamefont {Nicklas}}, \bibinfo
  {author} {\bibfnamefont {J.}~\bibnamefont {Est{\`e}ve}},\ \bibnamefont {and}\
  \bibinfo {author} {\bibfnamefont {M.~K.}\ \bibnamefont {Oberthaler}},\
  }\bibfield  {title} {\emph {\enquote {\bibinfo {title} {Nonlinear atom
  interferometer surpasses classical precision limit},}\ }}\href
  {https://doi.org/10.1038/nature08919} {\bibfield  {journal} {\bibinfo
  {journal} {Nature}\ }\textbf {\bibinfo {volume} {464}},\ \bibinfo {pages}
  {1165} (\bibinfo {year} {2010})}\BibitemShut {NoStop}%
\bibitem [{\citenamefont {Demkowicz-Dobrza{\'n}ski}\ \emph
  {et~al.}(2015)\citenamefont {Demkowicz-Dobrza{\'n}ski}, \citenamefont
  {Jarzyna},\ and\ \citenamefont
  {Ko{\l}ody{\'n}ski}}]{DemkowiczDobrzanski2015}%
  \BibitemOpen
  \bibfield  {author} {\bibinfo {author} {\bibfnamefont {R.}~\bibnamefont
  {Demkowicz-Dobrza{\'n}ski}}, \bibinfo {author} {\bibfnamefont
  {M.}~\bibnamefont {Jarzyna}},\ \bibnamefont {and}\ \bibinfo {author}
  {\bibfnamefont {J.}~\bibnamefont {Ko{\l}ody{\'n}ski}},\ }in\ \href
  {https://doi.org/10.1016/bs.po.2015.02.003} {\emph {\bibinfo {booktitle}
  {Progress in Optics}}},\ Vol.~\bibinfo {volume} {60},\ \bibinfo {editor}
  {edited by\ \bibinfo {editor} {\bibfnamefont {E.}~\bibnamefont {Wolf}}}\
  (\bibinfo  {publisher} {Elsevier},\ \bibinfo {year} {2015})\ pp.\ \bibinfo
  {pages} {345--435},\ \Eprint {https://arxiv.org/abs/1405.7703}
  {arXiv:1405.7703 [quant-ph]} \BibitemShut {NoStop}%
\bibitem [{\citenamefont {Bongs}\ \emph {et~al.}(2019)\citenamefont {Bongs},
  \citenamefont {Holynski}, \citenamefont {Vovrosh}, \citenamefont {Bouyer},
  \citenamefont {Condon}, \citenamefont {Rasel}, \citenamefont {Schubert},
  \citenamefont {Schleich},\ and\ \citenamefont {Roura}}]{Bongs2021}%
  \BibitemOpen
  \bibfield  {author} {\bibinfo {author} {\bibfnamefont {K.}~\bibnamefont
  {Bongs}}, \bibinfo {author} {\bibfnamefont {M.}~\bibnamefont {Holynski}},
  \bibinfo {author} {\bibfnamefont {J.}~\bibnamefont {Vovrosh}}, \bibinfo
  {author} {\bibfnamefont {P.}~\bibnamefont {Bouyer}}, \bibinfo {author}
  {\bibfnamefont {G.}~\bibnamefont {Condon}}, \bibinfo {author} {\bibfnamefont
  {E.}~\bibnamefont {Rasel}}, \bibinfo {author} {\bibfnamefont
  {C.}~\bibnamefont {Schubert}}, \bibinfo {author} {\bibfnamefont {W.~P.}\
  \bibnamefont {Schleich}},\ \bibnamefont {and}\ \bibinfo {author}
  {\bibfnamefont {A.}~\bibnamefont {Roura}},\ }\bibfield  {title} {\emph
  {\enquote {\bibinfo {title} {Taking atom interferometric quantum sensors from
  the laboratory to real-world applications},}\ }}\href
  {https://doi.org/10.1038/s42254-019-0117-4} {\bibfield  {journal} {\bibinfo
  {journal} {Nat. Rev. Phys.}\ }\textbf {\bibinfo {volume} {1}},\ \bibinfo
  {pages} {731} (\bibinfo {year} {2019})}\BibitemShut {NoStop}%
\bibitem [{\citenamefont {Tse}\ \emph {et~al.}(2019)\citenamefont {Tse},
  \citenamefont {Yu}, \citenamefont {Kijbunchoo}, \citenamefont
  {Fernandez-Galiana}, \citenamefont {Dupej}, \citenamefont {Barsotti},
  \citenamefont {Blair}, \citenamefont {Brown}, \citenamefont {Dwyer},
  \citenamefont {Effler}, \citenamefont {Evans}, \citenamefont {Fritschel},
  \citenamefont {Frolov}, \citenamefont {Green}, \citenamefont {Mansell},
  \citenamefont {Matichard}, \citenamefont {Mavalvala}, \citenamefont
  {McClelland}, \citenamefont {McCuller}, \citenamefont {McRae}, \citenamefont
  {Miller}, \citenamefont {Mullavey}, \citenamefont {Oelker}, \citenamefont
  {Phinney}, \citenamefont {Sigg}, \citenamefont {Slagmolen}, \citenamefont
  {Vo}, \citenamefont {Ward}, \citenamefont {Whittle}, \citenamefont {Abbott},
  \citenamefont {Adams}, \citenamefont {Adhikari}, \citenamefont {Ananyeva},
  \citenamefont {Appert}, \citenamefont {Arai}, \citenamefont {Areeda},
  \citenamefont {Asali}, \citenamefont {Aston}, \citenamefont {Austin},
  \citenamefont {Baer}, \citenamefont {Ball}, \citenamefont {Ballmer},
  \citenamefont {Banagiri}, \citenamefont {Barker}, \citenamefont {Bartlett},
  \citenamefont {Berger}, \citenamefont {Betzwieser}, \citenamefont
  {Bhattacharjee}, \citenamefont {Billingsley}, \citenamefont {Biscans},
  \citenamefont {Blair}, \citenamefont {Bode}, \citenamefont {Booker},
  \citenamefont {Bork}, \citenamefont {Bramley}, \citenamefont {Brooks},
  \citenamefont {Buikema}, \citenamefont {Cahillane}, \citenamefont {Cannon},
  \citenamefont {Chen}, \citenamefont {Ciobanu}, \citenamefont {Clara},
  \citenamefont {Cooper}, \citenamefont {Corley}, \citenamefont {Countryman},
  \citenamefont {Covas}, \citenamefont {Coyne}, \citenamefont {Datrier},
  \citenamefont {Davis}, \citenamefont {Di~Fronzo}, \citenamefont {Driggers},
  \citenamefont {Etzel}, \citenamefont {Evans}, \citenamefont {Feicht},
  \citenamefont {Fulda}, \citenamefont {Fyffe}, \citenamefont {Giaime},
  \citenamefont {Giardina}, \citenamefont {Godwin}, \citenamefont {Goetz},
  \citenamefont {Gras}, \citenamefont {Gray}, \citenamefont {Gray},
  \citenamefont {Gupta}, \citenamefont {Gustafson}, \citenamefont {Gustafson},
  \citenamefont {Hanks}, \citenamefont {Hanson}, \citenamefont {Hardwick},
  \citenamefont {Hasskew}, \citenamefont {Heintze}, \citenamefont
  {Helmling-Cornell}, \citenamefont {Holland}, \citenamefont {Jones},
  \citenamefont {Kandhasamy}, \citenamefont {Karki}, \citenamefont {Kasprzack},
  \citenamefont {Kawabe}, \citenamefont {King}, \citenamefont {Kissel},
  \citenamefont {Kumar}, \citenamefont {Landry}, \citenamefont {Lane},
  \citenamefont {Lantz}, \citenamefont {Laxen}, \citenamefont {Lecoeuche},
  \citenamefont {Leviton}, \citenamefont {Liu}, \citenamefont {Lormand},
  \citenamefont {Lundgren}, \citenamefont {Macas}, \citenamefont {MacInnis},
  \citenamefont {Macleod}, \citenamefont {M{\'a}rka}, \citenamefont
  {M{\'a}rka}, \citenamefont {Martynov}, \citenamefont {Mason}, \citenamefont
  {Massinger}, \citenamefont {McCarthy}, \citenamefont {McCormick},
  \citenamefont {McIver}, \citenamefont {Mendell}, \citenamefont {Merfeld},
  \citenamefont {Merilh}, \citenamefont {Meylahn}, \citenamefont {Mistry},
  \citenamefont {Mittleman}, \citenamefont {Moreno}, \citenamefont {Mow-Lowry},
  \citenamefont {Mozzon}, \citenamefont {Nelson}, \citenamefont {Nguyen},
  \citenamefont {Nuttall}, \citenamefont {Oberling}, \citenamefont {Oram},
  \citenamefont {O'Reilly}, \citenamefont {Osthelder}, \citenamefont {Ottaway},
  \citenamefont {Overmier}, \citenamefont {Palamos}, \citenamefont {Parker},
  \citenamefont {Payne}, \citenamefont {Pele}, \citenamefont {Perez},
  \citenamefont {Pirello}, \citenamefont {Radkins}, \citenamefont {Ramirez},
  \citenamefont {Richardson}, \citenamefont {Riles}, \citenamefont {Robertson},
  \citenamefont {Rollins}, \citenamefont {Romel}, \citenamefont {Romie},
  \citenamefont {Ross}, \citenamefont {Ryan}, \citenamefont {Sadecki},
  \citenamefont {Sanchez}, \citenamefont {Sanchez}, \citenamefont {Saravanan},
  \citenamefont {Savage}, \citenamefont {Schaetzl}, \citenamefont {Schnabel},
  \citenamefont {Schofield}, \citenamefont {Schwartz}, \citenamefont {Sellers},
  \citenamefont {Shaffer}, \citenamefont {Smith}, \citenamefont {Soni},
  \citenamefont {Sorazu}, \citenamefont {Spencer}, \citenamefont {Strain},
  \citenamefont {Sun}, \citenamefont {Szczepa{\'n}czyk}, \citenamefont
  {Thomas}, \citenamefont {Thomas}, \citenamefont {Thorne}, \citenamefont
  {Toland}, \citenamefont {Torrie}, \citenamefont {Traylor}, \citenamefont
  {Urban}, \citenamefont {Vajente}, \citenamefont {Valdes}, \citenamefont
  {Vander-Hyde}, \citenamefont {Veitch}, \citenamefont {Venkateswara},
  \citenamefont {Venugopalan}, \citenamefont {Viets}, \citenamefont {Vorvick},
  \citenamefont {Wade}, \citenamefont {Warner}, \citenamefont {Weaver},
  \citenamefont {Weiss}, \citenamefont {Willke}, \citenamefont {Wipf},
  \citenamefont {Xiao}, \citenamefont {Yamamoto}, \citenamefont {Yap},
  \citenamefont {Yu}, \citenamefont {Zhang}, \citenamefont {Zucker},\ and\
  \citenamefont {Zweizig}}]{Tse2019}%
  \BibitemOpen
  \bibfield  {author} {\bibinfo {author} {\bibfnamefont {M.}~\bibnamefont
  {Tse}}, \bibnamefont {et~al.},\ }\bibfield  {title} {\emph {\enquote
  {\bibinfo {title} {Quantum-enhanced advanced {LIGO} detectors in the era of
  gravitational-wave astronomy},}\ }}\href
  {https://doi.org/10.1103/physrevlett.123.231107} {\bibfield  {journal}
  {\bibinfo  {journal} {Phys. Rev. Lett.}\ }\textbf {\bibinfo {volume} {123}},\
  \bibinfo {pages} {231107} (\bibinfo {year} {2019})}\BibitemShut {NoStop}%
\bibitem [{\citenamefont {Ganapathy}\ \emph {et~al.}(2023)\citenamefont
  {Ganapathy}, \citenamefont {Jia}, \citenamefont {Nakano}, \citenamefont {Xu},
  \citenamefont {Aritomi}, \citenamefont {Cullen}, \citenamefont {Kijbunchoo},
  \citenamefont {Dwyer}, \citenamefont {Mullavey}, \citenamefont {McCuller},
  \citenamefont {Abbott}, \citenamefont {Abouelfettouh}, \citenamefont
  {Adhikari}, \citenamefont {Ananyeva}, \citenamefont {Appert}, \citenamefont
  {Arai}, \citenamefont {Aston}, \citenamefont {Ball}, \citenamefont {Ballmer},
  \citenamefont {Barker}, \citenamefont {Barsotti}, \citenamefont {Berger},
  \citenamefont {Betzwieser}, \citenamefont {Bhattacharjee}, \citenamefont
  {Billingsley}, \citenamefont {Biscans}, \citenamefont {Bode}, \citenamefont
  {Bonilla}, \citenamefont {Bossilkov}, \citenamefont {Branch}, \citenamefont
  {Brooks}, \citenamefont {Brown}, \citenamefont {Bryant}, \citenamefont
  {Cahillane}, \citenamefont {Cao}, \citenamefont {Capote}, \citenamefont
  {Clara}, \citenamefont {Collins}, \citenamefont {Compton}, \citenamefont
  {Cottingham}, \citenamefont {Coyne}, \citenamefont {Crouch}, \citenamefont
  {Csizmazia}, \citenamefont {Dartez}, \citenamefont {Demos}, \citenamefont
  {Dohmen}, \citenamefont {Driggers}, \citenamefont {Effler}, \citenamefont
  {Ejlli}, \citenamefont {Etzel}, \citenamefont {Evans}, \citenamefont
  {Feicht}, \citenamefont {Frey}, \citenamefont {Frischhertz}, \citenamefont
  {Fritschel}, \citenamefont {Frolov}, \citenamefont {Fulda}, \citenamefont
  {Fyffe}, \citenamefont {Gateley}, \citenamefont {Giaime}, \citenamefont
  {Giardina}, \citenamefont {Glanzer}, \citenamefont {Goetz}, \citenamefont
  {Goetz}, \citenamefont {Goodwin-Jones}, \citenamefont {Gras}, \citenamefont
  {Gray}, \citenamefont {Griffith}, \citenamefont {Grote}, \citenamefont
  {Guidry}, \citenamefont {Hall}, \citenamefont {Hanks}, \citenamefont
  {Hanson}, \citenamefont {Heintze}, \citenamefont {Helmling-Cornell},
  \citenamefont {Holland}, \citenamefont {Hoyland}, \citenamefont {Huang},
  \citenamefont {Inoue}, \citenamefont {James}, \citenamefont {Jennings},
  \citenamefont {Karat}, \citenamefont {Karki}, \citenamefont {Kasprzack},
  \citenamefont {Kawabe}, \citenamefont {King}, \citenamefont {Kissel},
  \citenamefont {Komori}, \citenamefont {Kontos}, \citenamefont {Kumar},
  \citenamefont {Kuns}, \citenamefont {Landry}, \citenamefont {Lantz},
  \citenamefont {Laxen}, \citenamefont {Lee}, \citenamefont {Lesovsky},
  \citenamefont {Llamas}, \citenamefont {Lormand}, \citenamefont {Loughlin},
  \citenamefont {Macas}, \citenamefont {MacInnis}, \citenamefont {Makarem},
  \citenamefont {Mannix}, \citenamefont {Mansell}, \citenamefont {Martin},
  \citenamefont {Mason}, \citenamefont {Matichard}, \citenamefont {Mavalvala},
  \citenamefont {Maxwell}, \citenamefont {McCarrol}, \citenamefont {McCarthy},
  \citenamefont {McClelland}, \citenamefont {McCormick}, \citenamefont {McRae},
  \citenamefont {Mera}, \citenamefont {Merilh}, \citenamefont {Meylahn},
  \citenamefont {Mittleman}, \citenamefont {Moraru}, \citenamefont {Moreno},
  \citenamefont {Nelson}, \citenamefont {Neunzert}, \citenamefont {Notte},
  \citenamefont {Oberling}, \citenamefont {O'Hanlon}, \citenamefont
  {Osthelder}, \citenamefont {Ottaway}, \citenamefont {Overmier}, \citenamefont
  {Parker}, \citenamefont {Pele}, \citenamefont {Pham}, \citenamefont
  {Pirello}, \citenamefont {Quetschke}, \citenamefont {Ramirez}, \citenamefont
  {Reyes}, \citenamefont {Richardson}, \citenamefont {Robinson}, \citenamefont
  {Rollins}, \citenamefont {Romel}, \citenamefont {Romie}, \citenamefont
  {Ross}, \citenamefont {Ryan}, \citenamefont {Sadecki}, \citenamefont
  {Sanchez}, \citenamefont {Sanchez}, \citenamefont {Sanchez}, \citenamefont
  {Savage}, \citenamefont {Schaetzl}, \citenamefont {Schiworski}, \citenamefont
  {Schnabel}, \citenamefont {Schofield}, \citenamefont {Schwartz},
  \citenamefont {Sellers}, \citenamefont {Shaffer}, \citenamefont {Short},
  \citenamefont {Sigg}, \citenamefont {Slagmolen}, \citenamefont {Soike},
  \citenamefont {Soni}, \citenamefont {Srivastava}, \citenamefont {Sun},
  \citenamefont {Tanner}, \citenamefont {Thomas}, \citenamefont {Thomas},
  \citenamefont {Thorne}, \citenamefont {Torrie}, \citenamefont {Traylor},
  \citenamefont {Ubhi}, \citenamefont {Vajente}, \citenamefont {Vanosky},
  \citenamefont {Vecchio}, \citenamefont {Veitch}, \citenamefont {Vibhute},
  \citenamefont {von Reis}, \citenamefont {Warner}, \citenamefont {Weaver},
  \citenamefont {Weiss}, \citenamefont {Whittle}, \citenamefont {Willke},
  \citenamefont {Wipf}, \citenamefont {Yamamoto}, \citenamefont {Zhang},\ and\
  \citenamefont {Zucker}}]{LIGO2023}%
  \BibitemOpen
  \bibfield  {author} {\bibinfo {author} {\bibfnamefont {D.}~\bibnamefont
  {Ganapathy}}, \bibnamefont {et~al.} (\bibinfo {collaboration} {LIGO O4
  Detector Collaboration}),\ }\bibfield  {title} {\emph {\enquote {\bibinfo
  {title} {Broadband quantum enhancement of the {LIGO} detectors with
  frequency-dependent squeezing},}\ }}\href
  {https://doi.org/10.1103/physrevx.13.041021} {\bibfield  {journal} {\bibinfo
  {journal} {Phys. Rev. X}\ }\textbf {\bibinfo {volume} {13}},\ \bibinfo
  {pages} {041021} (\bibinfo {year} {2023})}\BibitemShut {NoStop}%
\bibitem [{\citenamefont {Wieczorek}\ \emph {et~al.}(2015)\citenamefont
  {Wieczorek}, \citenamefont {Hofer}, \citenamefont {Hoelscher-Obermaier},
  \citenamefont {Riedinger}, \citenamefont {Hammerer},\ and\ \citenamefont
  {Aspelmeyer}}]{Wieczorek2015}%
  \BibitemOpen
  \bibfield  {author} {\bibinfo {author} {\bibfnamefont {W.}~\bibnamefont
  {Wieczorek}}, \bibinfo {author} {\bibfnamefont {S.~G.}\ \bibnamefont
  {Hofer}}, \bibinfo {author} {\bibfnamefont {J.}~\bibnamefont
  {Hoelscher-Obermaier}}, \bibinfo {author} {\bibfnamefont {R.}~\bibnamefont
  {Riedinger}}, \bibinfo {author} {\bibfnamefont {K.}~\bibnamefont
  {Hammerer}},\ \bibnamefont {and}\ \bibinfo {author} {\bibfnamefont
  {M.}~\bibnamefont {Aspelmeyer}},\ }\bibfield  {title} {\emph {\enquote
  {\bibinfo {title} {Optimal state estimation for cavity optomechanical
  systems},}\ }}\href {https://doi.org/10.1103/physrevlett.114.223601}
  {\bibfield  {journal} {\bibinfo  {journal} {Phys. Rev. Lett.}\ }\textbf
  {\bibinfo {volume} {114}},\ \bibinfo {pages} {223601} (\bibinfo {year}
  {2015})}\BibitemShut {NoStop}%
\bibitem [{\citenamefont {Rossi}\ \emph {et~al.}(2019)\citenamefont {Rossi},
  \citenamefont {Mason}, \citenamefont {Chen},\ and\ \citenamefont
  {Schliesser}}]{Rossi2019}%
  \BibitemOpen
  \bibfield  {author} {\bibinfo {author} {\bibfnamefont {M.}~\bibnamefont
  {Rossi}}, \bibinfo {author} {\bibfnamefont {D.}~\bibnamefont {Mason}},
  \bibinfo {author} {\bibfnamefont {J.}~\bibnamefont {Chen}},\ \bibnamefont
  {and}\ \bibinfo {author} {\bibfnamefont {A.}~\bibnamefont {Schliesser}},\
  }\bibfield  {title} {\emph {\enquote {\bibinfo {title} {Observing and
  verifying the quantum trajectory of a mechanical resonator},}\ }}\href
  {https://doi.org/10.1103/physrevlett.123.163601} {\bibfield  {journal}
  {\bibinfo  {journal} {Phys. Rev. Lett.}\ }\textbf {\bibinfo {volume} {123}},\
  \bibinfo {pages} {163601} (\bibinfo {year} {2019})}\BibitemShut {NoStop}%
\bibitem [{\citenamefont {Magrini}\ \emph {et~al.}(2021)\citenamefont
  {Magrini}, \citenamefont {Rosenzweig}, \citenamefont {Bach}, \citenamefont
  {Deutschmann-Olek}, \citenamefont {Hofer}, \citenamefont {Hong},
  \citenamefont {Kiesel}, \citenamefont {Kugi},\ and\ \citenamefont
  {Aspelmeyer}}]{Magrini2021}%
  \BibitemOpen
  \bibfield  {author} {\bibinfo {author} {\bibfnamefont {L.}~\bibnamefont
  {Magrini}}, \bibinfo {author} {\bibfnamefont {P.}~\bibnamefont {Rosenzweig}},
  \bibinfo {author} {\bibfnamefont {C.}~\bibnamefont {Bach}}, \bibinfo {author}
  {\bibfnamefont {A.}~\bibnamefont {Deutschmann-Olek}}, \bibinfo {author}
  {\bibfnamefont {S.~G.}\ \bibnamefont {Hofer}}, \bibinfo {author}
  {\bibfnamefont {S.}~\bibnamefont {Hong}}, \bibinfo {author} {\bibfnamefont
  {N.}~\bibnamefont {Kiesel}}, \bibinfo {author} {\bibfnamefont
  {A.}~\bibnamefont {Kugi}},\ \bibnamefont {and}\ \bibinfo {author}
  {\bibfnamefont {M.}~\bibnamefont {Aspelmeyer}},\ }\bibfield  {title} {\emph
  {\enquote {\bibinfo {title} {Real-time optimal quantum control of mechanical
  motion at room temperature},}\ }}\href
  {https://doi.org/10.1038/s41586-021-03602-3} {\bibfield  {journal} {\bibinfo
  {journal} {Nature}\ }\textbf {\bibinfo {volume} {595}},\ \bibinfo {pages}
  {373} (\bibinfo {year} {2021})}\BibitemShut {NoStop}%
\bibitem [{\citenamefont {Tebbenjohanns}\ \emph {et~al.}(2021)\citenamefont
  {Tebbenjohanns}, \citenamefont {Mattana}, \citenamefont {Rossi},
  \citenamefont {Frimmer},\ and\ \citenamefont {Novotny}}]{Tebbenjohanns2021}%
  \BibitemOpen
  \bibfield  {author} {\bibinfo {author} {\bibfnamefont {F.}~\bibnamefont
  {Tebbenjohanns}}, \bibinfo {author} {\bibfnamefont {M.~L.}\ \bibnamefont
  {Mattana}}, \bibinfo {author} {\bibfnamefont {M.}~\bibnamefont {Rossi}},
  \bibinfo {author} {\bibfnamefont {M.}~\bibnamefont {Frimmer}},\ \bibnamefont
  {and}\ \bibinfo {author} {\bibfnamefont {L.}~\bibnamefont {Novotny}},\
  }\bibfield  {title} {\emph {\enquote {\bibinfo {title} {Quantum control of a
  nanoparticle optically levitated in cryogenic free space},}\ }}\href
  {https://doi.org/10.1038/s41586-021-03617-w} {\bibfield  {journal} {\bibinfo
  {journal} {Nature}\ }\textbf {\bibinfo {volume} {595}},\ \bibinfo {pages}
  {378} (\bibinfo {year} {2021})}\BibitemShut {NoStop}%
\bibitem [{\citenamefont {Braginsky}\ and\ \citenamefont
  {Khalili}(1995)}]{Braginsky1995}%
  \BibitemOpen
  \bibfield  {author} {\bibinfo {author} {\bibfnamefont {V.~B.}\ \bibnamefont
  {Braginsky}}\ \bibnamefont {and}\ \bibinfo {author} {\bibfnamefont {F.~Y.}\
  \bibnamefont {Khalili}},\ }\href@noop {} {\emph {\bibinfo {title} {Quantum
  measurement}}}\ (\bibinfo  {publisher} {Cambridge University Press},\
  \bibinfo {year} {1995})\BibitemShut {NoStop}%
\bibitem [{\citenamefont {Wiseman}\ and\ \citenamefont
  {Milburn}(2009)}]{Wiseman2009}%
  \BibitemOpen
  \bibfield  {author} {\bibinfo {author} {\bibfnamefont {H.~M.}\ \bibnamefont
  {Wiseman}}\ \bibnamefont {and}\ \bibinfo {author} {\bibfnamefont {G.~J.}\
  \bibnamefont {Milburn}},\ }\href@noop {} {\emph {\bibinfo {title} {Quantum
  measurement and control}}}\ (\bibinfo  {publisher} {Cambridge University
  Press},\ \bibinfo {year} {2009})\BibitemShut {NoStop}%
\bibitem [{\citenamefont {Ferraro}\ \emph {et~al.}(2005)\citenamefont
  {Ferraro}, \citenamefont {Olivares},\ and\ \citenamefont
  {Paris}}]{Ferraro2005}%
  \BibitemOpen
  \bibfield  {author} {\bibinfo {author} {\bibfnamefont {A.}~\bibnamefont
  {Ferraro}}, \bibinfo {author} {\bibfnamefont {S.}~\bibnamefont {Olivares}},\
  \bibnamefont {and}\ \bibinfo {author} {\bibfnamefont {M.~G.~A.}\ \bibnamefont
  {Paris}},\ }\href@noop {} {\emph {\bibinfo {title} {{Gaussian} states in
  continuous variable quantum information}}}\ (\bibinfo  {publisher}
  {Bibliopolis},\ \bibinfo {address} {Napoli},\ \bibinfo {year} {2005})\
  \Eprint {https://arxiv.org/abs/quant-ph/0503237} {arXiv:quant-ph/0503237}
  \BibitemShut {NoStop}%
\bibitem [{\citenamefont {Weedbrook}\ \emph {et~al.}(2012)\citenamefont
  {Weedbrook}, \citenamefont {Pirandola}, \citenamefont
  {Garc{\'\i}a-Patr{\'o}n}, \citenamefont {Cerf}, \citenamefont {Ralph},
  \citenamefont {Shapiro},\ and\ \citenamefont {Lloyd}}]{Weedbrook2012}%
  \BibitemOpen
  \bibfield  {author} {\bibinfo {author} {\bibfnamefont {C.}~\bibnamefont
  {Weedbrook}}, \bibinfo {author} {\bibfnamefont {S.}~\bibnamefont
  {Pirandola}}, \bibinfo {author} {\bibfnamefont {R.}~\bibnamefont
  {Garc{\'\i}a-Patr{\'o}n}}, \bibinfo {author} {\bibfnamefont {N.~J.}\
  \bibnamefont {Cerf}}, \bibinfo {author} {\bibfnamefont {T.~C.}\ \bibnamefont
  {Ralph}}, \bibinfo {author} {\bibfnamefont {J.~H.}\ \bibnamefont {Shapiro}},\
  \bibnamefont {and}\ \bibinfo {author} {\bibfnamefont {S.}~\bibnamefont
  {Lloyd}},\ }\bibfield  {title} {\emph {\enquote {\bibinfo {title} {{Gaussian}
  quantum information},}\ }}\href {https://doi.org/10.1103/revmodphys.84.621}
  {\bibfield  {journal} {\bibinfo  {journal} {Rev. Mod. Phys.}\ }\textbf
  {\bibinfo {volume} {84}},\ \bibinfo {pages} {621} (\bibinfo {year}
  {2012})}\BibitemShut {NoStop}%
\bibitem [{\citenamefont {Hadfield}(2009)}]{Hadfield2009}%
  \BibitemOpen
  \bibfield  {author} {\bibinfo {author} {\bibfnamefont {R.~H.}\ \bibnamefont
  {Hadfield}},\ }\bibfield  {title} {\emph {\enquote {\bibinfo {title}
  {Single-photon detectors for optical quantum information applications},}\
  }}\href {https://doi.org/10.1038/nphoton.2009.230} {\bibfield  {journal}
  {\bibinfo  {journal} {Nat. Photonics}\ }\textbf {\bibinfo {volume} {3}},\
  \bibinfo {pages} {696} (\bibinfo {year} {2009})}\BibitemShut {NoStop}%
\bibitem [{\citenamefont {Becker}(2005)}]{Becker2005}%
  \BibitemOpen
  \bibfield  {author} {\bibinfo {author} {\bibfnamefont {W.}~\bibnamefont
  {Becker}},\ }in\ \href {https://doi.org/10.1007/3-540-28882-1_5} {\emph
  {\bibinfo {booktitle} {Advanced Time-Correlated Single Photon Counting
  Techniques}}},\ \bibinfo {series} {Springer Series in Chemical Physics},
  Vol.~\bibinfo {volume} {81}\ (\bibinfo  {publisher} {Springer},\ \bibinfo
  {address} {Berlin, Heidelberg},\ \bibinfo {year} {2005})\ pp.\ \bibinfo
  {pages} {61--212}\BibitemShut {NoStop}%
\bibitem [{\citenamefont {Cohen}\ \emph {et~al.}(2015)\citenamefont {Cohen},
  \citenamefont {Meenehan}, \citenamefont {MacCabe}, \citenamefont
  {Gr{\"o}blacher}, \citenamefont {Safavi-Naeini}, \citenamefont {Marsili},
  \citenamefont {Shaw},\ and\ \citenamefont {Painter}}]{Cohen2015}%
  \BibitemOpen
  \bibfield  {author} {\bibinfo {author} {\bibfnamefont {J.~D.}\ \bibnamefont
  {Cohen}}, \bibinfo {author} {\bibfnamefont {S.~M.}\ \bibnamefont {Meenehan}},
  \bibinfo {author} {\bibfnamefont {G.~S.}\ \bibnamefont {MacCabe}}, \bibinfo
  {author} {\bibfnamefont {S.}~\bibnamefont {Gr{\"o}blacher}}, \bibinfo
  {author} {\bibfnamefont {A.~H.}\ \bibnamefont {Safavi-Naeini}}, \bibinfo
  {author} {\bibfnamefont {F.}~\bibnamefont {Marsili}}, \bibinfo {author}
  {\bibfnamefont {M.~D.}\ \bibnamefont {Shaw}},\ \bibnamefont {and}\ \bibinfo
  {author} {\bibfnamefont {O.}~\bibnamefont {Painter}},\ }\bibfield  {title}
  {\emph {\enquote {\bibinfo {title} {Phonon counting and intensity
  interferometry of a nanomechanical resonator},}\ }}\href
  {https://doi.org/10.1038/nature14349} {\bibfield  {journal} {\bibinfo
  {journal} {Nature}\ }\textbf {\bibinfo {volume} {520}},\ \bibinfo {pages}
  {522} (\bibinfo {year} {2015})}\BibitemShut {NoStop}%
\bibitem [{\citenamefont {Riedinger}\ \emph {et~al.}(2018)\citenamefont
  {Riedinger}, \citenamefont {Wallucks}, \citenamefont {Marinkovi{\'c}},
  \citenamefont {L{\"o}schnauer}, \citenamefont {Aspelmeyer}, \citenamefont
  {Hong},\ and\ \citenamefont {Gr{\"o}blacher}}]{Riedinger2018}%
  \BibitemOpen
  \bibfield  {author} {\bibinfo {author} {\bibfnamefont {R.}~\bibnamefont
  {Riedinger}}, \bibinfo {author} {\bibfnamefont {A.}~\bibnamefont {Wallucks}},
  \bibinfo {author} {\bibfnamefont {I.}~\bibnamefont {Marinkovi{\'c}}},
  \bibinfo {author} {\bibfnamefont {C.}~\bibnamefont {L{\"o}schnauer}},
  \bibinfo {author} {\bibfnamefont {M.}~\bibnamefont {Aspelmeyer}}, \bibinfo
  {author} {\bibfnamefont {S.}~\bibnamefont {Hong}},\ \bibnamefont {and}\
  \bibinfo {author} {\bibfnamefont {S.}~\bibnamefont {Gr{\"o}blacher}},\
  }\bibfield  {title} {\emph {\enquote {\bibinfo {title} {Remote quantum
  entanglement between two micromechanical oscillators},}\ }}\href
  {https://doi.org/10.1038/s41586-018-0036-z} {\bibfield  {journal} {\bibinfo
  {journal} {Nature}\ }\textbf {\bibinfo {volume} {556}},\ \bibinfo {pages}
  {473} (\bibinfo {year} {2018})}\BibitemShut {NoStop}%
\bibitem [{\citenamefont {Galinskiy}\ \emph {et~al.}(2020)\citenamefont
  {Galinskiy}, \citenamefont {Tsaturyan}, \citenamefont {Parniak},\ and\
  \citenamefont {Polzik}}]{Galinskiy2020}%
  \BibitemOpen
  \bibfield  {author} {\bibinfo {author} {\bibfnamefont {I.}~\bibnamefont
  {Galinskiy}}, \bibinfo {author} {\bibfnamefont {Y.}~\bibnamefont
  {Tsaturyan}}, \bibinfo {author} {\bibfnamefont {M.}~\bibnamefont {Parniak}},\
  \bibnamefont {and}\ \bibinfo {author} {\bibfnamefont {E.~S.}\ \bibnamefont
  {Polzik}},\ }\bibfield  {title} {\emph {\enquote {\bibinfo {title} {Phonon
  counting thermometry of an ultracoherent membrane resonator near its motional
  ground state},}\ }}\href {https://doi.org/10.1364/optica.390939} {\bibfield
  {journal} {\bibinfo  {journal} {Optica}\ }\textbf {\bibinfo {volume} {7}},\
  \bibinfo {pages} {718} (\bibinfo {year} {2020})}\BibitemShut {NoStop}%
\bibitem [{\citenamefont {Fiaschi}\ \emph {et~al.}(2021)\citenamefont
  {Fiaschi}, \citenamefont {Hensen}, \citenamefont {Wallucks}, \citenamefont
  {Benevides}, \citenamefont {Li}, \citenamefont {Alegre},\ and\ \citenamefont
  {Gr{\"o}blacher}}]{Fiaschi2021}%
  \BibitemOpen
  \bibfield  {author} {\bibinfo {author} {\bibfnamefont {N.}~\bibnamefont
  {Fiaschi}}, \bibinfo {author} {\bibfnamefont {B.}~\bibnamefont {Hensen}},
  \bibinfo {author} {\bibfnamefont {A.}~\bibnamefont {Wallucks}}, \bibinfo
  {author} {\bibfnamefont {R.}~\bibnamefont {Benevides}}, \bibinfo {author}
  {\bibfnamefont {J.}~\bibnamefont {Li}}, \bibinfo {author} {\bibfnamefont
  {T.~P.~M.}\ \bibnamefont {Alegre}},\ \bibnamefont {and}\ \bibinfo {author}
  {\bibfnamefont {S.}~\bibnamefont {Gr{\"o}blacher}},\ }\bibfield  {title}
  {\emph {\enquote {\bibinfo {title} {Optomechanical quantum teleportation},}\
  }}\href {https://doi.org/10.1038/s41566-021-00866-z} {\bibfield  {journal}
  {\bibinfo  {journal} {Nat. Photonics}\ }\textbf {\bibinfo {volume} {15}},\
  \bibinfo {pages} {817} (\bibinfo {year} {2021})}\BibitemShut {NoStop}%
\bibitem [{\citenamefont {Ortolano}\ \emph {et~al.}(2021)\citenamefont
  {Ortolano}, \citenamefont {Losero}, \citenamefont {Pirandola}, \citenamefont
  {Genovese},\ and\ \citenamefont {Ruo-Berchera}}]{Ortolano2021}%
  \BibitemOpen
  \bibfield  {author} {\bibinfo {author} {\bibfnamefont {G.}~\bibnamefont
  {Ortolano}}, \bibinfo {author} {\bibfnamefont {E.}~\bibnamefont {Losero}},
  \bibinfo {author} {\bibfnamefont {S.}~\bibnamefont {Pirandola}}, \bibinfo
  {author} {\bibfnamefont {M.}~\bibnamefont {Genovese}},\ \bibnamefont {and}\
  \bibinfo {author} {\bibfnamefont {I.}~\bibnamefont {Ruo-Berchera}},\
  }\bibfield  {title} {\emph {\enquote {\bibinfo {title} {Experimental quantum
  reading with photon counting},}\ }}\href
  {https://doi.org/10.1126/sciadv.abc7796} {\bibfield  {journal} {\bibinfo
  {journal} {Sci. Adv.}\ }\textbf {\bibinfo {volume} {7}},\ \bibinfo {pages}
  {eabc7796} (\bibinfo {year} {2021})}\BibitemShut {NoStop}%
\bibitem [{\citenamefont {Galinskiy}\ \emph {et~al.}(2024)\citenamefont
  {Galinskiy}, \citenamefont {Enzian}, \citenamefont {Parniak},\ and\
  \citenamefont {Polzik}}]{Galinskiy2024}%
  \BibitemOpen
  \bibfield  {author} {\bibinfo {author} {\bibfnamefont {I.}~\bibnamefont
  {Galinskiy}}, \bibinfo {author} {\bibfnamefont {G.}~\bibnamefont {Enzian}},
  \bibinfo {author} {\bibfnamefont {M.}~\bibnamefont {Parniak}},\ \bibnamefont
  {and}\ \bibinfo {author} {\bibfnamefont {E.~S.}\ \bibnamefont {Polzik}},\
  }\bibfield  {title} {\emph {\enquote {\bibinfo {title} {Nonclassical
  correlations between photons and phonons of center-of-mass motion of a
  mechanical oscillator},}\ }}\href
  {https://doi.org/10.1103/physrevlett.133.173605} {\bibfield  {journal}
  {\bibinfo  {journal} {Phys. Rev. Lett.}\ }\textbf {\bibinfo {volume} {133}},\
  \bibinfo {pages} {173605} (\bibinfo {year} {2024})}\BibitemShut {NoStop}%
\bibitem [{\citenamefont {Kiilerich}\ and\ \citenamefont
  {M{\o}lmer}(2014)}]{Kiilerich2014}%
  \BibitemOpen
  \bibfield  {author} {\bibinfo {author} {\bibfnamefont {A.~H.}\ \bibnamefont
  {Kiilerich}}\ \bibnamefont {and}\ \bibinfo {author} {\bibfnamefont
  {K.}~\bibnamefont {M{\o}lmer}},\ }\bibfield  {title} {\emph {\enquote
  {\bibinfo {title} {Estimation of atomic interaction parameters by photon
  counting},}\ }}\href {https://doi.org/10.1103/physreva.89.052110} {\bibfield
  {journal} {\bibinfo  {journal} {Phys. Rev. A}\ }\textbf {\bibinfo {volume}
  {89}},\ \bibinfo {pages} {052110} (\bibinfo {year} {2014})}\BibitemShut
  {NoStop}%
\bibitem [{\citenamefont {Plenio}\ and\ \citenamefont
  {Knight}(1998)}]{Plenio1998}%
  \BibitemOpen
  \bibfield  {author} {\bibinfo {author} {\bibfnamefont {M.~B.}\ \bibnamefont
  {Plenio}}\ \bibnamefont {and}\ \bibinfo {author} {\bibfnamefont {P.~L.}\
  \bibnamefont {Knight}},\ }\bibfield  {title} {\emph {\enquote {\bibinfo
  {title} {The quantum-jump approach to dissipative dynamics in quantum
  optics},}\ }}\href {https://doi.org/10.1103/revmodphys.70.101} {\bibfield
  {journal} {\bibinfo  {journal} {Rev. Mod. Phys.}\ }\textbf {\bibinfo {volume}
  {70}},\ \bibinfo {pages} {101} (\bibinfo {year} {1998})}\BibitemShut
  {NoStop}%
\bibitem [{\citenamefont {Clark}\ and\ \citenamefont
  {Ko{\l}ody{\'n}ski}(2025)}]{Clark2025}%
  \BibitemOpen
  \bibfield  {author} {\bibinfo {author} {\bibfnamefont {L.~A.}\ \bibnamefont
  {Clark}}\ \bibnamefont {and}\ \bibinfo {author} {\bibfnamefont
  {J.}~\bibnamefont {Ko{\l}ody{\'n}ski}},\ }\bibfield  {title} {\emph {\enquote
  {\bibinfo {title} {Efficient inference of quantum system parameters by
  {Approximate Bayesian Computation}},}\ }}\href
  {https://doi.org/10.1103/physrevapplied.23.044040} {\bibfield  {journal}
  {\bibinfo  {journal} {Phys. Rev. Applied}\ }\textbf {\bibinfo {volume}
  {23}},\ \bibinfo {pages} {044040} (\bibinfo {year} {2025})}\BibitemShut
  {NoStop}%
\bibitem [{\citenamefont {Rinaldi}\ \emph {et~al.}(2024)\citenamefont
  {Rinaldi}, \citenamefont {Gonz{\'a}lez~Lastre}, \citenamefont
  {Garc{\'\i}a~Herreros}, \citenamefont {Ahmed}, \citenamefont {Khanahmadi},
  \citenamefont {Nori},\ and\ \citenamefont
  {S{\'a}nchez~Mu{\~n}oz}}]{Rinaldi2024}%
  \BibitemOpen
  \bibfield  {author} {\bibinfo {author} {\bibfnamefont {E.}~\bibnamefont
  {Rinaldi}}, \bibinfo {author} {\bibfnamefont {M.}~\bibnamefont
  {Gonz{\'a}lez~Lastre}}, \bibinfo {author} {\bibfnamefont {S.}~\bibnamefont
  {Garc{\'\i}a~Herreros}}, \bibinfo {author} {\bibfnamefont {S.}~\bibnamefont
  {Ahmed}}, \bibinfo {author} {\bibfnamefont {M.}~\bibnamefont {Khanahmadi}},
  \bibinfo {author} {\bibfnamefont {F.}~\bibnamefont {Nori}},\ \bibnamefont
  {and}\ \bibinfo {author} {\bibfnamefont {C.}~\bibnamefont
  {S{\'a}nchez~Mu{\~n}oz}},\ }\bibfield  {title} {\emph {\enquote {\bibinfo
  {title} {Parameter estimation from quantum-jump data using neural
  networks},}\ }}\href {https://doi.org/10.1088/2058-9565/ad3c68} {\bibfield
  {journal} {\bibinfo  {journal} {Quantum Sci. Technol.}\ }\textbf {\bibinfo
  {volume} {9}},\ \bibinfo {pages} {035018} (\bibinfo {year}
  {2024})}\BibitemShut {NoStop}%
\bibitem [{\citenamefont {Radaelli}\ \emph {et~al.}(2026)\citenamefont
  {Radaelli}, \citenamefont {Smiga}, \citenamefont {Landi},\ and\ \citenamefont
  {Binder}}]{Radaelli2024}%
  \BibitemOpen
  \bibfield  {author} {\bibinfo {author} {\bibfnamefont {M.}~\bibnamefont
  {Radaelli}}, \bibinfo {author} {\bibfnamefont {J.~A.}\ \bibnamefont {Smiga}},
  \bibinfo {author} {\bibfnamefont {G.~T.}\ \bibnamefont {Landi}},\
  \bibnamefont {and}\ \bibinfo {author} {\bibfnamefont {F.~C.}\ \bibnamefont
  {Binder}},\ }\bibfield  {title} {\emph {\enquote {\bibinfo {title} {Parameter
  estimation for quantum jump unraveling},}\ }}\href
  {https://doi.org/10.22331/q-2026-02-02-1993} {\bibfield  {journal} {\bibinfo
  {journal} {Quantum}\ }\textbf {\bibinfo {volume} {10}},\ \bibinfo {pages}
  {1993} (\bibinfo {year} {2026})}\BibitemShut {NoStop}%
\bibitem [{\citenamefont {Anteneh}(2025)}]{Anteneh2025}%
  \BibitemOpen
  \bibfield  {author} {\bibinfo {author} {\bibfnamefont {A.}~\bibnamefont
  {Anteneh}},\ }\bibfield  {title} {\emph {\enquote {\bibinfo {title} {Quantum
  parameter estimation with uncertainty quantification from continuous
  measurement data using neural network ensembles},}\ }}\bibfield  {journal}
  {\bibinfo  {journal} {arXiv preprint}\ }\href
  {https://doi.org/10.48550/arXiv.2509.10756} {10.48550/arXiv.2509.10756}
  (\bibinfo {year} {2025}),\ \Eprint {https://arxiv.org/abs/2509.10756}
  {arXiv:2509.10756 [quant-ph]} \BibitemShut {NoStop}%
\bibitem [{\citenamefont {Khan}\ \emph {et~al.}(2025)\citenamefont {Khan},
  \citenamefont {Albarelli},\ and\ \citenamefont {Datta}}]{Khan2025}%
  \BibitemOpen
  \bibfield  {author} {\bibinfo {author} {\bibfnamefont {A.}~\bibnamefont
  {Khan}}, \bibinfo {author} {\bibfnamefont {F.}~\bibnamefont {Albarelli}},\
  \bibnamefont {and}\ \bibinfo {author} {\bibfnamefont {A.}~\bibnamefont
  {Datta}},\ }\bibfield  {title} {\emph {\enquote {\bibinfo {title} {Tensor
  network approach to sensing quantum light-matter interactions},}\ }}\href
  {https://doi.org/10.1103/ljh3-3l4j} {\bibfield  {journal} {\bibinfo
  {journal} {PRX Quantum}\ }\textbf {\bibinfo {volume} {6}},\ \bibinfo {pages}
  {040343} (\bibinfo {year} {2025})}\BibitemShut {NoStop}%
\bibitem [{\citenamefont {Gebhart}\ \emph {et~al.}(2023)\citenamefont
  {Gebhart}, \citenamefont {Santagati}, \citenamefont {Gentile}, \citenamefont
  {Gauger}, \citenamefont {Craig}, \citenamefont {Ares}, \citenamefont
  {Banchi}, \citenamefont {Marquardt}, \citenamefont {Pezz{\`e}},\ and\
  \citenamefont {Bonato}}]{Gebhart2023}%
  \BibitemOpen
  \bibfield  {author} {\bibinfo {author} {\bibfnamefont {V.}~\bibnamefont
  {Gebhart}}, \bibinfo {author} {\bibfnamefont {R.}~\bibnamefont {Santagati}},
  \bibinfo {author} {\bibfnamefont {A.~A.}\ \bibnamefont {Gentile}}, \bibinfo
  {author} {\bibfnamefont {E.~M.}\ \bibnamefont {Gauger}}, \bibinfo {author}
  {\bibfnamefont {D.}~\bibnamefont {Craig}}, \bibinfo {author} {\bibfnamefont
  {N.}~\bibnamefont {Ares}}, \bibinfo {author} {\bibfnamefont {L.}~\bibnamefont
  {Banchi}}, \bibinfo {author} {\bibfnamefont {F.}~\bibnamefont {Marquardt}},
  \bibinfo {author} {\bibfnamefont {L.}~\bibnamefont {Pezz{\`e}}},\
  \bibnamefont {and}\ \bibinfo {author} {\bibfnamefont {C.}~\bibnamefont
  {Bonato}},\ }\bibfield  {title} {\emph {\enquote {\bibinfo {title} {Learning
  quantum systems},}\ }}\href {https://doi.org/10.1038/s42254-022-00552-1}
  {\bibfield  {journal} {\bibinfo  {journal} {Nat. Rev. Phys.}\ }\textbf
  {\bibinfo {volume} {5}},\ \bibinfo {pages} {141} (\bibinfo {year}
  {2023})}\BibitemShut {NoStop}%
\bibitem [{\citenamefont {Gilks}\ \emph {et~al.}(1995)\citenamefont {Gilks},
  \citenamefont {Richardson},\ and\ \citenamefont {Spiegelhalter}}]{Gilks1995}%
  \BibitemOpen
  \bibinfo {editor} {\bibfnamefont {W.~R.}\ \bibnamefont {Gilks}}, \bibinfo
  {editor} {\bibfnamefont {S.}~\bibnamefont {Richardson}},\ \bibnamefont {and}\
  \bibinfo {editor} {\bibfnamefont {D.~J.}\ \bibnamefont {Spiegelhalter}},\
  eds.,\ \href {https://doi.org/10.1201/b14835} {\emph {\bibinfo {title}
  {{Markov} chain {Monte Carlo} in practice}}}\ (\bibinfo  {publisher} {CRC
  Press},\ \bibinfo {year} {1995})\BibitemShut {NoStop}%
\bibitem [{\citenamefont {Murphy}(2012)}]{Murphy2012}%
  \BibitemOpen
  \bibfield  {author} {\bibinfo {author} {\bibfnamefont {K.~P.}\ \bibnamefont
  {Murphy}},\ }\href@noop {} {\emph {\bibinfo {title} {Machine learning: A
  probabilistic perspective}}}\ (\bibinfo  {publisher} {MIT Press},\ \bibinfo
  {year} {2012})\BibitemShut {NoStop}%
\bibitem [{\citenamefont {Granade}\ \emph {et~al.}(2017)\citenamefont
  {Granade}, \citenamefont {Ferrie}, \citenamefont {Hincks}, \citenamefont
  {Casagrande}, \citenamefont {Alexander}, \citenamefont {Gross}, \citenamefont
  {Kononenko},\ and\ \citenamefont {Sanders}}]{Granade2017}%
  \BibitemOpen
  \bibfield  {author} {\bibinfo {author} {\bibfnamefont {C.}~\bibnamefont
  {Granade}}, \bibinfo {author} {\bibfnamefont {C.}~\bibnamefont {Ferrie}},
  \bibinfo {author} {\bibfnamefont {I.}~\bibnamefont {Hincks}}, \bibinfo
  {author} {\bibfnamefont {S.}~\bibnamefont {Casagrande}}, \bibinfo {author}
  {\bibfnamefont {T.}~\bibnamefont {Alexander}}, \bibinfo {author}
  {\bibfnamefont {J.}~\bibnamefont {Gross}}, \bibinfo {author} {\bibfnamefont
  {M.}~\bibnamefont {Kononenko}},\ \bibnamefont {and}\ \bibinfo {author}
  {\bibfnamefont {Y.}~\bibnamefont {Sanders}},\ }\bibfield  {title} {\emph
  {\enquote {\bibinfo {title} {{QInfer}: Statistical inference software for
  quantum applications},}\ }}\href {https://doi.org/10.22331/q-2017-04-25-5}
  {\bibfield  {journal} {\bibinfo  {journal} {Quantum}\ }\textbf {\bibinfo
  {volume} {1}},\ \bibinfo {pages} {5} (\bibinfo {year} {2017})}\BibitemShut
  {NoStop}%
\bibitem [{\citenamefont {Sisson}\ \emph {et~al.}(2018)\citenamefont {Sisson},
  \citenamefont {Fan},\ and\ \citenamefont {Beaumont}}]{Sisson2018}%
  \BibitemOpen
  \bibinfo {editor} {\bibfnamefont {S.~A.}\ \bibnamefont {Sisson}}, \bibinfo
  {editor} {\bibfnamefont {Y.}~\bibnamefont {Fan}},\ \bibnamefont {and}\
  \bibinfo {editor} {\bibfnamefont {M.}~\bibnamefont {Beaumont}},\ eds.,\ \href
  {https://doi.org/10.1201/9781315117195} {\emph {\bibinfo {title} {Handbook of
  {Approximate Bayesian Computation}}}}\ (\bibinfo  {publisher} {CRC Press},\
  \bibinfo {year} {2018})\BibitemShut {NoStop}%
\bibitem [{\citenamefont {Turner}\ and\ \citenamefont
  {Van~Zandt}(2012)}]{Turner2012}%
  \BibitemOpen
  \bibfield  {author} {\bibinfo {author} {\bibfnamefont {B.~M.}\ \bibnamefont
  {Turner}}\ \bibnamefont {and}\ \bibinfo {author} {\bibfnamefont
  {T.}~\bibnamefont {Van~Zandt}},\ }\bibfield  {title} {\emph {\enquote
  {\bibinfo {title} {A tutorial on {Approximate Bayesian Computation}},}\
  }}\href {https://doi.org/10.1016/j.jmp.2012.02.005} {\bibfield  {journal}
  {\bibinfo  {journal} {J. Math. Psychol.}\ }\textbf {\bibinfo {volume} {56}},\
  \bibinfo {pages} {69} (\bibinfo {year} {2012})}\BibitemShut {NoStop}%
\bibitem [{\citenamefont {Beaumont}\ \emph {et~al.}(2002)\citenamefont
  {Beaumont}, \citenamefont {Zhang},\ and\ \citenamefont
  {Balding}}]{Beaumont2002}%
  \BibitemOpen
  \bibfield  {author} {\bibinfo {author} {\bibfnamefont {M.~A.}\ \bibnamefont
  {Beaumont}}, \bibinfo {author} {\bibfnamefont {W.}~\bibnamefont {Zhang}},\
  \bibnamefont {and}\ \bibinfo {author} {\bibfnamefont {D.~J.}\ \bibnamefont
  {Balding}},\ }\bibfield  {title} {\emph {\enquote {\bibinfo {title}
  {{Approximate Bayesian Computation} in population genetics},}\ }}\href
  {https://doi.org/10.1093/genetics/162.4.2025} {\bibfield  {journal} {\bibinfo
   {journal} {Genetics}\ }\textbf {\bibinfo {volume} {162}},\ \bibinfo {pages}
  {2025} (\bibinfo {year} {2002})}\BibitemShut {NoStop}%
\bibitem [{\citenamefont {Goodfellow}\ \emph {et~al.}(2016)\citenamefont
  {Goodfellow}, \citenamefont {Bengio},\ and\ \citenamefont
  {Courville}}]{Goodfellow-et-al-2016}%
  \BibitemOpen
  \bibfield  {author} {\bibinfo {author} {\bibfnamefont {I.}~\bibnamefont
  {Goodfellow}}, \bibinfo {author} {\bibfnamefont {Y.}~\bibnamefont {Bengio}},\
  \bibnamefont {and}\ \bibinfo {author} {\bibfnamefont {A.}~\bibnamefont
  {Courville}},\ }\href {http://www.deeplearningbook.org} {\emph {\bibinfo
  {title} {{Deep Learning}}}}\ (\bibinfo  {publisher} {MIT Press},\ \bibinfo
  {year} {2016})\BibitemShut {NoStop}%
\bibitem [{\citenamefont {Granade}\ \emph {et~al.}(2012)\citenamefont
  {Granade}, \citenamefont {Ferrie}, \citenamefont {Wiebe},\ and\ \citenamefont
  {Cory}}]{Granade2012}%
  \BibitemOpen
  \bibfield  {author} {\bibinfo {author} {\bibfnamefont {C.~E.}\ \bibnamefont
  {Granade}}, \bibinfo {author} {\bibfnamefont {C.}~\bibnamefont {Ferrie}},
  \bibinfo {author} {\bibfnamefont {N.}~\bibnamefont {Wiebe}},\ \bibnamefont
  {and}\ \bibinfo {author} {\bibfnamefont {D.~G.}\ \bibnamefont {Cory}},\
  }\bibfield  {title} {\emph {\enquote {\bibinfo {title} {Robust online
  {Hamiltonian} learning},}\ }}\href
  {https://doi.org/10.1088/1367-2630/14/10/103013} {\bibfield  {journal}
  {\bibinfo  {journal} {New J. Phys.}\ }\textbf {\bibinfo {volume} {14}},\
  \bibinfo {pages} {103013} (\bibinfo {year} {2012})}\BibitemShut {NoStop}%
\bibitem [{\citenamefont {Sergeevich}\ \emph {et~al.}(2011)\citenamefont
  {Sergeevich}, \citenamefont {Chandran}, \citenamefont {Combes}, \citenamefont
  {Bartlett},\ and\ \citenamefont {Wiseman}}]{Sergeevich2011}%
  \BibitemOpen
  \bibfield  {author} {\bibinfo {author} {\bibfnamefont {A.}~\bibnamefont
  {Sergeevich}}, \bibinfo {author} {\bibfnamefont {A.}~\bibnamefont
  {Chandran}}, \bibinfo {author} {\bibfnamefont {J.}~\bibnamefont {Combes}},
  \bibinfo {author} {\bibfnamefont {S.~D.}\ \bibnamefont {Bartlett}},\
  \bibnamefont {and}\ \bibinfo {author} {\bibfnamefont {H.~M.}\ \bibnamefont
  {Wiseman}},\ }\bibfield  {title} {\emph {\enquote {\bibinfo {title}
  {Characterization of a qubit {Hamiltonian} using adaptive measurements in a
  fixed basis},}\ }}\href {https://doi.org/10.1103/physreva.84.052315}
  {\bibfield  {journal} {\bibinfo  {journal} {Phys. Rev. A}\ }\textbf {\bibinfo
  {volume} {84}},\ \bibinfo {pages} {052315} (\bibinfo {year}
  {2011})}\BibitemShut {NoStop}%
\bibitem [{\citenamefont {Wiebe}\ and\ \citenamefont
  {Granade}(2016)}]{Wiebe2016}%
  \BibitemOpen
  \bibfield  {author} {\bibinfo {author} {\bibfnamefont {N.}~\bibnamefont
  {Wiebe}}\ \bibnamefont {and}\ \bibinfo {author} {\bibfnamefont
  {C.}~\bibnamefont {Granade}},\ }\bibfield  {title} {\emph {\enquote {\bibinfo
  {title} {Efficient {Bayesian} phase estimation},}\ }}\href
  {https://doi.org/10.1103/physrevlett.117.010503} {\bibfield  {journal}
  {\bibinfo  {journal} {Phys. Rev. Lett.}\ }\textbf {\bibinfo {volume} {117}},\
  \bibinfo {pages} {010503} (\bibinfo {year} {2016})}\BibitemShut {NoStop}%
\bibitem [{\citenamefont {Khanahmadi}\ and\ \citenamefont
  {M{\o}lmer}(2021)}]{Khanahmadi2021}%
  \BibitemOpen
  \bibfield  {author} {\bibinfo {author} {\bibfnamefont {M.}~\bibnamefont
  {Khanahmadi}}\ \bibnamefont {and}\ \bibinfo {author} {\bibfnamefont
  {K.}~\bibnamefont {M{\o}lmer}},\ }\bibfield  {title} {\emph {\enquote
  {\bibinfo {title} {Time-dependent atomic magnetometry with a recurrent neural
  network},}\ }}\href {https://doi.org/10.1103/physreva.103.032406} {\bibfield
  {journal} {\bibinfo  {journal} {Phys. Rev. A}\ }\textbf {\bibinfo {volume}
  {103}},\ \bibinfo {pages} {032406} (\bibinfo {year} {2021})}\BibitemShut
  {NoStop}%
\bibitem [{\citenamefont {Nolan}\ \emph {et~al.}(2021)\citenamefont {Nolan},
  \citenamefont {Smerzi},\ and\ \citenamefont {Pezz{\`e}}}]{Nolan2021}%
  \BibitemOpen
  \bibfield  {author} {\bibinfo {author} {\bibfnamefont {S.}~\bibnamefont
  {Nolan}}, \bibinfo {author} {\bibfnamefont {A.}~\bibnamefont {Smerzi}},\
  \bibnamefont {and}\ \bibinfo {author} {\bibfnamefont {L.}~\bibnamefont
  {Pezz{\`e}}},\ }\bibfield  {title} {\emph {\enquote {\bibinfo {title} {A
  machine learning approach to {Bayesian} parameter estimation},}\ }}\href
  {https://doi.org/10.1038/s41534-021-00497-w} {\bibfield  {journal} {\bibinfo
  {journal} {npj Quantum Inf.}\ }\textbf {\bibinfo {volume} {7}},\ \bibinfo
  {pages} {169} (\bibinfo {year} {2021})}\BibitemShut {NoStop}%
\bibitem [{\citenamefont {Huerta~Alderete}\ \emph {et~al.}(2022)\citenamefont
  {Huerta~Alderete}, \citenamefont {Gordon}, \citenamefont {Sauvage},
  \citenamefont {Sone}, \citenamefont {Sornborger}, \citenamefont {Coles},\
  and\ \citenamefont {Cerezo}}]{HuertaAlderete2022}%
  \BibitemOpen
  \bibfield  {author} {\bibinfo {author} {\bibfnamefont {C.}~\bibnamefont
  {Huerta~Alderete}}, \bibinfo {author} {\bibfnamefont {M.~H.}\ \bibnamefont
  {Gordon}}, \bibinfo {author} {\bibfnamefont {F.}~\bibnamefont {Sauvage}},
  \bibinfo {author} {\bibfnamefont {A.}~\bibnamefont {Sone}}, \bibinfo {author}
  {\bibfnamefont {A.~T.}\ \bibnamefont {Sornborger}}, \bibinfo {author}
  {\bibfnamefont {P.~J.}\ \bibnamefont {Coles}},\ \bibnamefont {and}\ \bibinfo
  {author} {\bibfnamefont {M.}~\bibnamefont {Cerezo}},\ }\bibfield  {title}
  {\emph {\enquote {\bibinfo {title} {Inference-based quantum sensing},}\
  }}\href {https://doi.org/10.1103/physrevlett.129.190501} {\bibfield
  {journal} {\bibinfo  {journal} {Phys. Rev. Lett.}\ }\textbf {\bibinfo
  {volume} {129}},\ \bibinfo {pages} {190501} (\bibinfo {year}
  {2022})}\BibitemShut {NoStop}%
\bibitem [{\citenamefont {Khalid}\ \emph {et~al.}(2023)\citenamefont {Khalid},
  \citenamefont {Weidner}, \citenamefont {Jonckheere}, \citenamefont
  {Schirmer},\ and\ \citenamefont {Langbein}}]{Khalid2023}%
  \BibitemOpen
  \bibfield  {author} {\bibinfo {author} {\bibfnamefont {I.}~\bibnamefont
  {Khalid}}, \bibinfo {author} {\bibfnamefont {C.~A.}\ \bibnamefont {Weidner}},
  \bibinfo {author} {\bibfnamefont {E.~A.}\ \bibnamefont {Jonckheere}},
  \bibinfo {author} {\bibfnamefont {S.~G.}\ \bibnamefont {Schirmer}},\
  \bibnamefont {and}\ \bibinfo {author} {\bibfnamefont {F.~C.}\ \bibnamefont
  {Langbein}},\ }\bibfield  {title} {\emph {\enquote {\bibinfo {title}
  {Sample-efficient model-based reinforcement learning for quantum control},}\
  }}\href {https://doi.org/10.1103/physrevresearch.5.043002} {\bibfield
  {journal} {\bibinfo  {journal} {Phys. Rev. Research}\ }\textbf {\bibinfo
  {volume} {5}},\ \bibinfo {pages} {043002} (\bibinfo {year}
  {2023})}\BibitemShut {NoStop}%
\bibitem [{\citenamefont {Nikoloska}\ \emph {et~al.}(2026)\citenamefont
  {Nikoloska}, \citenamefont {Joudeh}, \citenamefont {van Sloun},\ and\
  \citenamefont {Simeone}}]{Nikoloska2025}%
  \BibitemOpen
  \bibfield  {author} {\bibinfo {author} {\bibfnamefont {I.}~\bibnamefont
  {Nikoloska}}, \bibinfo {author} {\bibfnamefont {H.}~\bibnamefont {Joudeh}},
  \bibinfo {author} {\bibfnamefont {R.}~\bibnamefont {van Sloun}},\
  \bibnamefont {and}\ \bibinfo {author} {\bibfnamefont {O.}~\bibnamefont
  {Simeone}},\ }\bibfield  {title} {\emph {\enquote {\bibinfo {title} {Dynamic
  estimation loss control in variational quantum sensing via online conformal
  inference},}\ }}in\ \href {https://doi.org/10.1109/icassp55912.2026.11463311}
  {\emph {\bibinfo {booktitle} {2026 IEEE International Conference on
  Acoustics, Speech and Signal Processing (ICASSP)}}}\ (\bibinfo {year}
  {2026})\ pp.\ \bibinfo {pages} {21076--21080}\BibitemShut {NoStop}%
\bibitem [{\citenamefont {Zhou}\ \emph {et~al.}(2024)\citenamefont {Zhou},
  \citenamefont {Du}, \citenamefont {Yin}, \citenamefont {Zhao}, \citenamefont
  {Tian},\ and\ \citenamefont {Tao}}]{Zhou2024}%
  \BibitemOpen
  \bibfield  {author} {\bibinfo {author} {\bibfnamefont {Z.}~\bibnamefont
  {Zhou}}, \bibinfo {author} {\bibfnamefont {Y.}~\bibnamefont {Du}}, \bibinfo
  {author} {\bibfnamefont {X.-F.}\ \bibnamefont {Yin}}, \bibinfo {author}
  {\bibfnamefont {S.}~\bibnamefont {Zhao}}, \bibinfo {author} {\bibfnamefont
  {X.}~\bibnamefont {Tian}},\ \bibnamefont {and}\ \bibinfo {author}
  {\bibfnamefont {D.}~\bibnamefont {Tao}},\ }\bibfield  {title} {\emph
  {\enquote {\bibinfo {title} {Optical quantum sensing for agnostic
  environments via {Deep Learning}},}\ }}\href
  {https://doi.org/10.1103/physrevresearch.6.043267} {\bibfield  {journal}
  {\bibinfo  {journal} {Phys. Rev. Research}\ }\textbf {\bibinfo {volume}
  {6}},\ \bibinfo {pages} {043267} (\bibinfo {year} {2024})}\BibitemShut
  {NoStop}%
\bibitem [{\citenamefont {Xu}\ \emph {et~al.}(2021)\citenamefont {Xu},
  \citenamefont {Wang}, \citenamefont {Yuan},\ and\ \citenamefont
  {Wang}}]{Xu2021}%
  \BibitemOpen
  \bibfield  {author} {\bibinfo {author} {\bibfnamefont {H.}~\bibnamefont
  {Xu}}, \bibinfo {author} {\bibfnamefont {L.}~\bibnamefont {Wang}}, \bibinfo
  {author} {\bibfnamefont {H.}~\bibnamefont {Yuan}},\ \bibnamefont {and}\
  \bibinfo {author} {\bibfnamefont {X.}~\bibnamefont {Wang}},\ }\bibfield
  {title} {\emph {\enquote {\bibinfo {title} {Generalizable control for
  multiparameter quantum metrology},}\ }}\href
  {https://doi.org/10.1103/physreva.103.042615} {\bibfield  {journal} {\bibinfo
   {journal} {Phys. Rev. A}\ }\textbf {\bibinfo {volume} {103}},\ \bibinfo
  {pages} {042615} (\bibinfo {year} {2021})}\BibitemShut {NoStop}%
\bibitem [{\citenamefont {Xiao}\ \emph {et~al.}(2022)\citenamefont {Xiao},
  \citenamefont {Fan},\ and\ \citenamefont {Zeng}}]{Xiao2022}%
  \BibitemOpen
  \bibfield  {author} {\bibinfo {author} {\bibfnamefont {T.}~\bibnamefont
  {Xiao}}, \bibinfo {author} {\bibfnamefont {J.}~\bibnamefont {Fan}},\
  \bibnamefont {and}\ \bibinfo {author} {\bibfnamefont {G.}~\bibnamefont
  {Zeng}},\ }\bibfield  {title} {\emph {\enquote {\bibinfo {title} {Parameter
  estimation in quantum sensing based on deep reinforcement learning},}\
  }}\href {https://doi.org/10.1038/s41534-021-00513-z} {\bibfield  {journal}
  {\bibinfo  {journal} {npj Quantum Inf.}\ }\textbf {\bibinfo {volume} {8}},\
  \bibinfo {pages} {2} (\bibinfo {year} {2022})}\BibitemShut {NoStop}%
\bibitem [{\citenamefont {Fiderer}\ \emph {et~al.}(2021)\citenamefont
  {Fiderer}, \citenamefont {Schuff},\ and\ \citenamefont
  {Braun}}]{Fiderer2021}%
  \BibitemOpen
  \bibfield  {author} {\bibinfo {author} {\bibfnamefont {L.~J.}\ \bibnamefont
  {Fiderer}}, \bibinfo {author} {\bibfnamefont {J.}~\bibnamefont {Schuff}},\
  \bibnamefont {and}\ \bibinfo {author} {\bibfnamefont {D.}~\bibnamefont
  {Braun}},\ }\bibfield  {title} {\emph {\enquote {\bibinfo {title}
  {Neural-network heuristics for adaptive {Bayesian} quantum estimation},}\
  }}\href {https://doi.org/10.1103/prxquantum.2.020303} {\bibfield  {journal}
  {\bibinfo  {journal} {PRX Quantum}\ }\textbf {\bibinfo {volume} {2}},\
  \bibinfo {pages} {020303} (\bibinfo {year} {2021})}\BibitemShut {NoStop}%
\bibitem [{\citenamefont {Lumino}\ \emph {et~al.}(2018)\citenamefont {Lumino},
  \citenamefont {Polino}, \citenamefont {Rab}, \citenamefont {Milani},
  \citenamefont {Spagnolo}, \citenamefont {Wiebe},\ and\ \citenamefont
  {Sciarrino}}]{Lumino2018}%
  \BibitemOpen
  \bibfield  {author} {\bibinfo {author} {\bibfnamefont {A.}~\bibnamefont
  {Lumino}}, \bibinfo {author} {\bibfnamefont {E.}~\bibnamefont {Polino}},
  \bibinfo {author} {\bibfnamefont {A.~S.}\ \bibnamefont {Rab}}, \bibinfo
  {author} {\bibfnamefont {G.}~\bibnamefont {Milani}}, \bibinfo {author}
  {\bibfnamefont {N.}~\bibnamefont {Spagnolo}}, \bibinfo {author}
  {\bibfnamefont {N.}~\bibnamefont {Wiebe}},\ \bibnamefont {and}\ \bibinfo
  {author} {\bibfnamefont {F.}~\bibnamefont {Sciarrino}},\ }\bibfield  {title}
  {\emph {\enquote {\bibinfo {title} {Experimental phase estimation enhanced by
  machine learning},}\ }}\href
  {https://doi.org/10.1103/physrevapplied.10.044033} {\bibfield  {journal}
  {\bibinfo  {journal} {Phys. Rev. Applied}\ }\textbf {\bibinfo {volume}
  {10}},\ \bibinfo {pages} {044033} (\bibinfo {year} {2018})}\BibitemShut
  {NoStop}%
\bibitem [{\citenamefont {Cimini}\ \emph {et~al.}(2019)\citenamefont {Cimini},
  \citenamefont {Gianani}, \citenamefont {Spagnolo}, \citenamefont {Leccese},
  \citenamefont {Sciarrino},\ and\ \citenamefont {Barbieri}}]{Cimini2019}%
  \BibitemOpen
  \bibfield  {author} {\bibinfo {author} {\bibfnamefont {V.}~\bibnamefont
  {Cimini}}, \bibinfo {author} {\bibfnamefont {I.}~\bibnamefont {Gianani}},
  \bibinfo {author} {\bibfnamefont {N.}~\bibnamefont {Spagnolo}}, \bibinfo
  {author} {\bibfnamefont {F.}~\bibnamefont {Leccese}}, \bibinfo {author}
  {\bibfnamefont {F.}~\bibnamefont {Sciarrino}},\ \bibnamefont {and}\ \bibinfo
  {author} {\bibfnamefont {M.}~\bibnamefont {Barbieri}},\ }\bibfield  {title}
  {\emph {\enquote {\bibinfo {title} {Calibration of quantum sensors by neural
  networks},}\ }}\href {https://doi.org/10.1103/physrevlett.123.230502}
  {\bibfield  {journal} {\bibinfo  {journal} {Phys. Rev. Lett.}\ }\textbf
  {\bibinfo {volume} {123}},\ \bibinfo {pages} {230502} (\bibinfo {year}
  {2019})}\BibitemShut {NoStop}%
\bibitem [{\citenamefont {Cimini}\ \emph {et~al.}(2023)\citenamefont {Cimini},
  \citenamefont {Valeri}, \citenamefont {Polino}, \citenamefont {Piacentini},
  \citenamefont {Ceccarelli}, \citenamefont {Corrielli}, \citenamefont
  {Spagnolo}, \citenamefont {Osellame},\ and\ \citenamefont
  {Sciarrino}}]{Cimini2023}%
  \BibitemOpen
  \bibfield  {author} {\bibinfo {author} {\bibfnamefont {V.}~\bibnamefont
  {Cimini}}, \bibinfo {author} {\bibfnamefont {M.}~\bibnamefont {Valeri}},
  \bibinfo {author} {\bibfnamefont {E.}~\bibnamefont {Polino}}, \bibinfo
  {author} {\bibfnamefont {S.}~\bibnamefont {Piacentini}}, \bibinfo {author}
  {\bibfnamefont {F.}~\bibnamefont {Ceccarelli}}, \bibinfo {author}
  {\bibfnamefont {G.}~\bibnamefont {Corrielli}}, \bibinfo {author}
  {\bibfnamefont {N.}~\bibnamefont {Spagnolo}}, \bibinfo {author}
  {\bibfnamefont {R.}~\bibnamefont {Osellame}},\ \bibnamefont {and}\ \bibinfo
  {author} {\bibfnamefont {F.}~\bibnamefont {Sciarrino}},\ }\bibfield  {title}
  {\emph {\enquote {\bibinfo {title} {Deep reinforcement learning for quantum
  multiparameter estimation},}\ }}\href
  {https://doi.org/10.1117/1.ap.5.1.016005} {\bibfield  {journal} {\bibinfo
  {journal} {Adv. Photonics}\ }\textbf {\bibinfo {volume} {5}},\ \bibinfo
  {pages} {016005} (\bibinfo {year} {2023})}\BibitemShut {NoStop}%
\bibitem [{\citenamefont {Chang}\ \emph {et~al.}(2014)\citenamefont {Chang},
  \citenamefont {Vuleti{\'c}},\ and\ \citenamefont {Lukin}}]{Chang2014}%
  \BibitemOpen
  \bibfield  {author} {\bibinfo {author} {\bibfnamefont {D.~E.}\ \bibnamefont
  {Chang}}, \bibinfo {author} {\bibfnamefont {V.}~\bibnamefont {Vuleti{\'c}}},\
  \bibnamefont {and}\ \bibinfo {author} {\bibfnamefont {M.~D.}\ \bibnamefont
  {Lukin}},\ }\bibfield  {title} {\emph {\enquote {\bibinfo {title} {Quantum
  nonlinear optics---Photon by photon},}\ }}\href
  {https://doi.org/10.1038/nphoton.2014.192} {\bibfield  {journal} {\bibinfo
  {journal} {Nat. Photonics}\ }\textbf {\bibinfo {volume} {8}},\ \bibinfo
  {pages} {685} (\bibinfo {year} {2014})}\BibitemShut {NoStop}%
\bibitem [{\citenamefont {Aspelmeyer}\ \emph {et~al.}(2014)\citenamefont
  {Aspelmeyer}, \citenamefont {Kippenberg},\ and\ \citenamefont
  {Marquardt}}]{Aspelmeyer2014}%
  \BibitemOpen
  \bibfield  {author} {\bibinfo {author} {\bibfnamefont {M.}~\bibnamefont
  {Aspelmeyer}}, \bibinfo {author} {\bibfnamefont {T.~J.}\ \bibnamefont
  {Kippenberg}},\ \bibnamefont {and}\ \bibinfo {author} {\bibfnamefont
  {F.}~\bibnamefont {Marquardt}},\ }\bibfield  {title} {\emph {\enquote
  {\bibinfo {title} {Cavity optomechanics},}\ }}\href
  {https://doi.org/10.1103/revmodphys.86.1391} {\bibfield  {journal} {\bibinfo
  {journal} {Rev. Mod. Phys.}\ }\textbf {\bibinfo {volume} {86}},\ \bibinfo
  {pages} {1391} (\bibinfo {year} {2014})}\BibitemShut {NoStop}%
\bibitem [{\citenamefont {Paul}(1982)}]{Paul1982}%
  \BibitemOpen
  \bibfield  {author} {\bibinfo {author} {\bibfnamefont {H.}~\bibnamefont
  {Paul}},\ }\bibfield  {title} {\emph {\enquote {\bibinfo {title} {Photon
  antibunching},}\ }}\href {https://doi.org/10.1103/revmodphys.54.1061}
  {\bibfield  {journal} {\bibinfo  {journal} {Rev. Mod. Phys.}\ }\textbf
  {\bibinfo {volume} {54}},\ \bibinfo {pages} {1061} (\bibinfo {year}
  {1982})}\BibitemShut {NoStop}%
\bibitem [{\citenamefont {Lee}(1993)}]{Lee1993}%
  \BibitemOpen
  \bibfield  {author} {\bibinfo {author} {\bibfnamefont {C.~T.}\ \bibnamefont
  {Lee}},\ }\bibfield  {title} {\emph {\enquote {\bibinfo {title} {External
  photodetection of cavity radiation},}\ }}\href
  {https://doi.org/10.1103/physreva.48.2285} {\bibfield  {journal} {\bibinfo
  {journal} {Phys. Rev. A}\ }\textbf {\bibinfo {volume} {48}},\ \bibinfo
  {pages} {2285} (\bibinfo {year} {1993})}\BibitemShut {NoStop}%
\bibitem [{\citenamefont {Chru{\'s}ci{\'n}ski}\ and\ \citenamefont
  {Pascazio}(2017)}]{Chruscinski2017}%
  \BibitemOpen
  \bibfield  {author} {\bibinfo {author} {\bibfnamefont {D.}~\bibnamefont
  {Chru{\'s}ci{\'n}ski}}\ \bibnamefont {and}\ \bibinfo {author} {\bibfnamefont
  {S.}~\bibnamefont {Pascazio}},\ }\bibfield  {title} {\emph {\enquote
  {\bibinfo {title} {A brief history of the {GKLS} equation},}\ }}\href
  {https://doi.org/10.1142/s1230161217400017} {\bibfield  {journal} {\bibinfo
  {journal} {Open Syst. Inf. Dyn.}\ }\textbf {\bibinfo {volume} {24}},\
  \bibinfo {pages} {1740001} (\bibinfo {year} {2017})}\BibitemShut {NoStop}%
\bibitem [{Note1()}]{Note1}%
  \BibitemOpen
  \bibinfo {note} {Alternatively, one may consider Eq.~(\ref {eq:true_evol}) as
  a starting point to describe system dynamics.}\BibitemShut {Stop}%
\bibitem [{\citenamefont {Carmichael}(2008)}]{Carmichael2008}%
  \BibitemOpen
  \bibfield  {author} {\bibinfo {author} {\bibfnamefont {H.~J.}\ \bibnamefont
  {Carmichael}},\ }\href {https://doi.org/10.1007/978-3-540-71320-3} {\emph
  {\bibinfo {title} {Statistical methods in quantum optics 2}}},\ Theoretical
  and Mathematical Physics\ (\bibinfo  {publisher} {Springer},\ \bibinfo
  {address} {Berlin, Heidelberg},\ \bibinfo {year} {2008})\BibitemShut
  {NoStop}%
\bibitem [{\citenamefont {Gardiner}(1985)}]{Gardiner1985}%
  \BibitemOpen
  \bibfield  {author} {\bibinfo {author} {\bibfnamefont {C.~W.}\ \bibnamefont
  {Gardiner}},\ }\href {https://doi.org/10.1007/978-3-662-02452-2} {\emph
  {\bibinfo {title} {Handbook of stochastic methods for physics, chemistry and
  the natural sciences}}},\ Springer Series in Synergetics\ (\bibinfo
  {publisher} {Springer},\ \bibinfo {address} {Berlin, Heidelberg},\ \bibinfo
  {year} {1985})\BibitemShut {NoStop}%
\bibitem [{\citenamefont {Dalibard}\ \emph {et~al.}(1992)\citenamefont
  {Dalibard}, \citenamefont {Castin},\ and\ \citenamefont
  {M{\o}lmer}}]{Dalibard1992}%
  \BibitemOpen
  \bibfield  {author} {\bibinfo {author} {\bibfnamefont {J.}~\bibnamefont
  {Dalibard}}, \bibinfo {author} {\bibfnamefont {Y.}~\bibnamefont {Castin}},\
  \bibnamefont {and}\ \bibinfo {author} {\bibfnamefont {K.}~\bibnamefont
  {M{\o}lmer}},\ }\bibfield  {title} {\emph {\enquote {\bibinfo {title}
  {Wave-function approach to dissipative processes in quantum optics},}\
  }}\href {https://doi.org/10.1103/physrevlett.68.580} {\bibfield  {journal}
  {\bibinfo  {journal} {Phys. Rev. Lett.}\ }\textbf {\bibinfo {volume} {68}},\
  \bibinfo {pages} {580} (\bibinfo {year} {1992})}\BibitemShut {NoStop}%
\bibitem [{\citenamefont {M{\o}lmer}\ \emph {et~al.}(1993)\citenamefont
  {M{\o}lmer}, \citenamefont {Castin},\ and\ \citenamefont
  {Dalibard}}]{Moelmer1993}%
  \BibitemOpen
  \bibfield  {author} {\bibinfo {author} {\bibfnamefont {K.}~\bibnamefont
  {M{\o}lmer}}, \bibinfo {author} {\bibfnamefont {Y.}~\bibnamefont {Castin}},\
  \bibnamefont {and}\ \bibinfo {author} {\bibfnamefont {J.}~\bibnamefont
  {Dalibard}},\ }\bibfield  {title} {\emph {\enquote {\bibinfo {title} {{Monte
  Carlo} wave-function method in quantum optics},}\ }}\href
  {https://doi.org/10.1364/josab.10.000524} {\bibfield  {journal} {\bibinfo
  {journal} {J. Opt. Soc. Am. B}\ }\textbf {\bibinfo {volume} {10}},\ \bibinfo
  {pages} {524} (\bibinfo {year} {1993})}\BibitemShut {NoStop}%
\bibitem [{\citenamefont {Hegerfeldt}(1993)}]{Hegerfeldt1993}%
  \BibitemOpen
  \bibfield  {author} {\bibinfo {author} {\bibfnamefont {G.~C.}\ \bibnamefont
  {Hegerfeldt}},\ }\bibfield  {title} {\emph {\enquote {\bibinfo {title} {How
  to reset an atom after a photon detection: Applications to photon-counting
  processes},}\ }}\href {https://doi.org/10.1103/physreva.47.449} {\bibfield
  {journal} {\bibinfo  {journal} {Phys. Rev. A}\ }\textbf {\bibinfo {volume}
  {47}},\ \bibinfo {pages} {449} (\bibinfo {year} {1993})}\BibitemShut
  {NoStop}%
\bibitem [{\citenamefont {Daley}\ and\ \citenamefont
  {Vere-Jones}(2003)}]{Daley2005}%
  \BibitemOpen
  \bibfield  {author} {\bibinfo {author} {\bibfnamefont {D.~J.}\ \bibnamefont
  {Daley}}\ \bibnamefont {and}\ \bibinfo {author} {\bibfnamefont
  {D.}~\bibnamefont {Vere-Jones}},\ }\href@noop {} {\emph {\bibinfo {title} {An
  introduction to the theory of point processes. {Volume I}: Elementary theory
  and methods}}},\ \bibinfo {edition} {2nd}\ ed.\ (\bibinfo  {publisher}
  {Springer},\ \bibinfo {address} {New York},\ \bibinfo {year}
  {2003})\BibitemShut {NoStop}%
\bibitem [{\citenamefont {Carmichael}\ \emph {et~al.}(1989)\citenamefont
  {Carmichael}, \citenamefont {Singh}, \citenamefont {Vyas},\ and\
  \citenamefont {Rice}}]{Carmichael1989}%
  \BibitemOpen
  \bibfield  {author} {\bibinfo {author} {\bibfnamefont {H.~J.}\ \bibnamefont
  {Carmichael}}, \bibinfo {author} {\bibfnamefont {S.}~\bibnamefont {Singh}},
  \bibinfo {author} {\bibfnamefont {R.}~\bibnamefont {Vyas}},\ \bibnamefont
  {and}\ \bibinfo {author} {\bibfnamefont {P.~R.}\ \bibnamefont {Rice}},\
  }\bibfield  {title} {\emph {\enquote {\bibinfo {title} {Photoelectron waiting
  times and atomic state reduction in resonance fluorescence},}\ }}\href
  {https://doi.org/10.1103/physreva.39.1200} {\bibfield  {journal} {\bibinfo
  {journal} {Phys. Rev. A}\ }\textbf {\bibinfo {volume} {39}},\ \bibinfo
  {pages} {1200} (\bibinfo {year} {1989})}\BibitemShut {NoStop}%
\bibitem [{\citenamefont {Landi}\ \emph {et~al.}(2024)\citenamefont {Landi},
  \citenamefont {Kewming}, \citenamefont {Mitchison},\ and\ \citenamefont
  {Potts}}]{Landi2024}%
  \BibitemOpen
  \bibfield  {author} {\bibinfo {author} {\bibfnamefont {G.~T.}\ \bibnamefont
  {Landi}}, \bibinfo {author} {\bibfnamefont {M.~J.}\ \bibnamefont {Kewming}},
  \bibinfo {author} {\bibfnamefont {M.~T.}\ \bibnamefont {Mitchison}},\
  \bibnamefont {and}\ \bibinfo {author} {\bibfnamefont {P.~P.}\ \bibnamefont
  {Potts}},\ }\bibfield  {title} {\emph {\enquote {\bibinfo {title} {Current
  fluctuations in open quantum systems: Bridging the gap between quantum
  continuous measurements and full counting statistics},}\ }}\href
  {https://doi.org/10.1103/prxquantum.5.020201} {\bibfield  {journal} {\bibinfo
   {journal} {PRX Quantum}\ }\textbf {\bibinfo {volume} {5}},\ \bibinfo {pages}
  {020201} (\bibinfo {year} {2024})}\BibitemShut {NoStop}%
\bibitem [{\citenamefont {Burgarth}\ \emph {et~al.}(2015)\citenamefont
  {Burgarth}, \citenamefont {Giovannetti}, \citenamefont {Kato},\ and\
  \citenamefont {Yuasa}}]{Burgarth_2015}%
  \BibitemOpen
  \bibfield  {author} {\bibinfo {author} {\bibfnamefont {D.}~\bibnamefont
  {Burgarth}}, \bibinfo {author} {\bibfnamefont {V.}~\bibnamefont
  {Giovannetti}}, \bibinfo {author} {\bibfnamefont {A.~N.}\ \bibnamefont
  {Kato}},\ \bibnamefont {and}\ \bibinfo {author} {\bibfnamefont
  {K.}~\bibnamefont {Yuasa}},\ }\bibfield  {title} {\emph {\enquote {\bibinfo
  {title} {Quantum estimation via sequential measurements},}\ }}\href
  {https://doi.org/10.1088/1367-2630/17/11/113055} {\bibfield  {journal}
  {\bibinfo  {journal} {New J. Phys.}\ }\textbf {\bibinfo {volume} {17}},\
  \bibinfo {pages} {113055} (\bibinfo {year} {2015})}\BibitemShut {NoStop}%
\bibitem [{\citenamefont {Kiilerich}\ and\ \citenamefont
  {M{\o}lmer}(2016)}]{Kiilerich2016}%
  \BibitemOpen
  \bibfield  {author} {\bibinfo {author} {\bibfnamefont {A.~H.}\ \bibnamefont
  {Kiilerich}}\ \bibnamefont {and}\ \bibinfo {author} {\bibfnamefont
  {K.}~\bibnamefont {M{\o}lmer}},\ }\bibfield  {title} {\emph {\enquote
  {\bibinfo {title} {{Bayesian} parameter estimation by continuous homodyne
  detection},}\ }}\href {https://doi.org/10.1103/physreva.94.032103} {\bibfield
   {journal} {\bibinfo  {journal} {Phys. Rev. A}\ }\textbf {\bibinfo {volume}
  {94}},\ \bibinfo {pages} {032103} (\bibinfo {year} {2016})}\BibitemShut
  {NoStop}%
\bibitem [{\citenamefont {Clark}\ \emph {et~al.}(2019)\citenamefont {Clark},
  \citenamefont {Stokes},\ and\ \citenamefont {Beige}}]{Clark2019}%
  \BibitemOpen
  \bibfield  {author} {\bibinfo {author} {\bibfnamefont {L.~A.}\ \bibnamefont
  {Clark}}, \bibinfo {author} {\bibfnamefont {A.}~\bibnamefont {Stokes}},\
  \bibnamefont {and}\ \bibinfo {author} {\bibfnamefont {A.}~\bibnamefont
  {Beige}},\ }\bibfield  {title} {\emph {\enquote {\bibinfo {title} {Quantum
  jump metrology},}\ }}\href {https://doi.org/10.1103/physreva.99.022102}
  {\bibfield  {journal} {\bibinfo  {journal} {Phys. Rev. A}\ }\textbf {\bibinfo
  {volume} {99}},\ \bibinfo {pages} {022102} (\bibinfo {year}
  {2019})}\BibitemShut {NoStop}%
\bibitem [{\citenamefont {Clark}\ \emph {et~al.}(2022)\citenamefont {Clark},
  \citenamefont {Markowicz},\ and\ \citenamefont
  {Ko{\l}ody{\'n}ski}}]{Clark2022}%
  \BibitemOpen
  \bibfield  {author} {\bibinfo {author} {\bibfnamefont {L.~A.}\ \bibnamefont
  {Clark}}, \bibinfo {author} {\bibfnamefont {B.}~\bibnamefont {Markowicz}},\
  \bibnamefont {and}\ \bibinfo {author} {\bibfnamefont {J.}~\bibnamefont
  {Ko{\l}ody{\'n}ski}},\ }\bibfield  {title} {\emph {\enquote {\bibinfo {title}
  {Exploiting non-linear effects in optomechanical sensors with continuous
  photon-counting},}\ }}\href {https://doi.org/10.22331/q-2022-09-20-812}
  {\bibfield  {journal} {\bibinfo  {journal} {Quantum}\ }\textbf {\bibinfo
  {volume} {6}},\ \bibinfo {pages} {812} (\bibinfo {year} {2022})}\BibitemShut
  {NoStop}%
\bibitem [{\citenamefont {Mandel}\ and\ \citenamefont
  {Wolf}(1995)}]{Mandel1995}%
  \BibitemOpen
  \bibfield  {author} {\bibinfo {author} {\bibfnamefont {L.}~\bibnamefont
  {Mandel}}\ \bibnamefont {and}\ \bibinfo {author} {\bibfnamefont
  {E.}~\bibnamefont {Wolf}},\ }\href {https://doi.org/10.1017/cbo9781139644105}
  {\emph {\bibinfo {title} {Optical coherence and quantum optics}}}\ (\bibinfo
  {publisher} {Cambridge University Press},\ \bibinfo {address} {Cambridge},\
  \bibinfo {year} {1995})\BibitemShut {NoStop}%
\bibitem [{\citenamefont {Teich}\ \emph {et~al.}(1984)\citenamefont {Teich},
  \citenamefont {Saleh},\ and\ \citenamefont {Pe{\v{r}}ina}}]{Teich1984}%
  \BibitemOpen
  \bibfield  {author} {\bibinfo {author} {\bibfnamefont {M.~C.}\ \bibnamefont
  {Teich}}, \bibinfo {author} {\bibfnamefont {B.~E.~A.}\ \bibnamefont
  {Saleh}},\ \bibnamefont {and}\ \bibinfo {author} {\bibfnamefont
  {J.}~\bibnamefont {Pe{\v{r}}ina}},\ }\bibfield  {title} {\emph {\enquote
  {\bibinfo {title} {Role of primary excitation statistics in the generation of
  antibunched and sub-{Poisson} light},}\ }}\href
  {https://doi.org/10.1364/josab.1.000366} {\bibfield  {journal} {\bibinfo
  {journal} {J. Opt. Soc. Am. B}\ }\textbf {\bibinfo {volume} {1}},\ \bibinfo
  {pages} {366} (\bibinfo {year} {1984})}\BibitemShut {NoStop}%
\bibitem [{\citenamefont {Cao}(2006)}]{Cao2006}%
  \BibitemOpen
  \bibfield  {author} {\bibinfo {author} {\bibfnamefont {J.}~\bibnamefont
  {Cao}},\ }\bibfield  {title} {\emph {\enquote {\bibinfo {title} {Correlations
  in single molecule photon statistics: Renewal indicator},}\ }}\href
  {https://doi.org/10.1021/jp061302b} {\bibfield  {journal} {\bibinfo
  {journal} {J. Phys. Chem. B}\ }\textbf {\bibinfo {volume} {110}},\ \bibinfo
  {pages} {19040} (\bibinfo {year} {2006})}\BibitemShut {NoStop}%
\bibitem [{\citenamefont {Kay}(1993)}]{Kay1993}%
  \BibitemOpen
  \bibfield  {author} {\bibinfo {author} {\bibfnamefont {S.~M.}\ \bibnamefont
  {Kay}},\ }\href@noop {} {\emph {\bibinfo {title} {Fundamentals of statistical
  signal processing: Estimation theory}}}\ (\bibinfo  {publisher} {Prentice
  Hall},\ \bibinfo {year} {1993})\BibitemShut {NoStop}%
\bibitem [{\citenamefont {van~der Vaart}(1998)}]{Vaart1998}%
  \BibitemOpen
  \bibfield  {author} {\bibinfo {author} {\bibfnamefont {A.~W.}\ \bibnamefont
  {van~der Vaart}},\ }\href@noop {} {\emph {\bibinfo {title} {Asymptotic
  statistics}}}\ (\bibinfo  {publisher} {Cambridge University Press},\ \bibinfo
  {address} {Cambridge},\ \bibinfo {year} {1998})\BibitemShut {NoStop}%
\bibitem [{\citenamefont {Eldar}(2008)}]{Eldar2008}%
  \BibitemOpen
  \bibfield  {author} {\bibinfo {author} {\bibfnamefont {Y.~C.}\ \bibnamefont
  {Eldar}},\ }\bibfield  {title} {\emph {\enquote {\bibinfo {title} {Rethinking
  biased estimation: Improving maximum likelihood and the {Cram{\'e}r-Rao}
  bound},}\ }}\href {https://doi.org/10.1561/2000000008} {\bibfield  {journal}
  {\bibinfo  {journal} {Found. Trends Signal Process.}\ }\textbf {\bibinfo
  {volume} {1}},\ \bibinfo {pages} {305} (\bibinfo {year} {2008})}\BibitemShut
  {NoStop}%
\bibitem [{\citenamefont {Gammelmark}\ and\ \citenamefont
  {M{\o}lmer}(2013)}]{Gammelmark2013}%
  \BibitemOpen
  \bibfield  {author} {\bibinfo {author} {\bibfnamefont {S.}~\bibnamefont
  {Gammelmark}}\ \bibnamefont {and}\ \bibinfo {author} {\bibfnamefont
  {K.}~\bibnamefont {M{\o}lmer}},\ }\bibfield  {title} {\emph {\enquote
  {\bibinfo {title} {{Bayesian} parameter inference from continuously monitored
  quantum systems},}\ }}\href {https://doi.org/10.1103/physreva.87.032115}
  {\bibfield  {journal} {\bibinfo  {journal} {Phys. Rev. A}\ }\textbf {\bibinfo
  {volume} {87}},\ \bibinfo {pages} {032115} (\bibinfo {year}
  {2013})}\BibitemShut {NoStop}%
\bibitem [{\citenamefont {Albarelli}\ \emph {et~al.}(2018)\citenamefont
  {Albarelli}, \citenamefont {Rossi}, \citenamefont {Tamascelli},\ and\
  \citenamefont {Genoni}}]{Albarelli2018}%
  \BibitemOpen
  \bibfield  {author} {\bibinfo {author} {\bibfnamefont {F.}~\bibnamefont
  {Albarelli}}, \bibinfo {author} {\bibfnamefont {M.~A.~C.}\ \bibnamefont
  {Rossi}}, \bibinfo {author} {\bibfnamefont {D.}~\bibnamefont {Tamascelli}},\
  \bibnamefont {and}\ \bibinfo {author} {\bibfnamefont {M.~G.}\ \bibnamefont
  {Genoni}},\ }\bibfield  {title} {\emph {\enquote {\bibinfo {title} {Restoring
  {Heisenberg} scaling in noisy quantum metrology by monitoring the
  environment},}\ }}\href {https://doi.org/10.22331/q-2018-12-03-110}
  {\bibfield  {journal} {\bibinfo  {journal} {Quantum}\ }\textbf {\bibinfo
  {volume} {2}},\ \bibinfo {pages} {110} (\bibinfo {year} {2018})}\BibitemShut
  {NoStop}%
\bibitem [{\citenamefont {Lawless}(2003)}]{Lawless2003}%
  \BibitemOpen
  \bibfield  {author} {\bibinfo {author} {\bibfnamefont {J.~F.}\ \bibnamefont
  {Lawless}},\ }\href@noop {} {\emph {\bibinfo {title} {Statistical models and
  methods for lifetime data}}},\ \bibinfo {edition} {2nd}\ ed.,\ Wiley Series
  in Probability and Statistics\ (\bibinfo  {publisher} {Wiley-Interscience},\
  \bibinfo {address} {Hoboken, NJ},\ \bibinfo {year} {2003})\BibitemShut
  {NoStop}%
\bibitem [{\citenamefont {Dawid}\ \emph {et~al.}(2025)\citenamefont {Dawid},
  \citenamefont {Arnold}, \citenamefont {Requena}, \citenamefont {Gresch},
  \citenamefont {P{\l}odzie{\'n}}, \citenamefont {Donatella}, \citenamefont
  {Nicoli}, \citenamefont {Stornati}, \citenamefont {Koch}, \citenamefont
  {B{\"u}ttner}, \citenamefont {Oku{\l}a}, \citenamefont {Mu{\~n}oz-Gil},
  \citenamefont {Vargas-Hern{\'a}ndez}, \citenamefont {Cervera-Lierta},
  \citenamefont {Carrasquilla}, \citenamefont {Dunjko}, \citenamefont
  {Gabri{\'e}}, \citenamefont {Huembeli}, \citenamefont {van Nieuwenburg},
  \citenamefont {Vicentini}, \citenamefont {Wang}, \citenamefont {Wetzel},
  \citenamefont {Carleo}, \citenamefont {Greplov{\'a}}, \citenamefont {Krems},
  \citenamefont {Marquardt}, \citenamefont {Tomza}, \citenamefont
  {Lewenstein},\ and\ \citenamefont {Dauphin}}]{Dawid2025}%
  \BibitemOpen
  \bibfield  {author} {\bibinfo {author} {\bibfnamefont {A.}~\bibnamefont
  {Dawid}}, \bibnamefont {et~al.},\ }\href
  {https://doi.org/10.1017/9781009504942} {\emph {\bibinfo {title} {Machine
  learning in quantum sciences}}}\ (\bibinfo  {publisher} {Cambridge University
  Press},\ \bibinfo {address} {Cambridge},\ \bibinfo {year} {2025})\BibitemShut
  {NoStop}%
\bibitem [{\citenamefont {Palmieri}\ \emph {et~al.}(2024)\citenamefont
  {Palmieri}, \citenamefont {M{\"u}ller-Rigat}, \citenamefont {Srivastava},
  \citenamefont {Lewenstein}, \citenamefont {Rajchel-Mieldzio{\'c}},\ and\
  \citenamefont {P{\l}odzie{\'n}}}]{Palmieri2024}%
  \BibitemOpen
  \bibfield  {author} {\bibinfo {author} {\bibfnamefont {A.~M.}\ \bibnamefont
  {Palmieri}}, \bibinfo {author} {\bibfnamefont {G.}~\bibnamefont
  {M{\"u}ller-Rigat}}, \bibinfo {author} {\bibfnamefont {A.~K.}\ \bibnamefont
  {Srivastava}}, \bibinfo {author} {\bibfnamefont {M.}~\bibnamefont
  {Lewenstein}}, \bibinfo {author} {\bibfnamefont {G.}~\bibnamefont
  {Rajchel-Mieldzio{\'c}}},\ \bibnamefont {and}\ \bibinfo {author}
  {\bibfnamefont {M.}~\bibnamefont {P{\l}odzie{\'n}}},\ }\bibfield  {title}
  {\emph {\enquote {\bibinfo {title} {Enhancing quantum state tomography via
  resource-efficient attention-based neural networks},}\ }}\href
  {https://doi.org/10.1103/physrevresearch.6.033248} {\bibfield  {journal}
  {\bibinfo  {journal} {Phys. Rev. Research}\ }\textbf {\bibinfo {volume}
  {6}},\ \bibinfo {pages} {033248} (\bibinfo {year} {2024})}\BibitemShut
  {NoStop}%
\bibitem [{\citenamefont {Huembeli}\ \emph {et~al.}(2018)\citenamefont
  {Huembeli}, \citenamefont {Dauphin},\ and\ \citenamefont
  {Wittek}}]{Huembeli2018}%
  \BibitemOpen
  \bibfield  {author} {\bibinfo {author} {\bibfnamefont {P.}~\bibnamefont
  {Huembeli}}, \bibinfo {author} {\bibfnamefont {A.}~\bibnamefont {Dauphin}},\
  \bibnamefont {and}\ \bibinfo {author} {\bibfnamefont {P.}~\bibnamefont
  {Wittek}},\ }\bibfield  {title} {\emph {\enquote {\bibinfo {title}
  {Identifying quantum phase transitions with adversarial neural networks},}\
  }}\href {https://doi.org/10.1103/physrevb.97.134109} {\bibfield  {journal}
  {\bibinfo  {journal} {Phys. Rev. B}\ }\textbf {\bibinfo {volume} {97}},\
  \bibinfo {pages} {134109} (\bibinfo {year} {2018})}\BibitemShut {NoStop}%
\bibitem [{\citenamefont {Kottmann}\ \emph {et~al.}(2020)\citenamefont
  {Kottmann}, \citenamefont {Huembeli}, \citenamefont {Lewenstein},\ and\
  \citenamefont {Ac{\'\i}n}}]{Kottmann2020}%
  \BibitemOpen
  \bibfield  {author} {\bibinfo {author} {\bibfnamefont {K.}~\bibnamefont
  {Kottmann}}, \bibinfo {author} {\bibfnamefont {P.}~\bibnamefont {Huembeli}},
  \bibinfo {author} {\bibfnamefont {M.}~\bibnamefont {Lewenstein}},\
  \bibnamefont {and}\ \bibinfo {author} {\bibfnamefont {A.}~\bibnamefont
  {Ac{\'\i}n}},\ }\bibfield  {title} {\emph {\enquote {\bibinfo {title}
  {Unsupervised phase discovery with deep anomaly detection},}\ }}\href
  {https://doi.org/10.1103/physrevlett.125.170603} {\bibfield  {journal}
  {\bibinfo  {journal} {Phys. Rev. Lett.}\ }\textbf {\bibinfo {volume} {125}},\
  \bibinfo {pages} {170603} (\bibinfo {year} {2020})}\BibitemShut {NoStop}%
\bibitem [{\citenamefont {Duan}\ \emph {et~al.}(2025)\citenamefont {Duan},
  \citenamefont {Hu}, \citenamefont {Lu}, \citenamefont {Xiao}, \citenamefont
  {Jia}, \citenamefont {M{\o}lmer},\ and\ \citenamefont {Xiao}}]{Duan2025}%
  \BibitemOpen
  \bibfield  {author} {\bibinfo {author} {\bibfnamefont {J.}~\bibnamefont
  {Duan}}, \bibinfo {author} {\bibfnamefont {Z.}~\bibnamefont {Hu}}, \bibinfo
  {author} {\bibfnamefont {X.}~\bibnamefont {Lu}}, \bibinfo {author}
  {\bibfnamefont {L.}~\bibnamefont {Xiao}}, \bibinfo {author} {\bibfnamefont
  {S.}~\bibnamefont {Jia}}, \bibinfo {author} {\bibfnamefont {K.}~\bibnamefont
  {M{\o}lmer}},\ \bibnamefont {and}\ \bibinfo {author} {\bibfnamefont
  {Y.}~\bibnamefont {Xiao}},\ }\bibfield  {title} {\emph {\enquote {\bibinfo
  {title} {Concurrent spin squeezing and field tracking with machine
  learning},}\ }}\href {https://doi.org/10.1038/s41567-025-02855-3} {\bibfield
  {journal} {\bibinfo  {journal} {Nat. Phys.}\ }\textbf {\bibinfo {volume}
  {21}},\ \bibinfo {pages} {909} (\bibinfo {year} {2025})}\BibitemShut
  {NoStop}%
\bibitem [{\citenamefont {Fallani}\ \emph {et~al.}(2022)\citenamefont
  {Fallani}, \citenamefont {Rossi}, \citenamefont {Tamascelli},\ and\
  \citenamefont {Genoni}}]{Fallani2022}%
  \BibitemOpen
  \bibfield  {author} {\bibinfo {author} {\bibfnamefont {A.}~\bibnamefont
  {Fallani}}, \bibinfo {author} {\bibfnamefont {M.~A.~C.}\ \bibnamefont
  {Rossi}}, \bibinfo {author} {\bibfnamefont {D.}~\bibnamefont {Tamascelli}},\
  \bibnamefont {and}\ \bibinfo {author} {\bibfnamefont {M.~G.}\ \bibnamefont
  {Genoni}},\ }\bibfield  {title} {\emph {\enquote {\bibinfo {title} {Learning
  feedback control strategies for quantum metrology},}\ }}\href
  {https://doi.org/10.1103/prxquantum.3.020310} {\bibfield  {journal} {\bibinfo
   {journal} {PRX Quantum}\ }\textbf {\bibinfo {volume} {3}},\ \bibinfo {pages}
  {020310} (\bibinfo {year} {2022})}\BibitemShut {NoStop}%
\bibitem [{\citenamefont {Vaidhyanathan}\ \emph {et~al.}(2026)\citenamefont
  {Vaidhyanathan}, \citenamefont {Marquardt}, \citenamefont {Mitchison},\ and\
  \citenamefont {Ares}}]{Vaidhyanathan2024}%
  \BibitemOpen
  \bibfield  {author} {\bibinfo {author} {\bibfnamefont {P.}~\bibnamefont
  {Vaidhyanathan}}, \bibinfo {author} {\bibfnamefont {F.}~\bibnamefont
  {Marquardt}}, \bibinfo {author} {\bibfnamefont {M.~T.}\ \bibnamefont
  {Mitchison}},\ \bibnamefont {and}\ \bibinfo {author} {\bibfnamefont
  {N.}~\bibnamefont {Ares}},\ }\bibfield  {title} {\emph {\enquote {\bibinfo
  {title} {Quantum feedback control with a transformer neural network
  architecture},}\ }}\href {https://doi.org/10.1103/m429-jy1j} {\bibfield
  {journal} {\bibinfo  {journal} {Phys. Rev. Research}\ }\textbf {\bibinfo
  {volume} {8}},\ \bibinfo {pages} {L012043} (\bibinfo {year}
  {2026})}\BibitemShut {NoStop}%
\bibitem [{\citenamefont {Mingard}\ \emph {et~al.}(2021)\citenamefont
  {Mingard}, \citenamefont {Valle-P{\'e}rez}, \citenamefont {Skalse},\ and\
  \citenamefont {Louis}}]{Mingard2020}%
  \BibitemOpen
  \bibfield  {author} {\bibinfo {author} {\bibfnamefont {C.}~\bibnamefont
  {Mingard}}, \bibinfo {author} {\bibfnamefont {G.}~\bibnamefont
  {Valle-P{\'e}rez}}, \bibinfo {author} {\bibfnamefont {J.}~\bibnamefont
  {Skalse}},\ \bibnamefont {and}\ \bibinfo {author} {\bibfnamefont {A.~A.}\
  \bibnamefont {Louis}},\ }\bibfield  {title} {\emph {\enquote {\bibinfo
  {title} {Is {SGD} a {Bayesian} sampler? Well, almost},}\ }}\href@noop {}
  {\bibfield  {journal} {\bibinfo  {journal} {J. Mach. Learn. Res.}\ }\textbf
  {\bibinfo {volume} {22}},\ \bibinfo {pages} {1} (\bibinfo {year}
  {2021})}\BibitemShut {NoStop}%
\bibitem [{\citenamefont {Ruder}(2016)}]{Ruder2016}%
  \BibitemOpen
  \bibfield  {author} {\bibinfo {author} {\bibfnamefont {S.}~\bibnamefont
  {Ruder}},\ }\bibfield  {title} {\emph {\enquote {\bibinfo {title} {An
  overview of gradient descent optimization algorithms},}\ }}\bibfield
  {journal} {\bibinfo  {journal} {arXiv preprint}\ }\href
  {https://doi.org/10.48550/arXiv.1609.04747} {10.48550/arXiv.1609.04747}
  (\bibinfo {year} {2016}),\ \Eprint {https://arxiv.org/abs/1609.04747}
  {arXiv:1609.04747 [cs.LG]} \BibitemShut {NoStop}%
\bibitem [{\citenamefont {Kneissl}\ \emph {et~al.}(2025)\citenamefont
  {Kneissl}, \citenamefont {B{\"u}lte}, \citenamefont {Scholl},\ and\
  \citenamefont {Kutyniok}}]{Kneissl2025}%
  \BibitemOpen
  \bibfield  {author} {\bibinfo {author} {\bibfnamefont {C.}~\bibnamefont
  {Kneissl}}, \bibinfo {author} {\bibfnamefont {C.}~\bibnamefont {B{\"u}lte}},
  \bibinfo {author} {\bibfnamefont {P.}~\bibnamefont {Scholl}},\ \bibnamefont
  {and}\ \bibinfo {author} {\bibfnamefont {G.}~\bibnamefont {Kutyniok}},\
  }\bibfield  {title} {\emph {\enquote {\bibinfo {title} {Improved
  probabilistic regression using diffusion models},}\ }}in\ \href
  {https://openreview.net/forum?id=dUT5AgbIUT} {\emph {\bibinfo {booktitle}
  {Northern Lights Deep Learning Conference Abstracts 2026}}}\ (\bibinfo {year}
  {2025})\BibitemShut {NoStop}%
\bibitem [{\citenamefont {Nix}\ and\ \citenamefont {Weigend}(1994)}]{Nix1994}%
  \BibitemOpen
  \bibfield  {author} {\bibinfo {author} {\bibfnamefont {D.~A.}\ \bibnamefont
  {Nix}}\ \bibnamefont {and}\ \bibinfo {author} {\bibfnamefont {A.~S.}\
  \bibnamefont {Weigend}},\ }\bibfield  {title} {\emph {\enquote {\bibinfo
  {title} {Estimating the mean and variance of the target probability
  distribution},}\ }}in\ \href {https://doi.org/10.1109/icnn.1994.374138}
  {\emph {\bibinfo {booktitle} {Proceedings of 1994 IEEE International
  Conference on Neural Networks ({ICNN}'94)}}},\ Vol.~\bibinfo {volume} {1}\
  (\bibinfo {year} {1994})\ pp.\ \bibinfo {pages} {55--60}\BibitemShut
  {NoStop}%
\bibitem [{\citenamefont {Lakshminarayanan}\ \emph {et~al.}(2017)\citenamefont
  {Lakshminarayanan}, \citenamefont {Pritzel},\ and\ \citenamefont
  {Blundell}}]{Lakshminarayanan2017}%
  \BibitemOpen
  \bibfield  {author} {\bibinfo {author} {\bibfnamefont {B.}~\bibnamefont
  {Lakshminarayanan}}, \bibinfo {author} {\bibfnamefont {A.}~\bibnamefont
  {Pritzel}},\ \bibnamefont {and}\ \bibinfo {author} {\bibfnamefont
  {C.}~\bibnamefont {Blundell}},\ }\bibfield  {title} {\emph {\enquote
  {\bibinfo {title} {Simple and scalable predictive uncertainty estimation
  using deep ensembles},}\ }}in\ \href
  {https://proceedings.neurips.cc/paper_files/paper/2017/file/9ef2ed4b7fd2c810847ffa5fa85bce38-Paper.pdf}
  {\emph {\bibinfo {booktitle} {Advances in Neural Information Processing
  Systems}}},\ Vol.~\bibinfo {volume} {30},\ \bibinfo {editor} {edited by\
  \bibinfo {editor} {\bibfnamefont {I.}~\bibnamefont {Guyon}}, \bibinfo
  {editor} {\bibfnamefont {U.}~\bibnamefont {von Luxburg}}, \bibinfo {editor}
  {\bibfnamefont {S.}~\bibnamefont {Bengio}}, \bibinfo {editor} {\bibfnamefont
  {H.}~\bibnamefont {Wallach}}, \bibinfo {editor} {\bibfnamefont
  {R.}~\bibnamefont {Fergus}}, \bibinfo {editor} {\bibfnamefont
  {S.}~\bibnamefont {Vishwanathan}},\ \bibnamefont {and}\ \bibinfo {editor}
  {\bibfnamefont {R.}~\bibnamefont {Garnett}}}\ (\bibinfo  {publisher} {Curran
  Associates, Inc.},\ \bibinfo {year} {2017})\ pp.\ \bibinfo {pages}
  {6402--6413}\BibitemShut {NoStop}%
\bibitem [{\citenamefont {Hochreiter}\ and\ \citenamefont
  {Schmidhuber}(1997)}]{Hochreiter1997}%
  \BibitemOpen
  \bibfield  {author} {\bibinfo {author} {\bibfnamefont {S.}~\bibnamefont
  {Hochreiter}}\ \bibnamefont {and}\ \bibinfo {author} {\bibfnamefont
  {J.}~\bibnamefont {Schmidhuber}},\ }\bibfield  {title} {\emph {\enquote
  {\bibinfo {title} {Long short-term memory},}\ }}\href
  {https://doi.org/10.1162/neco.1997.9.8.1735} {\bibfield  {journal} {\bibinfo
  {journal} {Neural Comput.}\ }\textbf {\bibinfo {volume} {9}},\ \bibinfo
  {pages} {1735} (\bibinfo {year} {1997})}\BibitemShut {NoStop}%
\bibitem [{\citenamefont {Breuer}\ and\ \citenamefont
  {Petruccione}(2007)}]{Breuer2007}%
  \BibitemOpen
  \bibfield  {author} {\bibinfo {author} {\bibfnamefont {H.-P.}\ \bibnamefont
  {Breuer}}\ \bibnamefont {and}\ \bibinfo {author} {\bibfnamefont
  {F.}~\bibnamefont {Petruccione}},\ }\href
  {https://doi.org/10.1093/acprof:oso/9780199213900.001.0001} {\emph {\bibinfo
  {title} {The theory of open quantum systems}}}\ (\bibinfo  {publisher}
  {Oxford University Press},\ \bibinfo {address} {Oxford},\ \bibinfo {year}
  {2007})\BibitemShut {NoStop}%
\bibitem [{\citenamefont {Kronwald}\ \emph {et~al.}(2013)\citenamefont
  {Kronwald}, \citenamefont {Ludwig},\ and\ \citenamefont
  {Marquardt}}]{Kronwald2013}%
  \BibitemOpen
  \bibfield  {author} {\bibinfo {author} {\bibfnamefont {A.}~\bibnamefont
  {Kronwald}}, \bibinfo {author} {\bibfnamefont {M.}~\bibnamefont {Ludwig}},\
  \bibnamefont {and}\ \bibinfo {author} {\bibfnamefont {F.}~\bibnamefont
  {Marquardt}},\ }\bibfield  {title} {\emph {\enquote {\bibinfo {title} {Full
  photon statistics of a light beam transmitted through an optomechanical
  system},}\ }}\href {https://doi.org/10.1103/physreva.87.013847} {\bibfield
  {journal} {\bibinfo  {journal} {Phys. Rev. A}\ }\textbf {\bibinfo {volume}
  {87}},\ \bibinfo {pages} {013847} (\bibinfo {year} {2013})}\BibitemShut
  {NoStop}%
\bibitem [{\citenamefont {Guha}\ \emph {et~al.}(2020)\citenamefont {Guha},
  \citenamefont {Allain}, \citenamefont {Lema{\^\i}tre}, \citenamefont {Leo},\
  and\ \citenamefont {Favero}}]{Guha2020}%
  \BibitemOpen
  \bibfield  {author} {\bibinfo {author} {\bibfnamefont {B.}~\bibnamefont
  {Guha}}, \bibinfo {author} {\bibfnamefont {P.~E.}\ \bibnamefont {Allain}},
  \bibinfo {author} {\bibfnamefont {A.}~\bibnamefont {Lema{\^\i}tre}}, \bibinfo
  {author} {\bibfnamefont {G.}~\bibnamefont {Leo}},\ \bibnamefont {and}\
  \bibinfo {author} {\bibfnamefont {I.}~\bibnamefont {Favero}},\ }\bibfield
  {title} {\emph {\enquote {\bibinfo {title} {Force sensing with an
  optomechanical self-oscillator},}\ }}\href
  {https://doi.org/10.1103/physrevapplied.14.024079} {\bibfield  {journal}
  {\bibinfo  {journal} {Phys. Rev. Applied}\ }\textbf {\bibinfo {volume}
  {14}},\ \bibinfo {pages} {024079} (\bibinfo {year} {2020})}\BibitemShut
  {NoStop}%
\bibitem [{\citenamefont {H{\"a}lg}\ \emph {et~al.}(2021)\citenamefont
  {H{\"a}lg}, \citenamefont {Gisler}, \citenamefont {Tsaturyan}, \citenamefont
  {Catalini}, \citenamefont {Grob}, \citenamefont {Krass}, \citenamefont
  {H{\'e}ritier}, \citenamefont {Mattiat}, \citenamefont {Thamm}, \citenamefont
  {Schirhagl}, \citenamefont {Langman}, \citenamefont {Schliesser},
  \citenamefont {Degen},\ and\ \citenamefont {Eichler}}]{Haelg2021}%
  \BibitemOpen
  \bibfield  {author} {\bibinfo {author} {\bibfnamefont {D.}~\bibnamefont
  {H{\"a}lg}}, \bibnamefont {et~al.},\ }\bibfield  {title} {\emph {\enquote
  {\bibinfo {title} {Membrane-based scanning force microscopy},}\ }}\href
  {https://doi.org/10.1103/physrevapplied.15.l021001} {\bibfield  {journal}
  {\bibinfo  {journal} {Phys. Rev. Applied}\ }\textbf {\bibinfo {volume}
  {15}},\ \bibinfo {pages} {L021001} (\bibinfo {year} {2021})}\BibitemShut
  {NoStop}%
\bibitem [{\citenamefont {Kampel}\ \emph {et~al.}(2017)\citenamefont {Kampel},
  \citenamefont {Peterson}, \citenamefont {Fischer}, \citenamefont {Yu},
  \citenamefont {Cicak}, \citenamefont {Simmonds}, \citenamefont {Lehnert},\
  and\ \citenamefont {Regal}}]{Kampel2017}%
  \BibitemOpen
  \bibfield  {author} {\bibinfo {author} {\bibfnamefont {N.~S.}\ \bibnamefont
  {Kampel}}, \bibinfo {author} {\bibfnamefont {R.~W.}\ \bibnamefont
  {Peterson}}, \bibinfo {author} {\bibfnamefont {R.}~\bibnamefont {Fischer}},
  \bibinfo {author} {\bibfnamefont {P.-L.}\ \bibnamefont {Yu}}, \bibinfo
  {author} {\bibfnamefont {K.}~\bibnamefont {Cicak}}, \bibinfo {author}
  {\bibfnamefont {R.~W.}\ \bibnamefont {Simmonds}}, \bibinfo {author}
  {\bibfnamefont {K.~W.}\ \bibnamefont {Lehnert}},\ \bibnamefont {and}\
  \bibinfo {author} {\bibfnamefont {C.~A.}\ \bibnamefont {Regal}},\ }\bibfield
  {title} {\emph {\enquote {\bibinfo {title} {Improving broadband displacement
  detection with quantum correlations},}\ }}\href
  {https://doi.org/10.1103/physrevx.7.021008} {\bibfield  {journal} {\bibinfo
  {journal} {Phys. Rev. X}\ }\textbf {\bibinfo {volume} {7}},\ \bibinfo {pages}
  {021008} (\bibinfo {year} {2017})}\BibitemShut {NoStop}%
\bibitem [{\citenamefont {Mason}\ \emph {et~al.}(2019)\citenamefont {Mason},
  \citenamefont {Chen}, \citenamefont {Rossi}, \citenamefont {Tsaturyan},\ and\
  \citenamefont {Schliesser}}]{Mason2019}%
  \BibitemOpen
  \bibfield  {author} {\bibinfo {author} {\bibfnamefont {D.}~\bibnamefont
  {Mason}}, \bibinfo {author} {\bibfnamefont {J.}~\bibnamefont {Chen}},
  \bibinfo {author} {\bibfnamefont {M.}~\bibnamefont {Rossi}}, \bibinfo
  {author} {\bibfnamefont {Y.}~\bibnamefont {Tsaturyan}},\ \bibnamefont {and}\
  \bibinfo {author} {\bibfnamefont {A.}~\bibnamefont {Schliesser}},\ }\bibfield
   {title} {\emph {\enquote {\bibinfo {title} {Continuous force and
  displacement measurement below the standard quantum limit},}\ }}\href
  {https://doi.org/10.1038/s41567-019-0533-5} {\bibfield  {journal} {\bibinfo
  {journal} {Nat. Phys.}\ }\textbf {\bibinfo {volume} {15}},\ \bibinfo {pages}
  {745} (\bibinfo {year} {2019})}\BibitemShut {NoStop}%
\bibitem [{\citenamefont {Xia}\ \emph {et~al.}(2023)\citenamefont {Xia},
  \citenamefont {Agrawal}, \citenamefont {Pluchar}, \citenamefont {Brady},
  \citenamefont {Liu}, \citenamefont {Zhuang}, \citenamefont {Wilson},\ and\
  \citenamefont {Zhang}}]{Xia2023}%
  \BibitemOpen
  \bibfield  {author} {\bibinfo {author} {\bibfnamefont {Y.}~\bibnamefont
  {Xia}}, \bibinfo {author} {\bibfnamefont {A.~R.}\ \bibnamefont {Agrawal}},
  \bibinfo {author} {\bibfnamefont {C.~M.}\ \bibnamefont {Pluchar}}, \bibinfo
  {author} {\bibfnamefont {A.~J.}\ \bibnamefont {Brady}}, \bibinfo {author}
  {\bibfnamefont {Z.}~\bibnamefont {Liu}}, \bibinfo {author} {\bibfnamefont
  {Q.}~\bibnamefont {Zhuang}}, \bibinfo {author} {\bibfnamefont {D.~J.}\
  \bibnamefont {Wilson}},\ \bibnamefont {and}\ \bibinfo {author} {\bibfnamefont
  {Z.}~\bibnamefont {Zhang}},\ }\bibfield  {title} {\emph {\enquote {\bibinfo
  {title} {Entanglement-enhanced optomechanical sensing},}\ }}\href
  {https://doi.org/10.1038/s41566-023-01178-0} {\bibfield  {journal} {\bibinfo
  {journal} {Nat. Photonics}\ }\textbf {\bibinfo {volume} {17}},\ \bibinfo
  {pages} {470} (\bibinfo {year} {2023})}\BibitemShut {NoStop}%
\bibitem [{\citenamefont {Brady}\ \emph {et~al.}(2023)\citenamefont {Brady},
  \citenamefont {Chen}, \citenamefont {Xia}, \citenamefont {Manley},
  \citenamefont {Dey~Chowdhury}, \citenamefont {Xiao}, \citenamefont {Liu},
  \citenamefont {Harnik}, \citenamefont {Wilson}, \citenamefont {Zhang},\ and\
  \citenamefont {Zhuang}}]{Brady2023}%
  \BibitemOpen
  \bibfield  {author} {\bibinfo {author} {\bibfnamefont {A.~J.}\ \bibnamefont
  {Brady}}, \bibnamefont {et~al.},\ }\bibfield  {title} {\emph {\enquote
  {\bibinfo {title} {Entanglement-enhanced optomechanical sensor array with
  application to dark matter searches},}\ }}\href
  {https://doi.org/10.1038/s42005-023-01357-z} {\bibfield  {journal} {\bibinfo
  {journal} {Commun. Phys.}\ }\textbf {\bibinfo {volume} {6}},\ \bibinfo
  {pages} {237} (\bibinfo {year} {2023})}\BibitemShut {NoStop}%
\bibitem [{\citenamefont {Hebestreit}\ \emph {et~al.}(2018)\citenamefont
  {Hebestreit}, \citenamefont {Frimmer}, \citenamefont {Reimann},\ and\
  \citenamefont {Novotny}}]{Hebestreit2018}%
  \BibitemOpen
  \bibfield  {author} {\bibinfo {author} {\bibfnamefont {E.}~\bibnamefont
  {Hebestreit}}, \bibinfo {author} {\bibfnamefont {M.}~\bibnamefont {Frimmer}},
  \bibinfo {author} {\bibfnamefont {R.}~\bibnamefont {Reimann}},\ \bibnamefont
  {and}\ \bibinfo {author} {\bibfnamefont {L.}~\bibnamefont {Novotny}},\
  }\bibfield  {title} {\emph {\enquote {\bibinfo {title} {Sensing static forces
  with free-falling nanoparticles},}\ }}\href
  {https://doi.org/10.1103/physrevlett.121.063602} {\bibfield  {journal}
  {\bibinfo  {journal} {Phys. Rev. Lett.}\ }\textbf {\bibinfo {volume} {121}},\
  \bibinfo {pages} {063602} (\bibinfo {year} {2018})}\BibitemShut {NoStop}%
\bibitem [{\citenamefont {Tseng}\ \emph {et~al.}(2025)\citenamefont {Tseng},
  \citenamefont {Penny}, \citenamefont {Siegel}, \citenamefont {Wang},\ and\
  \citenamefont {Moore}}]{Tseng2025}%
  \BibitemOpen
  \bibfield  {author} {\bibinfo {author} {\bibfnamefont {Y.-H.}\ \bibnamefont
  {Tseng}}, \bibinfo {author} {\bibfnamefont {T.~W.}\ \bibnamefont {Penny}},
  \bibinfo {author} {\bibfnamefont {B.}~\bibnamefont {Siegel}}, \bibinfo
  {author} {\bibfnamefont {J.}~\bibnamefont {Wang}},\ \bibnamefont {and}\
  \bibinfo {author} {\bibfnamefont {D.~C.}\ \bibnamefont {Moore}},\ }\bibfield
  {title} {\emph {\enquote {\bibinfo {title} {Search for dark matter scattering
  from optically levitated nanoparticles},}\ }}\href
  {https://doi.org/10.1103/j76m-gcp1} {\bibfield  {journal} {\bibinfo
  {journal} {PRX Quantum}\ }\textbf {\bibinfo {volume} {6}},\ \bibinfo {pages}
  {040367} (\bibinfo {year} {2025})}\BibitemShut {NoStop}%
\bibitem [{\citenamefont {Krause}\ \emph {et~al.}(2012)\citenamefont {Krause},
  \citenamefont {Winger}, \citenamefont {Blasius}, \citenamefont {Lin},\ and\
  \citenamefont {Painter}}]{Krause2012}%
  \BibitemOpen
  \bibfield  {author} {\bibinfo {author} {\bibfnamefont {A.~G.}\ \bibnamefont
  {Krause}}, \bibinfo {author} {\bibfnamefont {M.}~\bibnamefont {Winger}},
  \bibinfo {author} {\bibfnamefont {T.~D.}\ \bibnamefont {Blasius}}, \bibinfo
  {author} {\bibfnamefont {Q.}~\bibnamefont {Lin}},\ \bibnamefont {and}\
  \bibinfo {author} {\bibfnamefont {O.}~\bibnamefont {Painter}},\ }\bibfield
  {title} {\emph {\enquote {\bibinfo {title} {A high-resolution microchip
  optomechanical accelerometer},}\ }}\href
  {https://doi.org/10.1038/nphoton.2012.245} {\bibfield  {journal} {\bibinfo
  {journal} {Nat. Photonics}\ }\textbf {\bibinfo {volume} {6}},\ \bibinfo
  {pages} {768} (\bibinfo {year} {2012})}\BibitemShut {NoStop}%
\bibitem [{\citenamefont {Guo}\ \emph {et~al.}(2017)\citenamefont {Guo},
  \citenamefont {Norte},\ and\ \citenamefont {Gr{\"o}blacher}}]{Guo2017}%
  \BibitemOpen
  \bibfield  {author} {\bibinfo {author} {\bibfnamefont {J.}~\bibnamefont
  {Guo}}, \bibinfo {author} {\bibfnamefont {R.~A.}\ \bibnamefont {Norte}},\
  \bibnamefont {and}\ \bibinfo {author} {\bibfnamefont {S.}~\bibnamefont
  {Gr{\"o}blacher}},\ }\bibfield  {title} {\emph {\enquote {\bibinfo {title}
  {Integrated optical force sensors using focusing photonic crystal arrays},}\
  }}\href {https://doi.org/10.1364/oe.25.009196} {\bibfield  {journal}
  {\bibinfo  {journal} {Opt. Express}\ }\textbf {\bibinfo {volume} {25}},\
  \bibinfo {pages} {9196} (\bibinfo {year} {2017})}\BibitemShut {NoStop}%
\bibitem [{\citenamefont {Hong}\ \emph {et~al.}(2017)\citenamefont {Hong},
  \citenamefont {Riedinger}, \citenamefont {Marinkovi{\'c}}, \citenamefont
  {Wallucks}, \citenamefont {Hofer}, \citenamefont {Norte}, \citenamefont
  {Aspelmeyer},\ and\ \citenamefont {Gr{\"o}blacher}}]{Hong2017}%
  \BibitemOpen
  \bibfield  {author} {\bibinfo {author} {\bibfnamefont {S.}~\bibnamefont
  {Hong}}, \bibinfo {author} {\bibfnamefont {R.}~\bibnamefont {Riedinger}},
  \bibinfo {author} {\bibfnamefont {I.}~\bibnamefont {Marinkovi{\'c}}},
  \bibinfo {author} {\bibfnamefont {A.}~\bibnamefont {Wallucks}}, \bibinfo
  {author} {\bibfnamefont {S.~G.}\ \bibnamefont {Hofer}}, \bibinfo {author}
  {\bibfnamefont {R.~A.}\ \bibnamefont {Norte}}, \bibinfo {author}
  {\bibfnamefont {M.}~\bibnamefont {Aspelmeyer}},\ \bibnamefont {and}\ \bibinfo
  {author} {\bibfnamefont {S.}~\bibnamefont {Gr{\"o}blacher}},\ }\bibfield
  {title} {\emph {\enquote {\bibinfo {title} {{Hanbury Brown} and {Twiss}
  interferometry of single phonons from an optomechanical resonator},}\ }}\href
  {https://doi.org/10.1126/science.aan7939} {\bibfield  {journal} {\bibinfo
  {journal} {Science}\ }\textbf {\bibinfo {volume} {358}},\ \bibinfo {pages}
  {203} (\bibinfo {year} {2017})}\BibitemShut {NoStop}%
\bibitem [{\citenamefont {Arjovsky}\ \emph {et~al.}(2017)\citenamefont
  {Arjovsky}, \citenamefont {Chintala},\ and\ \citenamefont
  {Bottou}}]{Arjovsky2017}%
  \BibitemOpen
  \bibfield  {author} {\bibinfo {author} {\bibfnamefont {M.}~\bibnamefont
  {Arjovsky}}, \bibinfo {author} {\bibfnamefont {S.}~\bibnamefont {Chintala}},\
  \bibnamefont {and}\ \bibinfo {author} {\bibfnamefont {L.}~\bibnamefont
  {Bottou}},\ }\bibfield  {title} {\emph {\enquote {\bibinfo {title}
  {{Wasserstein} generative adversarial networks},}\ }}in\ \href@noop {} {\emph
  {\bibinfo {booktitle} {Proceedings of the 34th International Conference on
  Machine Learning}}},\ \bibinfo {series} {Proceedings of Machine Learning
  Research}, Vol.~\bibinfo {volume} {70}\ (\bibinfo  {publisher} {PMLR},\
  \bibinfo {year} {2017})\ pp.\ \bibinfo {pages} {214--223}\BibitemShut
  {NoStop}%
\bibitem [{\citenamefont {Fearnhead}\ and\ \citenamefont
  {Prangle}(2012)}]{Fearnhead2010}%
  \BibitemOpen
  \bibfield  {author} {\bibinfo {author} {\bibfnamefont {P.}~\bibnamefont
  {Fearnhead}}\ \bibnamefont {and}\ \bibinfo {author} {\bibfnamefont
  {D.}~\bibnamefont {Prangle}},\ }\bibfield  {title} {\emph {\enquote {\bibinfo
  {title} {Constructing summary statistics for {Approximate Bayesian
  Computation}: Semi-automatic {Approximate Bayesian Computation}},}\ }}\href
  {https://doi.org/10.1111/j.1467-9868.2011.01010.x} {\bibfield  {journal}
  {\bibinfo  {journal} {J. R. Stat. Soc. Ser. B Stat. Methodol.}\ }\textbf
  {\bibinfo {volume} {74}},\ \bibinfo {pages} {419} (\bibinfo {year}
  {2012})}\BibitemShut {NoStop}%
\bibitem [{\citenamefont {Wong}\ \emph {et~al.}(2018)\citenamefont {Wong},
  \citenamefont {Jiang}, \citenamefont {Wu},\ and\ \citenamefont
  {Zheng}}]{Wong2018}%
  \BibitemOpen
  \bibfield  {author} {\bibinfo {author} {\bibfnamefont {W.}~\bibnamefont
  {Wong}}, \bibinfo {author} {\bibfnamefont {B.}~\bibnamefont {Jiang}},
  \bibinfo {author} {\bibfnamefont {T.-y.}\ \bibnamefont {Wu}},\ \bibnamefont
  {and}\ \bibinfo {author} {\bibfnamefont {C.}~\bibnamefont {Zheng}},\
  }\bibfield  {title} {\emph {\enquote {\bibinfo {title} {Learning summary
  statistic for {Approximate Bayesian Computation} via deep neural network},}\
  }}\bibfield  {journal} {\bibinfo  {journal} {Stat. Sin.}\ }\href
  {https://doi.org/10.5705/ss.202015.0340} {10.5705/ss.202015.0340} (\bibinfo
  {year} {2018})\BibitemShut {NoStop}%
\bibitem [{\citenamefont {{\AA}kesson}\ \emph {et~al.}(2022)\citenamefont
  {{\AA}kesson}, \citenamefont {Singh}, \citenamefont {Wrede},\ and\
  \citenamefont {Hellander}}]{Aakesson2020}%
  \BibitemOpen
  \bibfield  {author} {\bibinfo {author} {\bibfnamefont {M.}~\bibnamefont
  {{\AA}kesson}}, \bibinfo {author} {\bibfnamefont {P.}~\bibnamefont {Singh}},
  \bibinfo {author} {\bibfnamefont {F.}~\bibnamefont {Wrede}},\ \bibnamefont
  {and}\ \bibinfo {author} {\bibfnamefont {A.}~\bibnamefont {Hellander}},\
  }\bibfield  {title} {\emph {\enquote {\bibinfo {title} {Convolutional neural
  networks as summary statistics for {Approximate Bayesian Computation}},}\
  }}\href {https://doi.org/10.1109/tcbb.2021.3108695} {\bibfield  {journal}
  {\bibinfo  {journal} {IEEE/ACM Trans. Comput. Biol. Bioinform.}\ }\textbf
  {\bibinfo {volume} {19}},\ \bibinfo {pages} {3353} (\bibinfo {year}
  {2022})}\BibitemShut {NoStop}%
\bibitem [{\citenamefont {Macieszczak}\ \emph {et~al.}(2016)\citenamefont
  {Macieszczak}, \citenamefont {Gu{\c{t}}{\u{a}}}, \citenamefont {Lesanovsky},\
  and\ \citenamefont {Garrahan}}]{Macieszczak2016}%
  \BibitemOpen
  \bibfield  {author} {\bibinfo {author} {\bibfnamefont {K.}~\bibnamefont
  {Macieszczak}}, \bibinfo {author} {\bibfnamefont {M.}~\bibnamefont
  {Gu{\c{t}}{\u{a}}}}, \bibinfo {author} {\bibfnamefont {I.}~\bibnamefont
  {Lesanovsky}},\ \bibnamefont {and}\ \bibinfo {author} {\bibfnamefont {J.~P.}\
  \bibnamefont {Garrahan}},\ }\bibfield  {title} {\emph {\enquote {\bibinfo
  {title} {Dynamical phase transitions as a resource for quantum enhanced
  metrology},}\ }}\href {https://doi.org/10.1103/physreva.93.022103} {\bibfield
   {journal} {\bibinfo  {journal} {Phys. Rev. A}\ }\textbf {\bibinfo {volume}
  {93}},\ \bibinfo {pages} {022103} (\bibinfo {year} {2016})}\BibitemShut
  {NoStop}%
\bibitem [{\citenamefont {Yang}\ \emph {et~al.}(2026)\citenamefont {Yang},
  \citenamefont {Ketkar}, \citenamefont {Audenaert}, \citenamefont {Huelga},\
  and\ \citenamefont {Plenio}}]{Yang2025}%
  \BibitemOpen
  \bibfield  {author} {\bibinfo {author} {\bibfnamefont {D.}~\bibnamefont
  {Yang}}, \bibinfo {author} {\bibfnamefont {M.}~\bibnamefont {Ketkar}},
  \bibinfo {author} {\bibfnamefont {K.}~\bibnamefont {Audenaert}}, \bibinfo
  {author} {\bibfnamefont {S.~F.}\ \bibnamefont {Huelga}},\ \bibnamefont {and}\
  \bibinfo {author} {\bibfnamefont {M.~B.}\ \bibnamefont {Plenio}},\ }\bibfield
   {title} {\emph {\enquote {\bibinfo {title} {Quantum {Cram{\'e}r-Rao}
  precision limit of noisy continuous sensing},}\ }}\href
  {https://doi.org/10.1103/ltfw-4fwn} {\bibfield  {journal} {\bibinfo
  {journal} {Phys. Rev. Lett.}\ }\textbf {\bibinfo {volume} {136}},\ \bibinfo
  {pages} {070802} (\bibinfo {year} {2026})}\BibitemShut {NoStop}%
\bibitem [{\citenamefont {Barzanjeh}\ \emph {et~al.}(2022)\citenamefont
  {Barzanjeh}, \citenamefont {Xuereb}, \citenamefont {Gr{\"o}blacher},
  \citenamefont {Paternostro}, \citenamefont {Regal},\ and\ \citenamefont
  {Weig}}]{Barzanjeh2021}%
  \BibitemOpen
  \bibfield  {author} {\bibinfo {author} {\bibfnamefont {S.}~\bibnamefont
  {Barzanjeh}}, \bibinfo {author} {\bibfnamefont {A.}~\bibnamefont {Xuereb}},
  \bibinfo {author} {\bibfnamefont {S.}~\bibnamefont {Gr{\"o}blacher}},
  \bibinfo {author} {\bibfnamefont {M.}~\bibnamefont {Paternostro}}, \bibinfo
  {author} {\bibfnamefont {C.~A.}\ \bibnamefont {Regal}},\ \bibnamefont {and}\
  \bibinfo {author} {\bibfnamefont {E.~M.}\ \bibnamefont {Weig}},\ }\bibfield
  {title} {\emph {\enquote {\bibinfo {title} {Optomechanics for quantum
  technologies},}\ }}\href {https://doi.org/10.1038/s41567-021-01402-0}
  {\bibfield  {journal} {\bibinfo  {journal} {Nat. Phys.}\ }\textbf {\bibinfo
  {volume} {18}},\ \bibinfo {pages} {15} (\bibinfo {year} {2022})}\BibitemShut
  {NoStop}%
\bibitem [{\citenamefont {Cripe}\ \emph {et~al.}(2019)\citenamefont {Cripe},
  \citenamefont {Aggarwal}, \citenamefont {Lanza}, \citenamefont {Libson},
  \citenamefont {Singh}, \citenamefont {Heu}, \citenamefont {Follman},
  \citenamefont {Cole}, \citenamefont {Mavalvala},\ and\ \citenamefont
  {Corbitt}}]{Cripe2019}%
  \BibitemOpen
  \bibfield  {author} {\bibinfo {author} {\bibfnamefont {J.}~\bibnamefont
  {Cripe}}, \bibinfo {author} {\bibfnamefont {N.}~\bibnamefont {Aggarwal}},
  \bibinfo {author} {\bibfnamefont {R.}~\bibnamefont {Lanza}}, \bibinfo
  {author} {\bibfnamefont {A.}~\bibnamefont {Libson}}, \bibinfo {author}
  {\bibfnamefont {R.}~\bibnamefont {Singh}}, \bibinfo {author} {\bibfnamefont
  {P.}~\bibnamefont {Heu}}, \bibinfo {author} {\bibfnamefont {D.}~\bibnamefont
  {Follman}}, \bibinfo {author} {\bibfnamefont {G.~D.}\ \bibnamefont {Cole}},
  \bibinfo {author} {\bibfnamefont {N.}~\bibnamefont {Mavalvala}},\
  \bibnamefont {and}\ \bibinfo {author} {\bibfnamefont {T.}~\bibnamefont
  {Corbitt}},\ }\bibfield  {title} {\emph {\enquote {\bibinfo {title}
  {Measurement of quantum back action in the audio band at room temperature},}\
  }}\href {https://doi.org/10.1038/s41586-019-1051-4} {\bibfield  {journal}
  {\bibinfo  {journal} {Nature}\ }\textbf {\bibinfo {volume} {568}},\ \bibinfo
  {pages} {364} (\bibinfo {year} {2019})}\BibitemShut {NoStop}%
\bibitem [{\citenamefont {Belliardo}\ \emph
  {et~al.}(2024{\natexlab{a}})\citenamefont {Belliardo}, \citenamefont
  {Zoratti},\ and\ \citenamefont {Giovannetti}}]{Belliardo2024pra}%
  \BibitemOpen
  \bibfield  {author} {\bibinfo {author} {\bibfnamefont {F.}~\bibnamefont
  {Belliardo}}, \bibinfo {author} {\bibfnamefont {F.}~\bibnamefont {Zoratti}},\
  \bibnamefont {and}\ \bibinfo {author} {\bibfnamefont {V.}~\bibnamefont
  {Giovannetti}},\ }\bibfield  {title} {\emph {\enquote {\bibinfo {title}
  {Applications of model-aware reinforcement learning in {Bayesian} quantum
  metrology},}\ }}\href {https://doi.org/10.1103/physreva.109.062609}
  {\bibfield  {journal} {\bibinfo  {journal} {Phys. Rev. A}\ }\textbf {\bibinfo
  {volume} {109}},\ \bibinfo {pages} {062609} (\bibinfo {year}
  {2024}{\natexlab{a}})}\BibitemShut {NoStop}%
\bibitem [{\citenamefont {Belliardo}\ \emph
  {et~al.}(2024{\natexlab{b}})\citenamefont {Belliardo}, \citenamefont
  {Zoratti}, \citenamefont {Marquardt},\ and\ \citenamefont
  {Giovannetti}}]{Belliardo2024quantum}%
  \BibitemOpen
  \bibfield  {author} {\bibinfo {author} {\bibfnamefont {F.}~\bibnamefont
  {Belliardo}}, \bibinfo {author} {\bibfnamefont {F.}~\bibnamefont {Zoratti}},
  \bibinfo {author} {\bibfnamefont {F.}~\bibnamefont {Marquardt}},\
  \bibnamefont {and}\ \bibinfo {author} {\bibfnamefont {V.}~\bibnamefont
  {Giovannetti}},\ }\bibfield  {title} {\emph {\enquote {\bibinfo {title}
  {Model-aware reinforcement learning for high-performance {Bayesian}
  experimental design in quantum metrology},}\ }}\href
  {https://doi.org/10.22331/q-2024-12-10-1555} {\bibfield  {journal} {\bibinfo
  {journal} {Quantum}\ }\textbf {\bibinfo {volume} {8}},\ \bibinfo {pages}
  {1555} (\bibinfo {year} {2024}{\natexlab{b}})}\BibitemShut {NoStop}%
\bibitem [{\citenamefont {Qvarfort}\ \emph {et~al.}(2018)\citenamefont
  {Qvarfort}, \citenamefont {Serafini}, \citenamefont {Barker},\ and\
  \citenamefont {Bose}}]{Qvarfort2018}%
  \BibitemOpen
  \bibfield  {author} {\bibinfo {author} {\bibfnamefont {S.}~\bibnamefont
  {Qvarfort}}, \bibinfo {author} {\bibfnamefont {A.}~\bibnamefont {Serafini}},
  \bibinfo {author} {\bibfnamefont {P.~F.}\ \bibnamefont {Barker}},\
  \bibnamefont {and}\ \bibinfo {author} {\bibfnamefont {S.}~\bibnamefont
  {Bose}},\ }\bibfield  {title} {\emph {\enquote {\bibinfo {title} {Gravimetry
  through non-linear optomechanics},}\ }}\href
  {https://doi.org/10.1038/s41467-018-06037-z} {\bibfield  {journal} {\bibinfo
  {journal} {Nat. Commun.}\ }\textbf {\bibinfo {volume} {9}},\ \bibinfo {pages}
  {3690} (\bibinfo {year} {2018})}\BibitemShut {NoStop}%
\bibitem [{\citenamefont {Van~Trees}(1968)}]{Trees1968}%
  \BibitemOpen
  \bibfield  {author} {\bibinfo {author} {\bibfnamefont {H.~L.}\ \bibnamefont
  {Van~Trees}},\ }\href@noop {} {\emph {\bibinfo {title} {Detection,
  estimation, and modulation theory, part {I}}}}\ (\bibinfo  {publisher}
  {Wiley},\ \bibinfo {address} {New York},\ \bibinfo {year} {1968})\BibitemShut
  {NoStop}%
\bibitem [{\citenamefont {Radaelli}\ \emph {et~al.}(2024)\citenamefont
  {Radaelli}, \citenamefont {Landi},\ and\ \citenamefont
  {Binder}}]{Radaelli2024b}%
  \BibitemOpen
  \bibfield  {author} {\bibinfo {author} {\bibfnamefont {M.}~\bibnamefont
  {Radaelli}}, \bibinfo {author} {\bibfnamefont {G.~T.}\ \bibnamefont
  {Landi}},\ \bibnamefont {and}\ \bibinfo {author} {\bibfnamefont {F.~C.}\
  \bibnamefont {Binder}},\ }\bibfield  {title} {\emph {\enquote {\bibinfo
  {title} {{Gillespie} algorithm for quantum jump trajectories},}\ }}\href
  {https://doi.org/10.1103/physreva.110.062212} {\bibfield  {journal} {\bibinfo
   {journal} {Phys. Rev. A}\ }\textbf {\bibinfo {volume} {110}},\ \bibinfo
  {pages} {062212} (\bibinfo {year} {2024})}\BibitemShut {NoStop}%
\bibitem [{\citenamefont {Gammelmark}\ and\ \citenamefont
  {M{\o}lmer}(2014)}]{Gammelmark2014}%
  \BibitemOpen
  \bibfield  {author} {\bibinfo {author} {\bibfnamefont {S.}~\bibnamefont
  {Gammelmark}}\ \bibnamefont {and}\ \bibinfo {author} {\bibfnamefont
  {K.}~\bibnamefont {M{\o}lmer}},\ }\bibfield  {title} {\emph {\enquote
  {\bibinfo {title} {{Fisher} information and the quantum {Cram{\'e}r-Rao}
  sensitivity limit of continuous measurements},}\ }}\href
  {https://doi.org/10.1103/physrevlett.112.170401} {\bibfield  {journal}
  {\bibinfo  {journal} {Phys. Rev. Lett.}\ }\textbf {\bibinfo {volume} {112}},\
  \bibinfo {pages} {170401} (\bibinfo {year} {2014})}\BibitemShut {NoStop}%
\bibitem [{\citenamefont {Ilias}\ \emph {et~al.}(2022)\citenamefont {Ilias},
  \citenamefont {Yang}, \citenamefont {Huelga},\ and\ \citenamefont
  {Plenio}}]{Ilias2022}%
  \BibitemOpen
  \bibfield  {author} {\bibinfo {author} {\bibfnamefont {T.}~\bibnamefont
  {Ilias}}, \bibinfo {author} {\bibfnamefont {D.}~\bibnamefont {Yang}},
  \bibinfo {author} {\bibfnamefont {S.~F.}\ \bibnamefont {Huelga}},\
  \bibnamefont {and}\ \bibinfo {author} {\bibfnamefont {M.~B.}\ \bibnamefont
  {Plenio}},\ }\bibfield  {title} {\emph {\enquote {\bibinfo {title}
  {Criticality-enhanced quantum sensing via continuous measurement},}\ }}\href
  {https://doi.org/10.1103/prxquantum.3.010354} {\bibfield  {journal} {\bibinfo
   {journal} {PRX Quantum}\ }\textbf {\bibinfo {volume} {3}},\ \bibinfo {pages}
  {010354} (\bibinfo {year} {2022})}\BibitemShut {NoStop}%
\bibitem [{\citenamefont {Yang}\ \emph {et~al.}(2023)\citenamefont {Yang},
  \citenamefont {Huelga},\ and\ \citenamefont {Plenio}}]{Yang2023}%
  \BibitemOpen
  \bibfield  {author} {\bibinfo {author} {\bibfnamefont {D.}~\bibnamefont
  {Yang}}, \bibinfo {author} {\bibfnamefont {S.~F.}\ \bibnamefont {Huelga}},\
  \bibnamefont {and}\ \bibinfo {author} {\bibfnamefont {M.~B.}\ \bibnamefont
  {Plenio}},\ }\bibfield  {title} {\emph {\enquote {\bibinfo {title} {Efficient
  information retrieval for sensing via continuous measurement},}\ }}\href
  {https://doi.org/10.1103/physrevx.13.031012} {\bibfield  {journal} {\bibinfo
  {journal} {Phys. Rev. X}\ }\textbf {\bibinfo {volume} {13}},\ \bibinfo
  {pages} {031012} (\bibinfo {year} {2023})}\BibitemShut {NoStop}%
\bibitem [{\citenamefont {Cirac}\ \emph {et~al.}(2021)\citenamefont {Cirac},
  \citenamefont {P{\'e}rez-Garc{\'\i}a}, \citenamefont {Schuch},\ and\
  \citenamefont {Verstraete}}]{Cirac2021}%
  \BibitemOpen
  \bibfield  {author} {\bibinfo {author} {\bibfnamefont {J.~I.}\ \bibnamefont
  {Cirac}}, \bibinfo {author} {\bibfnamefont {D.}~\bibnamefont
  {P{\'e}rez-Garc{\'\i}a}}, \bibinfo {author} {\bibfnamefont {N.}~\bibnamefont
  {Schuch}},\ \bibnamefont {and}\ \bibinfo {author} {\bibfnamefont
  {F.}~\bibnamefont {Verstraete}},\ }\bibfield  {title} {\emph {\enquote
  {\bibinfo {title} {Matrix product states and projected entangled pair states:
  Concepts, symmetries, theorems},}\ }}\href
  {https://doi.org/10.1103/revmodphys.93.045003} {\bibfield  {journal}
  {\bibinfo  {journal} {Rev. Mod. Phys.}\ }\textbf {\bibinfo {volume} {93}},\
  \bibinfo {pages} {045003} (\bibinfo {year} {2021})}\BibitemShut {NoStop}%
\bibitem [{git()}]{github}%
  \BibitemOpen
  \href@noop {} {\enquote {\bibinfo {title}
  {Nonlinear-optomechanical-parameter-estimation ({GitHub} repository)},}\
  }\bibinfo {howpublished}
  {\url{https://github.com/QI2-lab/Nonlinear-optomechanical-parameter-estimation}}\BibitemShut
  {NoStop}%
\bibitem [{\citenamefont {Grimmett}\ and\ \citenamefont
  {Stirzaker}(2001)}]{Grimmett2001}%
  \BibitemOpen
  \bibfield  {author} {\bibinfo {author} {\bibfnamefont {G.~R.}\ \bibnamefont
  {Grimmett}}\ \bibnamefont {and}\ \bibinfo {author} {\bibfnamefont {D.~R.}\
  \bibnamefont {Stirzaker}},\ }\href
  {https://doi.org/10.1093/oso/9780198572237.001.0001} {\emph {\bibinfo {title}
  {Probability and random processes}}},\ \bibinfo {edition} {3rd}\ ed.\
  (\bibinfo  {publisher} {Oxford University Press},\ \bibinfo {address}
  {Oxford},\ \bibinfo {year} {2001})\BibitemShut {NoStop}%
\bibitem [{\citenamefont {Wackerly}\ \emph {et~al.}(2001)\citenamefont
  {Wackerly}, \citenamefont {Mendenhall},\ and\ \citenamefont
  {Scheaffer}}]{Wackerly2001}%
  \BibitemOpen
  \bibfield  {author} {\bibinfo {author} {\bibfnamefont {D.~D.}\ \bibnamefont
  {Wackerly}}, \bibinfo {author} {\bibfnamefont {W.}~\bibnamefont
  {Mendenhall}},\ \bibnamefont {and}\ \bibinfo {author} {\bibfnamefont {R.~L.}\
  \bibnamefont {Scheaffer}},\ }\href@noop {} {\emph {\bibinfo {title}
  {Mathematical statistics with applications}}}\ (\bibinfo  {publisher}
  {Thomson Brooks/Cole},\ \bibinfo {year} {2001})\BibitemShut {NoStop}%
\end{thebibliography}%

\end{document}